\documentclass[%
 reprint,
 superscriptaddress,
 amsmath,amssymb,
 aps,
 prx,
floatfix
]{revtex4-2}
\setcitestyle{super}
\usepackage[table, dvipsnames]{xcolor}
\usepackage{newtxtext,newtxmath} % alternative to times

\usepackage{graphicx}% Include figure files
\usepackage{dcolumn}% Align table columns on decimal point
\usepackage{bm}% bold math
\usepackage[colorlinks=true, linkcolor = blue, urlcolor  = blue, citecolor = blue, anchorcolor = blue]{hyperref}
\usepackage{amssymb}
\usepackage{enumitem}   
\usepackage{siunitx}
\usepackage{braket}
\usepackage{upgreek} % For for non-italic greek letters
\usepackage{comment}
\usepackage{color} % For using colors in the correction of the manuscript
\definecolor{blue}{rgb}{0,0,0.8}
\definecolor{black}{rgb}{0,0,0}
\definecolor{red}{rgb}{0.8,0,0}
\definecolor{green}{rgb}{0,0.4,0}

\definecolor{violet}{rgb}{0.8,0.2,0.8}
\definecolor{darkgreen}{RGB}{9, 90, 6}
\newcommand{\shorteq}{%
  \settowidth{\@tempdima}{-}% Width of hyphen
  \resizebox{\@tempdima}{\height}{=}%
}
\definecolor{darkPurple}{RGB}{150,0,150}
\definecolor{lightBrown}{rgb}{0.7,0.7,0.5}
\usepackage{lineno}
\usepackage{xspace}
\usepackage{multirow}
\usepackage{xfrac}
\usepackage{tensor}

\usepackage[tikz]{bclogo}

\usepackage{float}

\newcommand\scalemath[2]{\scalebox{#1}{\mbox{\ensuremath{\displaystyle #2}}}}

\renewcommand{\d}{\mathrm{d}}

\newcommand{\n}{\hat{\mathbf{n}}}

\newcommand{\g}{\hat{\mathbf{g}}}
\newcommand{\q}{{\mathbf{q}}}
\newcommand{\Gg}{{\mathbf{g}}}
\newcommand{\z}{\hat{\mathbf{z}}}
\newcommand{\V}{\mathbf{v}}
\newcommand{\e}{\hat{\mathbf{e}}}

\newcommand{\Y}{\mathcal{Y}}
\renewcommand{\P}{\mathcal{P}}
\newcommand{\B}{\mathsf{B}}
\newcommand{\D}{\mathsf{D}}
\newcommand{\W}{\mathsf{W}}
\newcommand{\C}{\mathsf{C}}
\renewcommand{\S}{\mathsf{S}}
\newcommand{\A}{\mathsf{A}}
\renewcommand{\L}{\mathcal{L}}
\renewcommand{\O}{\mathcal{O}}
\newcommand{\T}{\mathsf{T}}
\newcommand{\R}{\mathcal{R}}
\newcommand{\Q}{\mathsf{Q}}
\newcommand{\F}{\mathsf{F}}

\newcommand{\Sl}{\mathsf{S}^{(\ell)}}
\newcommand{\Sz}{\mathsf{S}^{(0)}}
\newcommand{\St}{\mathsf{S}^{(2)}}
\newcommand{\Sf}{\mathsf{S}^{(4)}}

\newcommand{\Dz}{\mathsf{D}^{(0)}}
\newcommand{\Dt}{\mathsf{D}^{(2)}}
\newcommand{\Az}{\mathsf{A}^{(0)}}
\newcommand{\At}{\mathsf{A}^{(2)}}
\newcommand{\Al}{\mathsf{A}^{(\ell)}}
\newcommand{\Dc}{{D}}

\newcommand{\Dcz}{D^{(0)}}
\newcommand{\Dct}{D^{(2)}}
\newcommand{\E}{\mathsf{E}}
\newcommand{\Tz}{\mathsf{T}^{(0)}}
\newcommand{\Tt}{\mathsf{T}^{(2)}}
\newcommand{\Tf}{\mathsf{T}^{(4)}}
\newcommand{\Tl}{\mathsf{T}^{(\ell)}}
\newcommand{\Qz}{\mathsf{Q}^{(0)}}
\newcommand{\Qt}{\mathsf{Q}^{(2)}}
\newcommand{\Ql}{\mathsf{Q}^{(\ell)}}
\newcommand{\Smu}{\mathsf{S}_\mu}

\newcommand{\iDt}[1]{\D_{2|{#1}}}

\newcommand{\iSf}[1]{\S_{4|{#1}}}

\newcommand{\Sig}{\mathcal{S}}

\newcommand{\mFA}{$\upmu$FA\xspace}

\newcommand{\la}{\left<}
\newcommand{\ra}{\right>}
\newcommand{\lla}{\left<\!\left<}
\newcommand{\rra}{\right>\!\right>}
\newcommand{\lb}{\left[}
\newcommand{\rb}{\right]}
\newcommand{\lp}{\left(}
\newcommand{\rp}{\right)}
\DeclareMathSymbol{\shortminus}{\mathbin}{AMSa}{"39}
\renewcommand{\unit}[1]{\xspace\mathrm{#1}\xspace}
\newcommand\tr{\mathop{\xspace\text{tr}\xspace}}

\usepackage{xcolor}
\usepackage{marginnote}

\newif\ifannotated

\annotatedfalse                 % default = CLEAN
\ifdefined\ANNOTATED
\annotatedtrue
\fi

\definecolor{rvblue}{RGB}{0,80,200}
\newcommand{\rvbadge}[1]{%
	\textsf{\footnotesize\color{rvblue}\fboxsep=.2ex\fbox{\textbf{#1}}}%
}

\newcommand{\rv}[2][]{%
	\ifannotated
	\marginnote{\rvbadge{#1}\;{\scriptsize\color{rvblue}#2}}[0cm]%
	\fi
}

\newcommand{\rvleft}[2][]{%
	\ifannotated
	{\reversemarginpar
		\marginnote{\rvbadge{#1}\;{\scriptsize\color{rvblue}#2}}[0cm]}%
	\fi
}

\ifannotated
\newcommand{\update}[1]{{\color{red}#1}}
\else
\newcommand{\update}[1]{{\color{black}#1}}
\fi

\begin{document}
	
\ifannotated
% restart numbering each page
\pagewiselinenumbers
% alternate margins
%\switchlinenumbers
\linenumbers
\fi

\title{{Geometry of the cumulant series in diffusion MRI}}

\author{Santiago Coelho}
\affiliation{Center for Biomedical Imaging and Center for Advanced Imaging Innovation and Research (CAI$\,^2\!$R), Department of Radiology, New York University School of Medicine, New York, USA}
\author{Jenny Chen}
\affiliation{Center for Biomedical Imaging and Center for Advanced Imaging Innovation and Research (CAI$\,^2\!$R), Department of Radiology, New York University School of Medicine, New York, USA}
\author{Filip Szczepankiewicz}
\affiliation{Department of Medical Radiation Physics, Lund University, Lund, Sweden}%
\author{Els Fieremans}%
\affiliation{Center for Biomedical Imaging and Center for Advanced Imaging Innovation and Research (CAI$\,^2\!$R), Department of Radiology, New York University School of Medicine, New York, USA}
\author{Dmitry S. Novikov}
\affiliation{Center for Biomedical Imaging and Center for Advanced Imaging Innovation and Research (CAI$\,^2\!$R), Department of Radiology, New York University School of Medicine, New York, USA}

% Today's date
%\date{\today}

\begin{abstract}
\noindent
%Water diffusion gives rise to micron-scale sensitivity of diffusion MRI (dMRI) to cellular-level tissue structure. 
%Precision medicine and quantitative imaging depend on uncovering the information content of dMRI and establishing its parsimonious hardware-independent fingerprint.
%\update{Based on the rotational SO(3) symmetry, we study the geometry of dMRI signal and the topology of its acquisition. 
%We identify irreducible components and a full set of invariants for the cumulant tensors}, 
%and relate them to tissue properties. Including all kurtosis  invariants improves multiple sclerosis classification in a cohort of 1189 subjects. 
%We design the shortest acquisitions based on icosahedral vertices to determine the most used invariants in only 1--2 minutes for the whole brain. Representing dMRI  via scalar invariant maps with definite symmetries will underpin machine learning classifiers of  pathology, development, and aging, while fast protocols will enable translation of advanced dMRI into clinic.
Water diffusion gives rise to micron-scale sensitivity of diffusion MRI (dMRI) to cellular-level tissue structure. Precision medicine and quantitative imaging depend on uncovering the information content of dMRI and establishing its parsimonious hardware-independent fingerprint. \update{Based on the rotational SO(3) symmetry, we study the geometry of the dMRI signal and the topology of its acquisition, identify irreducible components and a full set of invariants for the cumulant tensors}, and relate them to tissue properties. Including all kurtosis invariants improves multiple sclerosis classification in a cohort of 1189 subjects. We design the shortest acquisitions based on icosahedral vertices to determine the most used invariants in only 1–2 minutes for whole brain. Representing dMRI via scalar invariant maps with definite symmetries will underpin machine learning classifiers of pathology, development, and aging, while fast protocols will enable translation of advanced dMRI into clinic.
\end{abstract}

% \begin{description}
% Secondary publications and information retrieval purposes.
% You may use the \texttt{description} environment to structure your abstract;
% use the optional argument of the \verb+\item+ command to give the category of each item. 
% \end{description}

\maketitle

%\tableofcontents

% \section{\label{sec:level1}First-level heading:\protect\\ The line break was forced \lowercase{via} \textbackslash\textbackslash}

%\section*{Introduction}\label{s:Intro}
\noindent
Diffusion NMR or MRI (dMRI) measures a propagator of micrometer-scale displacements of spin-carrying molecules in an NMR sample or in every imaging voxel \cite{CALLAGHAN1991,JONES2010}. 
Porous media \cite{CORY1990,CALLAGHAN1991} and biological tissues \cite{TANNER1979,STANISZ1997}  can be  noninvasively probed by quantifying diffusion propagator for water molecules. 
Mapping its lowest-order cumulant, the voxel-wise diffusion tensor $\D$, via diffusion tensor imaging (DTI) \cite{BASSER1994}, has become an integral part of most human brain MRI clinical protocols and research studies. 
The information-rich signal beyond DTI forms the foundation of tissue microstructure imaging  \cite{JELESCU2017,NOVIKOV2019,ALEXANDER2019,NOVIKOV2021,KISELEV2021,WEISKOPF2021,LAMPINEN2023} --- a combination of biophysics, condensed matter physics and bioengineering. This allows MRI to become specific to disease processes at the scale of cells and organelles, and to provide non-invasive markers of development, aging and pathology \citep{ASSAF2008b,NILSSON2018}.

%%%%%%%%%%%%%%%%%%%%%%%%%%%%%%%%%%%%%%%%%%%%%%%%%%%%
\begin{figure*}[ht!!]
\centering
\includegraphics[width=\textwidth]{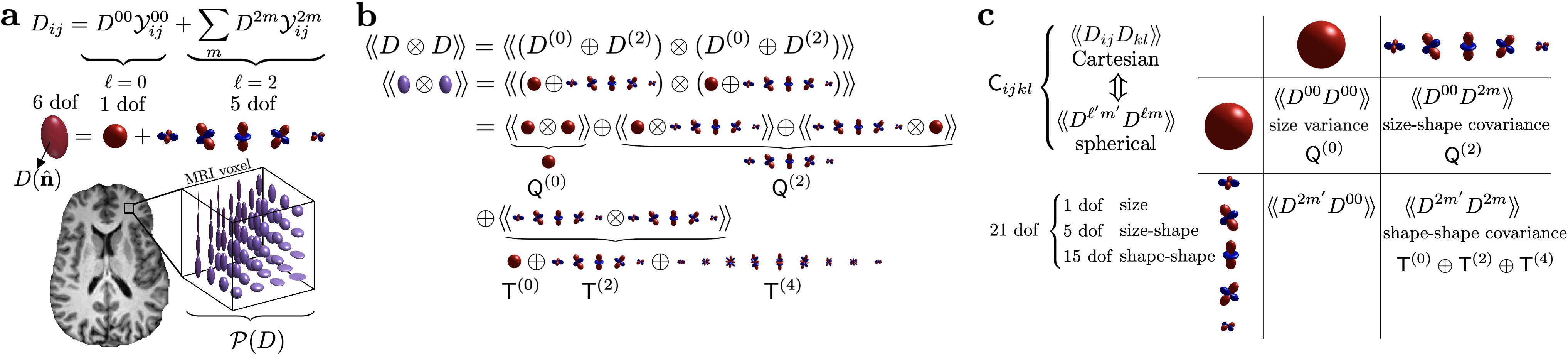}
\caption[caption FIG]{Outline: 
(a) An MRI voxel is represented via the distribution $\P(\Dc)$ of compartment diffusion tensors $\Dc$ decomposed in the spherical tensor basis. 
(b) The QT decomposition of the covariance tensor $\C$ arises from the addition of ``angular momenta'', Eq.~(\ref{DxD}): 
Central moments (\ref{eq:DCdefinitions}) of compartment tensors $\Dc$ correspond to tensor products,  subsequently averaged over the voxel-wise distribution $\P(\Dc)$.  
(c) Irreducible components of the QT decomposition represent 1 size-size, 5 size-shape, and $1+5+9$ shape-shape covariances in the 21 DOF of $\C$.}
\label{fig:outline}
\end{figure*}
%%%%%%%%%%%%%%%%%%%%%%%%%%%%%%%%%%%%%%%%%%%%%%%%%%%%

The information content of dMRI signal depends on the degree of coarse-graining \citep{NOVIKOV2019,NOVIKOV2021} over the diffusion length controlled by the diffusion time $t$. 
%\dn{changes: well-defined; perivascular}
For  long $t$, each {\it compartment} (a well-defined water population, e.g., intra-axonal, extra-cellular, perivascular, 
etc) can be  considered as fully coarse-grained, such that diffusion within  
it becomes Gaussian \cite{NOVIKOV2014,KISELEV2017,NOVIKOV2019}. 
This is typically the case in brain dMRI, with $t\gtrsim 50\,\unit{ms}$ in human scans. 
%Assuming (anisotropic) Gaussian diffusion in each compartment drastically simplifies the analysis. Indeed, 
The signal $\Sig=\exp(-\tr \B\Dc)$  (normalized to 1 in the absence of diffusion weighting) from any such ``Gaussian compartment'' defined by its diffusion tensor $\Dc$, becomes fully encoded by a $3\times 3$ symmetric $\B$-tensor \citep{CORY1990,MITRA1995,MORI1995,SHEMESH2015,WESTIN2016,TOPGAARD2017}. 
The signal $\la \Sig \ra$  from a distribution of Gaussian compartments within a voxel 
is then represented by the cumulant expansion \citep{VANKAMPEN1981,KISELEV2010,WESTIN2016}
\begin{equation}\label{eq:CumExpinB}
\ln \la \Sig\ra = \ln  \la e^{-\B_{ij} \Dc_{ij}} \ra = - \B_{ij} \,\D_{ij} + \tfrac12 \B_{ij} \B_{kl} \C_{ijkl} + {\cal O}(b^3) \,,
\end{equation}
where  $b=\tr\B$ is the so-called $b$-value\cite{JONES2010}, the summation over repeated Cartesian indices is assumed hereon, and 
\begin{equation}\label{eq:DCdefinitions}
\begin{aligned}
\D_{ij} &=   \la \Dc_{ij} \ra ,
\\  
\C_{ijkl} &= \lla \Dc_{ij} \, \Dc_{kl} \rra \equiv 
\la \vphantom{a^2}\! \lp \Dc_{ij} - \langle \Dc_{ij}\rangle \rp  \lp \vphantom{\la \Dc_{ij}\ra}\Dc_{kl} -\la \Dc_{kl}\ra\rp \ra 
\end{aligned}
\end{equation}
are the voxel-wise diffusion and covariance tensors. 
% $\D$ and $\C$.  
The averages $\la\dots\ra$ are taken over the distribution $\P(\Dc)$ of diffusion tensors % in a  voxel  
(Fig.~\ref{fig:outline}a), while $\lla \dots \rra$ denote its cumulants \citep{VANKAMPEN1981}.

\update{For shorter $t$, e.g., relevant for human body dMRI, \rvleft[R2]{}
the coarse-graining is incomplete, and diffusion in individual compartments remains non-Gaussian, characterized by time-dependent tensors $D(t)$ and higher-order cumulants, such as the kurtosis tensor $S(t)$.\cite{NOVIKOV2019} 
The $\B$-tensor then does not fully specify the measurement. Instead, the time-dependent cumulant tensors of the form (\ref{eq:DCdefinitions}), 
$\D(t)=\la \Dc(t)\ra$ and $\C(t)=\textstyle{\lla \Dc(t) \otimes \Dc(t)\rra}$, as well as the overall microscopic kurtosis tensor $\Smu(t)$, can be accessed \cite{JESPERSEN2013,NETOHENRIQUES2020} with double diffusion encoding (DDE)\cite{SHEMESH2015}. }
In any case, understanding the nature of tensors of the form (\ref{eq:DCdefinitions}) has become indispensable for dMRI.

%In general, the {\it time-dependent} $\D_{ij}(t)=\la \Dc_{ij}(t)\ra$ and $\C_{ijkl}(t)=\textstyle{\lla \Dc_{ij}(t) \Dc_{kl}(t)\rra}$ are relevant for shorter $t$, when diffusion in individual compartments remains non-Gaussian (incomplete coarse-graining); these tensors can be accessed \cite{JESPERSEN2013,NETOHENRIQUES2020} with double diffusion encoding\cite{SHEMESH2015}. Therefore, understanding the nature of tensors of the form (\ref{eq:DCdefinitions}) has become indispensable for dMRI.  

A fundamental problem is to classify symmetries of the cumulant tensors ubiquitous in representing a dMRI signal, and define their {\it tensor invariants}, i.e., basis-independent combinations of components  (such as the trace $\D_{ii}$). With the advent of precision medicine and quantitative imaging \cite{DESMOND2012}, the invariants form a basis- and hardware-independent fingerprint of MRI signals 
\cite{FRANK2002,KAZHDAN2003,GUTMAN2007,FUSTER2011,MIRZAALIAN2016,REISERT2017,SKIBBE2017,NOVIKOV2018,REISERT2018b,ZUCCHELLI2020,HERBERTHSON2021b,MOSS2025}; 
their information content will underpin classifiers of pathology, development and aging \cite{SMITH2021}. A practical problem is to relate tensors and their invariants to tissue properties \cite{JELESCU2017,ALEXANDER2019}, and to design fast unbiased measurements.%\sc{Add here (R@): FUSTER2011, ZUCCHELLI2020, HERBERTHSON2021}

Here we study the geometry of the dMRI signal by viewing the cumulant tensors via the addition of ``angular momenta'' coming from the compartment tensors $\Dc$. 
We decompose tensors $\D$ and $\C$ with respectively 6 and 21 degrees of freedom (DOF) according to irreducible representations of the SO(3) group of rotations. 
We then construct and classify a complete set of $3+18$ rotational invariants of the cumulant expansion (\ref{eq:CumExpinB}), or RICE, up to ${\cal O}(b^2)$. \rv[R2]{}
\update{The group-theory approach generalizes to the DDE-estimated  $\D(t)$, $\C(t)$, and $\Smu(t)$, yielding $3+18+12$ $t$-dependent invariants, respectively,}
reveals the cumulant tensors' symmetries and geometric meaning, 
\update{relates the topology of dMRI acquisitions to that of SO(3) group manifold,}
and connects with tissue biophysics embodied in the distribution of compartment tensors.  
We express all  known dMRI contrasts up to ${\cal O}(b^2)$ --- mean diffusivity (MD), fractional anisotropy (FA), mean, radial and axial kurtosis (MK, RK, AK), microscopic FA (\mFA), isotropic/anisotropic variance, etc --- via just 7 RICE invariants.  
Besides uncovering the 14 unexplored invariants \rv[R2]{} \update{of $\C$, and the 12 invariants of $\Smu(t)$ for non-Gaussian compartments,} constructing RICE according to symmetries makes them maximally independent, and thus potentially more specific to tissue microstructure changes.

Our geometric approach applies to hundreds of thousands publicly available human dMRI brain datasets \cite{CLIFFORD2008,JAHANSHAD2013,GLASSER2016,MILLER2016,VOLKOW2018,TAYLOR2017}. Using clinical dMRI of 1189 patients \cite{LIAO2024,CHEN2024}, we show that the complete set of invariants for the kurtosis tensor outperforms the subset of 6 conventional invariants for classification of multiple sclerosis. 
We also derive the shortest (``instant'') iRICE protocols for mapping MD, FA, MK in 1 minute, and MD, FA, MK, \mFA in 2 minutes on a clinical scanner, making advanced dMRI clinically feasible. 
The methodology further applies to higher-order tensors in Eq.~(\ref{eq:CumExpinB}), 
as well as to the 4th-order elasticity tensor in materials science and geology \cite{BACKUS1970,BETTEN1987,BONA2004,MOAKHER2008}. % removed BETTEN1987b
%\update{Finally, \rv[R2]{}the present framework applies to time-dependent $\D(t)$ and $C(t)$ from non-Gaussian compartments.}

\section*{Results}\label{s:Results}
%\update{\subsection*{Outline}}
\noindent
\update{
Our \rvleft[EDITOR]{} main method of investigation will be  theory of the SO(3) group,   which we introduce as a warm-up, supply along the way, and systematize in  {\it Methods}  and {\it Supplementary} sections. \rvleft[R2]{}
Our central QT decomposition viewpoint (\ref{DxD})--(\ref{QTLM}) on $\C$ as an addition of ``angular momenta" from compartment tensors $\Dc$  applies both for Gaussian and non-Gaussian compartments. 
For Gaussian compartments, axially symmetric $\B$-tensors yield all $\C$-tensor components and realize the Hopf mapping of the SO(3) group manifold onto conventional spherical shell-based dMRI acquisitions. 
For non-Gaussian compartments, the  $\B$-tensor representation (\ref{eq:CumExpinB}) 
of the cumulant series is inconsistent.  
The signal depends on the frequency spectrum of the diffusion weighting rather than just its shape $\B$; also, extra $\O(b^2)$ terms arise due to the intrinsic (microscopic) kurtosis $S(t)$ in individual compartments. Estimating these contributions requires a longer DDE acquisition. 
Therefore, we first develop our formalism assuming Gaussian compartments, Eqs.~(\ref{eq:CumExpinB})--(\ref{eq:DCdefinitions}), where $\C$ is the only  $\O(b^2)$ contribution, and establish its SA decomposition motivated by the kurtosis imaging. The relations (\ref{QT_SA_relation}) between QT and SA decompositions, their $3+18$ invariants, the relation to previous dMRI contrasts, and the fast iRICE protocols, all assume Gaussian compartments. We apply this formalism to the classification of multiple sclerosis from human brain dMRI.  
Finally, we come back to non-Gaussian compartments, consider the analog (\ref{eq:DDE_cumExp}) of the series (\ref{eq:CumExpinB}) for  DDE, and in Eq.~(\ref{eq:SH_DDE}) identify all the irreducible $t$-dependent   components  (\ref{DxD})--(\ref{QTLM}) of $\C(t)$, and of the  microscopic kurtosis $\Smu(t)$, Eq.~(\ref{Smu}), thus fully solving the problem up to $\O(b^2)$. 
}

\subsection*{Warm-up: Irreducible decomposition of the diffusion tensor}\label{s:D}
\noindent
%Our main method of investigation will be the representation theory of the SO(3) group applied to cumulant tensors (\ref{eq:DCdefinitions}). 
To set the stage, recall that a $3\times3$ symmetric matrix, such as a diffusion tensor $\Dc_{ij}$, splits into an isotropic component, $\Dcz_{ij}$, and a symmetric trace-free (STF) component, $\Dct_{ij}$:
\begin{equation} \label{D=Cart}
	\begin{aligned}
		\Dc_{ij} &= \Dcz_{ij} + \Dct_{ij} = \Dc^{00}  \Y^{00}_{ij}+ \sum_{m=-2}^2 \Dc^{2m} \Y^{2m}_{ij} \,,
	\end{aligned}
\end{equation}
where the matrices $\Y^{\ell m}$ form the standard spherical tensor basis, 
so that $Y^{\ell m}(\n) = \Y^{\ell m}_{ij} n_i n_j$ are the spherical harmonics on a unit sphere $|\n|=1$.\cite{THORNE1980} 
Here we use the Racah normalization for spherical harmonics (see {\it Methods}), such that  
%$\int_{\mathbb{S}^2}\d \n\, Y^{\ell m}(\n)Y^{\ell' m'}(\n) = \tfrac{4\pi}{2\ell+1} \delta_{\ell \ell'} \delta_{mm'}$.  
 $\Y^{00}_{ij}  = \delta_{ij}$ (unit matrix).
The $\ell=0$ component is the mean diffusivity $\Dc^{00} = \tfrac13 \tr \Dc $ setting the overall size of the diffusion tensor ellipsoid $\Dc(\n) = \Dc_{ij} n_i n_j$,  while the five $\ell=2$ components $\Dc^{2 m}$ are responsible for its shape and orientation.  

Upon rotations, the spherical tensor components $\Dc^{\ell m}$ with different $\ell$ do not mix: $\Dc^{2 m}$ transform amongst themselves; $\Dc^{00}$ is an invariant. Mathematically, they belong to different irreducible representations of the SO(3) group \citep{THORNE1980,Tinkham,HALL2015}. Hence, Eq.~(\ref{D=Cart}) realizes the {\it irreducible decomposition}\cite{FRANK2002}
\begin{equation} \label{D=irrep}
	\Dc = \Dcz \oplus \Dct \, ,\quad  \Dc^{(\ell)}\, \equiv \, \{\Dc^{\ell m}\}\,,\quad m=-\ell,\hdots,\ell \,, \ \ 
\end{equation}
where each irreducible component of degree $\ell$ has $2\ell+1$ DOF, such that the  6 DOF of a symmetric $3\times3$ matrix $\Dc$ splits into $6 = 1 + 5$, cf. Fig.~\ref{fig:outline}a. 
Spherical tensors  will be our workhorse, as they represent high-dimensional objects using the minimal number of DOF. 
Furthermore,  irreducible components with different $\ell$ point at distinct symmetries and physical origins.

The diffusion tensor in Eq.~(\ref{D=Cart}) may describe a microscopic tissue compartment. 
Consider the invariants of $\Dc$: the ellipsoid's shape is defined by 3 semi-axes (the eigenvalues). The remaining 3 DOF define its orientation in space (e.g., 3 Euler's angles), and depend on the coordinate frame. The irreducible decomposition (\ref{D=irrep}) allows one to construct tensor invariants symmetric with respect to the eigenvalues, and to assign the degree $\ell$ to them. The simplest, $\ell=0$ invariant, is the trace 
$\tr D =3 D^{00}$. 
The remaining two independent $\ell=2$ invariants are also given by traces: $\tr (\Dct)^2$ is proportional to the variance of the eigenvalues, and  $\tr (\Dct)^3=3 \,\det \Dct$ --- to their product. The latter relation is readily seen in the eigenbasis, where $\Dct=\mathrm{diag}[\lambda_1,\,\lambda_2,\,-(\lambda_1+\lambda_2)]$. 
Note we drop parentheses after `$\tr$' for readability, cf. {\it Methods/Notations}.

The overall diffusion tensor $\D_{ij}$ is the mean of the compartment tensors (\ref{D=Cart}) over $\P(\Dc)$, 
cf. Eq.~(\ref{eq:DCdefinitions}). 
Invariants of $\D$ give rise to common DTI metrics, such as MD from $\ell=0$, 
\begin{equation} \label{MD}
\mathrm{MD} = \D_0 \equiv  \D^{00} = \overline{\D(\n)}  =  \tfrac13 \D_{ii} \,, 
\end{equation}
%$\D_0 = \tfrac13 \tr \D$ ($\ell=0$), 
and FA (from both $\ell=0$ and $\ell=2$ sectors, cf. Eq.~(\ref{FA}) below). 
The $0 0$ spherical harmonics component equals the directional average (overbar) 
by virtue of the Racah normalization avoiding $\sqrt{4\pi}$ factors. 

The  covariance tensor $\C$  involves averages of the direct tensor products  $\Dc \otimes \Dc$, Eq.~(\ref{eq:DCdefinitions}). 
As such, it has a major symmetry (swapping first and second pairs of indices), and a minor symmetry (in each of the index pairs). 
Below we will formally map $\C$ onto the addition of quantum angular momenta with $\ell=0$ or $2$, construct all its irreducible components and  invariants. 
 
%This will allow us to construct all irreducible components of $\C$ with distinct symmetries and geometric meaning, define all tensor invariants, reveal the redundancy in existing $\O(b^2)$ dMRI metrics and uncover 14 unexplored invariants, as well as relate the irreducible components to dMRI measurements and find ways of fast estimation of the most common invariants.

%%%%%%%%%%%%%%%%%%%%%%%%%%%%%%%%%%%%%%%%%%%%%%%%%%%%%%%%%%%%%
\subsection*{Size and shape covariances: QT decomposition}
\noindent
Using the irreducible decomposition (\ref{D=irrep}) in Eq.~(\ref{eq:DCdefinitions}) results in
\begin{equation}\label{DxD}
	\begin{aligned}
		\C & = \lla \Dc \otimes \Dc \rra = \lla \left(\Dcz \oplus \Dct \right) \otimes  \left(\Dcz \oplus \Dct \right)  \rra \\
		&=\Qz \oplus \Qt \oplus \Tz \oplus \Tt \oplus \Tf,
	\end{aligned}
\end{equation}
where
\begin{subequations}\label{QT}
	\begin{align}\label{Qz}
		\Qz&\equiv \lla \Dcz \otimes \Dcz \rra  ,
		\\ \label{Qt}
		\Qt &\equiv 2 \lla \Dcz \otimes \Dct \rra   ,
		\\ \label{T}
		\T &\equiv \lla \Dct \otimes \Dct \rra =  \Tz \oplus \Tt \oplus \Tf  .
	\end{align}
\end{subequations}
Formally, this corresponds to the addition of two quantum angular momenta with $\ell=0$ or $\ell=2$,  see graphical representation in Fig.~\ref{fig:outline}b. As it is known from quantum mechanics \citep{Tinkham}, addition of angular momenta $\ell_1$ and $\ell_2$ yields all possible states with momenta between $|\ell_1-\ell_2|$ and $\ell_1+\ell_2$. 
Mathematically speaking \citep{HALL2015}, the tensor product of irreducible representations is reducible, and decomposes into a direct sum of irreducible representations.  %according to the Clebsch-Gordan rule. 
%:
%\begin{equation}
%	V_{\ell_1} \otimes V_{\ell_2} \;\cong\; 
%	\bigoplus_{\ell = |\ell_1 - \ell_2|}^{\ell_1+\ell_2} V_\ell ,
%\end{equation}
%where $V_\ell$ denotes the $(2\ell+1)$-dimensional irreducible representation of $\mathrm{SO}(3)$ labeled by angular momentum $\ell$. 
Tensors $\Q^{(\ell)}$ and $\T^{(\ell)}$ in Eqs.~(\ref{DxD})--(\ref{QT}) belong to the $(2\ell+1)$-dimensional irreducible representations $V_\ell$  of $\mathrm{SO}(3)$ with the angular momenta $\ell = 0, \, 2, \, 4$, according to 
$V_0 \otimes V_0 \cong V_0$,  $V_0 \otimes V_2 \cong V_2$, and $\mathrm{Sym\,}(V_2 \otimes V_2) \cong V_0 \oplus V_2 \oplus V_4$. 
%\begin{equation}
%	\begin{aligned}
%	\mathrm{Sym\,}(V_0 \otimes V_0) \;&\cong\; V_0\., \\
%	\,\,   
%	\mathrm{Sym\,}(V_0 \otimes V_2) \;&\cong\; V_2\,, \\
%	\,\,   
%	\mathrm{Sym\,}(V_2 \otimes V_2) \;&\cong\; V_0 \oplus V_2 \oplus V_4 \,,
%	\end{aligned}
%\end{equation}
%where $V_\ell$ denotes the $(2\ell+1)$-dimensional irreducible representation of $\mathrm{SO}(3)$ labeled by angular momentum $\ell$. 
Here $\mathrm{Sym\,}(\cdot)$ indicates that we consider the symmetric part of the tensor product due to the major symmetry of $\C$, such that 
the  representations with odd $\ell=1,\, 3$ in the $V_2 \otimes V_2$ case do not contribute. 
%they are forbidden by parity that allows $(-1)^\ell = +1$.  Parity restriction is consistent with the time-reversal invariance of Brownian motion.
This irreducible decomposition yields the DOF count for $\C$ tensor: $21 = 1+5+ 1+5 + 9$.

The physical intuition behind this formalism is as follows. In a distribution $\P(D)$ of compartmental diffusion tensors of Fig.~\ref{fig:outline}a, 
each ellipsoid (\ref{D=irrep}) has an isotropic (``size'') part $\Dcz$ and trace-free anisotropic (``shape'') part $\Dct$. Thus, the direct products in Eq.~(\ref{QT}) define the size variance $\Qz$, the size-shape covariance $\Qt$, and the shape-shape covariance $\T$, Fig.~\ref{fig:outline}b,c. The latter covariance is reducible and further splits into three irreducible spherical tensors 
$\Tz$, $\Tt$ and $\Tf$.

Besides the explicit classification based on symmetries, the benefit of using spherical tensors is in having the minimal number of DOF in each of them, as compared to the highly redundant Cartesian objects such as $\C_{ijkl}$. Therefore, one expects that the covariances $\lla \Dc^{\ell_1 m_1} \Dc^{\ell_2 m_2}\rra $ of the spherical-tensor components (\ref{D=irrep}) of compartmental diffusivities should be related to the corresponding spherical-tensor components of $\Q$ and $\T$, Fig.~\ref{fig:outline}c. We derive these explicit relations based on the spherical tensor algebra in Supplementary Section \ref{SM:CGcoeffsSTF}. In brief, the relations for $\Q$, 
\begin{subequations} \label{QTLM}
\begin{equation}\label{QLM}
	\begin{aligned}
		\Q^{00} =   \lla ( \Dc^{00} )^2 \rra,   \quad \Q^{2m} =  2  \lla \Dc^{00}  \Dc^{2m} \rra,
	\end{aligned}
\end{equation}
follow trivially from Eqs.~(\ref{Qz}) and (\ref{Qt}). However, Eq.~(\ref{T})
\begin{equation}\label{TLM}
	\begin{aligned}
%		\T^{LM} &= \braket{2,0,2,0|L,0}  \\ &\times 
%		\sum_{m+m'=M}  \lla \Dc^{2m}\Dc^{2m'}\rra \braket{2,m,2,m'|L,M} 
		\T^{\ell m} = \braket{2020|\ell 0}  \!\!
		\sum_{m_1+m_2=m} \! \braket{2m_1 2m_2 |\ell m} \textstyle{\lla \Dc^{2m_1}\Dc^{2m_2}\rra} 
	\end{aligned}
\end{equation}
\end{subequations}
involves the Clebsch-Gordan coefficients \citep{Tinkham} $\braket{\ell_1 m_1 \ell_2 m_2 | \ell m }$ which obey the selection rule $m_1 + m_2 = m$ and vanish otherwise. 
This rule makes the system (\ref{QTLM}) so sparse that it can be inverted by hand, cf. Supplementary Eq.~(\ref{eq:DDsolved}).  
This provides %a full 
an analytical solution for the tissue properties $\lla \Dc^{\ell_1 m_1} \Dc^{\ell_2 m_2}\rra$ in terms of the spherical  components of $\T$ and $\Q$, i.e., the irreducible components of  covariance tensor $\C$. 
We call Eqs.~(\ref{DxD})--(\ref{QTLM}) the {\it QT decomposition} of  covariance tensor.

\update{ 
Equations (\ref{DxD})--(\ref{QTLM}) straightforwardly extend for time-dependent $\Dc(t)$, with the $\Q(t)$ and $\T(t)$ tensors picking up the diffusion time dependence via $\lla \Dc^{\ell_1 m_1}(t) \Dc^{\ell_2 m_2}(t)\rra$. 
\rv[R2]{}
This general case will be considered in Eqs.~(\ref{eq:DDE_cumExp})--(\ref{eq:SH_DDE}). 
In the meantime, we assume Gaussian compartments, such that the cumulant series is fully determined by the $\B$-tensors, Eqs.~(\ref{eq:CumExpinB})--(\ref{eq:DCdefinitions}).
}

\begin{figure*}[th!!]
	\centering
	\includegraphics[width=0.85\textwidth]{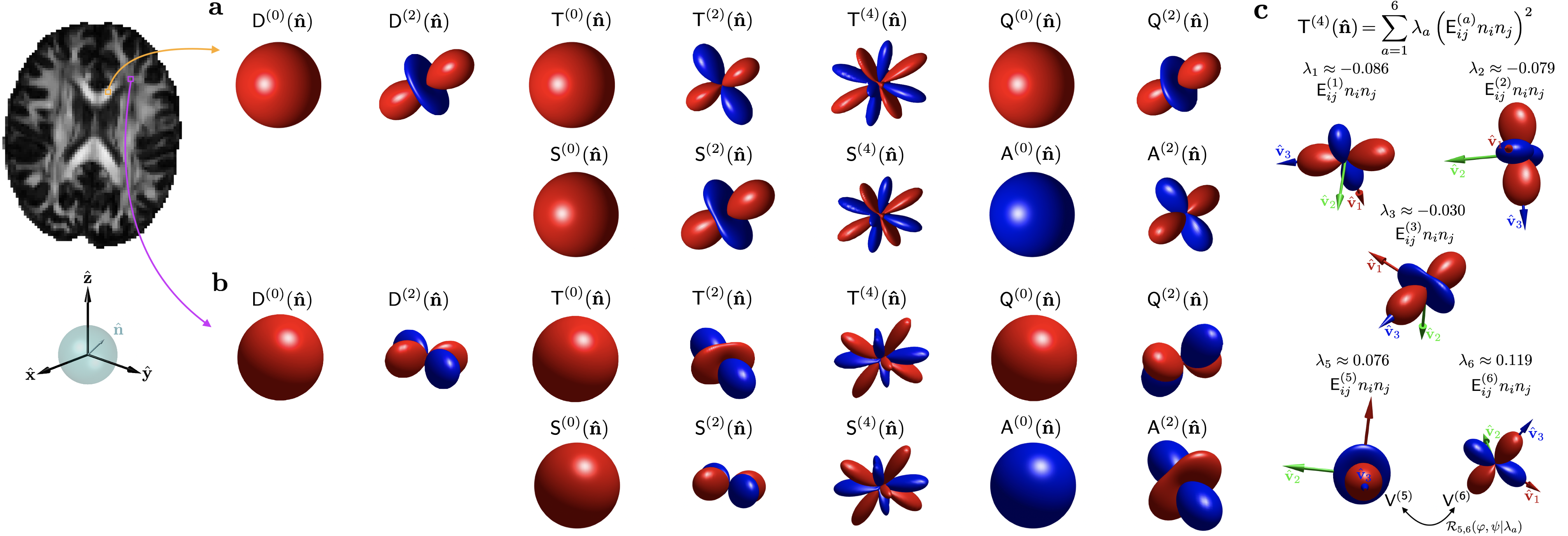}
	\caption{Irreducible decompositions of $\D$ and $\C$ tensors, Eqs.~(\ref{D=irrep}), (\ref{QT}), and (\ref{SA}), for two white matter voxels: (a) corpus callosum --- highly aligned fibers, and 
		(b) longitudinal superior fasciculus --- crossing fibers. Glyphs are color-coded by the sign (red = positive) while the radius represents the absolute value (rescaled to similar sizes).
		(c) Representation of the eigentensor decomposition, Eq.~(\ref{eq:eigtensordecomp}) in {\it Methods}, of $\Tf$ for the crossing fiber voxel shown in (b). 
		The 6 invariants of $\Tf=\Sf$ (cf. Fig.~\ref{fig:RotInvsALL}) correspond to 4 DOF from $\lambda_a$ (here $\lambda_4 = 0$, $\E_{ij}^{(4)}\propto \delta_{ij} $, and $\sum_{a=1}^6 \lambda_a = 0$), 
		and 2 DOF defining the relative orientations among any pair of eigentensors $\E_{ij}^{(a)}$. 
	}\label{fig:glyphsQTSA}
\end{figure*}

\subsection*{$\Q$ and $\T$ tensor components from $\B$-tensor shells}
\noindent
%\update{We now \rv[R2]{} assume Gaussian compartments, such that we can probe the cumulant series by the $\B$-tensors, Eqs.~(\ref{eq:CumExpinB})--(\ref{eq:DCdefinitions}).} 
The rank of the $\B$ tensor reflects how many dimensions of the diffusion process are being probed simultaneously; $\mbox{rank}\,\B > 1$ means probing the diffusion along more than one dimension. 
An example is the axially symmetric family \citep{ERIKSSON2015,TOPGAARD2017}
\begin{equation}\label{eq:AxSymB}
	\B_{ij}(b,\beta,\g)=b\,\Big( \,\beta \, g_i g_j + \tfrac{1-\beta}{3} \, \delta_{ij} \Big) 
\end{equation}
parametrized by its trace $b$ setting the overall scale; by the unit vector $\g$ along the symmetry axis; 
and by shape parameter $\beta$. 

Historically, diffusion weightings in MR have mostly probed a single direction $\g$ per measurement, using pulsed gradients  \citep{STEJSKAL&TANNER1965}, such that $\B_{ij} = b\, g_i g_j$; this so-called linear tensor encoding (LTE, $\beta=1$) corresponds to $\mbox{rank}\,\B = 1$. 
Varying $\beta$ changes the $\B$-tensor shape, e.g.,  $\beta=0$ for spherical tensor encoding (STE, isotropic $\B$-tensor),  and $\beta=-\tfrac12$ for planar tensor encoding (PTE, two equal nonzero eigenvalues) \citep{CORY1990,MITRA1995,MORI1995,SHEMESH2015,WESTIN2016,TOPGAARD2017}. 

Imaging voxels in the brain have vastly different fiber orientation dispersion and, as a result, distinct orientations and degrees of anisotropy of the underlying $\P(\Dc)$. It is therefore natural for a dMRI measurement to probe all directions uniformly. The most general $\B$-tensor is defined by its shape (3 eigenvalues summing up to the diffusion weighting $b$), and the 3 orientational DOF. Uniformly sampling all the orientations of $\B$ amounts to acting on it by all possible elements of the SO(3) group. Thus, a fixed-$b$ ``shell" for a generic $\B$-tensor is the group manifold of SO(3), which is the 3-dimensional sphere $\mathbb{S}^3$ (the group manifold of its universal cover SU(2)) with the antipodal points identified \cite{HALL2015}. In other words, a fixed-$b$ shell is a quotient space $\mathbb{S}^3/\mathbb{Z}_2$, where the factorization modulo group $\mathbb{Z}_2 = \{1, \, -1\}$ corresponds to gluing the antipodal points. Maximally isotropic coverage with $\B$-tensors amounts to a maximally uniform sampling of such a 3-sphere. 

Acquisitions with axially symmetric $\B$-tensors (\ref{eq:AxSymB})  are invariant under SO(2) rotations around $\g$. Factorization of the group manifold over 
$\mathrm{SO(2)} \cong \mathbb{S}^1$ (a unit circle) results in a familiar 2-dimensional spherical shell 
$\mathbb{S}^2 \cong \mathrm{SO}(3)/\mathrm{SO}(2)$. 
\update{
Topologically, it is a quotient space $\mathbb{S}^2 \cong \mathbb{S}^3/\mathbb{S}^1$ of a 3-sphere over a circle, known as Hopf fibration\cite{Hopf1931}.} 
(As $\mathbb{Z}_2$ is a subgroup of SO(2), the factorization over it is contained in the quotient.)  

Hence, the signal (\ref{eq:CumExpinB}) for axially symmetric $\B$-tensors (\ref{eq:AxSymB}), as function of the unit $\g \in \mathbb{S}^2$, can be represented in the canonical spherical harmonics basis on $\mathbb{S}^2$. The spherical harmonics components of its logarithm $\L=\ln \langle \Sig\rangle$ up to $\O(b^2)$  are  found by averaging the compartment signal 
$\Sig = \exp[-b\Dc(\g)] = \exp[- b \Dc^{00} - \beta b \sum \Dc^{2m}Y^{2m}(\g)]$ over $\P(\Dc)$: 
\begin{subequations} \label{eq:MGC_SH}
%\begin{equation}
\begin{align} 
	\label{eq:MGC_SH0}
	\L_{\rm}^{00}(b,\beta)   &=   -b \D^{00} + \tfrac12 \, b^2 \lp \Q^{00} + \beta^2\,\T^{00} \rp , \\
	\label{eq:MGC_SH2}
	\L_{\rm}^{2m}(b,\beta)   &=   -b\,\beta\, \D^{2m} + \tfrac12 \, b^2 \lp \beta\, \Q^{2m} + \beta^2\,\T^{2m} \rp , \\
	\label{eq:MGC_SH4}
	\L_{\rm}^{4m}(b,\beta)   &=    \tfrac12 \, b^2 \,\beta^2\, \T^{4m}  \,.
\end{align}
%\end{equation}
\end{subequations}
While components of $\D$ and $\Q$ appear naturally using Eqs.~(\ref{D=Cart}) and (\ref{QLM}), 
the components (\ref{TLM}) of $\T$ arise via the triple spherical harmonics integration, Supplementary  Eq.~(\ref{eq:tripleSHint}). 
 The corollaries of Eqs.~(\ref{eq:MGC_SH}) are as follows:  
\begin{enumerate}[label=(\roman*),noitemsep]
\item {\it Sufficiency:} Axially symmetric family (\ref{eq:AxSymB}) on the quotient space $\mathbb{S}^2$ is sufficient for estimating all 21 DOF of $\C$; 
the entire SO(3) manifold (non-axial $\B$) is not necessary. 

\item {\it Necessity:} To estimate all tensor components, one needs 3 distinct combinations $(b,\,\beta)$, with at least 2 nonzero $b$ and at least 2 nonzero $\beta$ (i.e., 2 non-STE $\B$ tensors). 

\item {\it STE} ($\beta=0$) can be used to isolate $\Q^{00}$, as $\L_{\rm STE} = -b\D^{00} + \frac12 b^2 \Q^{00} $ including %up to
$\O(b^2)$ \cite{LASIC2014,WESTIN2016,TOPGAARD2017,NILSSON2018}. 

\item {\it Conventional LTE} ($\beta=1$) 
alone cannot disentangle $\Ql$ and $\Tl$, for both $\ell = 0$ and $2$, from their sums.  

\end{enumerate}
As LTE plays a major role in everyday dMRI and in publicly available datasets \citep{CLIFFORD2008,JAHANSHAD2013,GLASSER2016,MILLER2016,VOLKOW2018}, we now consider it in detail.

%%%%%%%%%%%%%%%%%%%%%%%%%%%%%%%%%%%%%%%%%%%%%%%%%%%%%%
\subsection*{LTE versus other $\B$-tensor shapes: SA decomposition}

%%%%%%%%%%%%%%%%%%%%%%%%%%%%%%%%
\begin{figure*}[th!!!]
	\centering
	\includegraphics[width=0.92\textwidth]{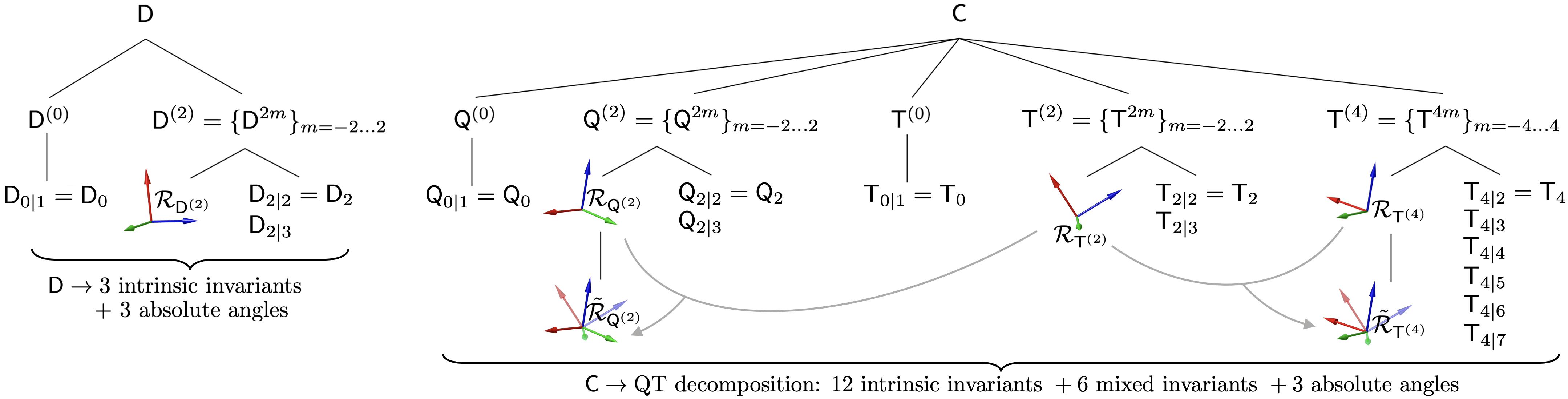}
	\caption[caption FIG]{Irreducible decompositions of tensors $\D$ and $\C$ (into $\Q$ and $\T$), Eqs.~(\ref{D=irrep})--(\ref{DxD}). 
		Each irreducible component has its intrinsic invariants (1 for $\ell=0$; 2 for $\ell=2$; and 6 for $\ell=4$). Together with these $1+2 + 1+2+6 = 12$ intrinsic invariants,  
		$\C$ has $3\cdot 3=9$ basis-dependent absolute angles defining the orientations of its $\Tt$, $\Tf$, and $\Qt$ via the rotation matrices $\mathcal{R}_{\Qt}$, $\mathcal{R}_{\Tt}$, and $\mathcal{R}_{\Tf}$, such that total DOF count of $\C$ is $21 = 12 + 9$. Out of these 9 absolute angles, 6 DOF are mixed invariants since they correspond to relative angles between $\Qt$, $\Tt$ and $\Tf$. As an example, we take $\mathcal{R}_{\Tt}$ as a reference, and compute the relative rotations $\tilde{\mathcal{R}}_{\Tf}$ and $\tilde{\mathcal{R}}_{\Qt}$. Maps of these invariants are shown in Fig.~\ref{fig:maps_QT}.
	}
	\label{fig:RotInvsALL}
\end{figure*}%
%%%%%%%%%%%%%%%%%%%%%%%%%%%%%%%%

\noindent
For LTE, $\B_{ij}\B_{kl}\to b^2 g_i g_j g_k g_l$ becomes symmetric in all indices \cite{LIU2003}, and 
Eq.~(\ref{eq:CumExpinB}) turns into the diffusion kurtosis imaging (DKI) signal representation \citep{JENSEN2005}:
\begin{equation} \label{DKI}
	\ln \la \Sig\ra = - b \D(\g) +   {b^2\over 2}  \S(\g) + {\cal O}(b^3) \,, \quad \S(\g) = \S_{ijkl}\, g_i g_j g_k g_l \,, 
%	\ln \la S\ra = - b\, g_i g_j \,\D_{ij} + \tfrac12  b^2 g_i g_j g_k g_l\, \S_{ijkl} + {\cal O}(b^3) \,,
\end{equation}
%where $\S$, the fully symmetric part of $\C$, 
where $\S$, the \textit{fully symmetric} part of $\C$, 
\begin{equation} \label{W}
 \S_{ijkl} = \C_{(ijkl)} = \tfrac13 (\C_{ijkl}+ \C_{iljk} + \C_{ikjl}) \equiv \tfrac13 {\D_0}^2 \, \W_{ijkl} \,, 
 %	 \W_{ijkl} &=   \frac{3}{\D_0^2}\,\S_{ijkl}\,,
\end{equation}
is proportional to the dimensionless {\it kurtosis tensor} $\W$ --- a 3d generalization of the kurtosis excess for the probability distribution of molecular displacements \citep{JENSEN2005}. 
The tensor-algebra 
notation  \citep{THORNE1980} for symmetrization over tensor indices between parentheses is assumed henceforth, 
e.g., $\D_{(ij)} = \tfrac12 (\D_{ij} + \D_{ji})$. 
%The kurtosis tensor is made dimensionless by normalizing $\S$ with mean diffusivity squared. 
Since $\W$ and $\S$ contain  the same information, we will focus on $\S$, and refer to both $\S$ and $\W$ as kurtosis.

Kurtosis splits into irreducible components with $\ell=0,\ 2,\ 4$: 
\begin{equation} \label{S=S0+S2+S4}
\S_{ijkl} = \S^{00} \delta_{(ij}\delta_{kl)} +\! \sum_{m=-2}^2\! \S^{2m} \Y^{2m}_{(ij}\delta^{}_{kl)} + \!\sum_{m=-4}^4\! \S^{4m} \Y^{4m}_{ijkl}
\ \ 
\end{equation}
such that $\S(\g) = \Sz + \St(\g) + \Sf(\g)$, where  spherical tensors $\Sz$, $\St$ and $\Sf$ are parametrized by $2\ell+1 = 1$, $5$ and $9$ spherical harmonics coefficients, respectively, totaling $1+5+9 = 15$ DOF, found from $\S_{ijkl}$ via Eqs.~(\ref{eq_Slm}) in {\it Methods}.

The remaining $21-15=6$ DOF of $\C$ are contained in the {\it asymmetric} (not antisymmetric!) part $\A$ of $\C$:
\begin{equation} \label{A}
	\A_{ijkl} = \C_{ijkl} -  \C_{(ijkl)} \, = \tfrac13 (2\C_{ijkl} - \C_{iljk} - \C_{ikjl})\,.
\end{equation}
To measure all DOF of $\C$, and thereby to access $\A$, the necessary and sufficient condition is the corollary (ii) after Eq.~(\ref{eq:MGC_SH}), which makes the requirement\cite{WESTIN2016}  $\mbox{rank}\,\B > 1$ more precise. 
Although $\A$ is a 4th-order Cartesian tensor, it has only 6 DOF, and thus it is equivalent to a symmetric $3\times3$ tensor \citep{ITIN2013,ITIN2015}:
\begin{subequations}
	\label{A=A}
	\begin{align}\label{Apq}
		%		\A_{ijkl}      &= \C_{ijkl} - \C_{(ijkl)}, \\
		\A_{pq}      &= \epsilon_{ikp}\epsilon_{jl q} \A_{ijkl} 
		\\ \label{Apq=}
		&= \delta_{pq} (\A_{iikk} - \A_{ikik}) + 2 (\A_{pkqk} -  \A_{pqkk})\,,
		\\ \label{Aijkl}
		\A_{ijkl} &= \tfrac16 (\epsilon_{ikp} \epsilon_{jl q} + \epsilon_{i l p} \epsilon_{j k q}) \A_{pq} \,,
	\end{align}
\end{subequations}
where $\epsilon_{ijk}$ is the fully antisymmetric Levi-Civita tensor. Thus, analogously to Eq.~(\ref{D=irrep}), tensor $\A$  splits into the irreducible components with $\ell=0$ and $2$, cf. Eqs.~(\ref{eq_Slm}).

%To sum up, we obtained the {\it SA decomposition} of $\C$ tensor 
We define the above irreducible components as the {\it SA decomposition} of $\C$ (which stands for symmetric-asymmetric):
\begin{equation}\label{SA}%
	\begin{aligned}
		\C &= \S \oplus \A \,, \\
		\S &= \Sz \oplus \St \oplus \Sf\,, \quad \A = \Az \oplus \At  \,,
	\end{aligned}
\end{equation}
where the DOF count holds: $21_\C = (1 + 5 + 9)_{\S} + (1+5)_{\A}$. The separation of information accessible through LTE ($\S$) vs beyond-LTE  ($\A$) implies that the SA decomposition is {\it LTE-driven}. Hundreds of thousands datasets acquired with LTE and moderate diffusion weightings are available \citep{CLIFFORD2008,JAHANSHAD2013,GLASSER2016,MILLER2016,VOLKOW2018}, and thus are sensitive to the information present in $\S$ only.

The contributions of SA decomposition to $\O(b^2)$ term of the cumulant expansion~(\ref{eq:CumExpinB}) for $\B$ tensor family (\ref{eq:AxSymB}) are  
\begin{subequations} \label{BBC}
	\begin{align}
	\label{BBS}
	 \B_{ij}\B_{kl}\S_{ijkl} & = 
	b^2  \lb \tfrac{5+4\beta^2}9 \Sz + \tfrac{7\beta+2\beta^2}9 \St(\g) + \beta^2 \Sf(\g) \rb , 
\\
	\label{BBA}
	 \B_{ij}\B_{kl}\A_{ijkl} & = 
	\tfrac29 b^2    \lb (1-\beta^2) \Az - \beta(1-\beta)  \At(\g) \rb .
	%\tfrac{b^2(1-\beta)(1+2\beta)}{9} \, \A^{00} - \tfrac{b^2\beta(1-\beta)}9 \, \A(\g) 
	\end{align}
\end{subequations}
To obtain Eqs.~(\ref{BBC}), we used 
%$ \S_{ijkl} \delta_{ij}\delta_{kl} = 5S^{00} \equiv 5\Sz$ and $\Y^{2m}_{(ij}\delta^{}_{kl)} \delta^{}_{kl} = \tfrac76 \Y^{2m}_{ij}$, 
Eq.~(\ref{A=A}) and Supplementary Eq.~(\ref{eq:rank24_CART2STF}),  such that 
$\Sz\equiv \S^{00}$,  
$\Az \equiv \A^{00} = \tfrac{1}{3} \A_{pp}=\tfrac{1}{3}(\A_{iikk}-\A_{ikik})$, and $ \At(\g) = \At_{pq} g_p g_q$. 
As expected, STE  is sensitive only to the isotropic parts $\Sz$ and $\Az$, while 
LTE couples to $\S(\g)$ and is insensitive to $\A(\g)$ (Eq.~(\ref{BBA}) vanishes). 
PTE  couples to the $\A$ tensor ellipsoid, Eq.~(\ref{BBA}) yielding $+\tfrac{b^2}{6} \A(\g)$, and to the combination 
$b^2\big(\frac23 \Sz - \frac13 \St(\g) + \frac14 \Sf(\g)\big)$ via Eq.~(\ref{BBS}). 
At fixed $b$, LTE maximizes the sensitivity to $\Sz$, $\St$, $\Sf$; 
STE --- to $\Az$; and PTE --- to $\At$.

Comparing $\O(b^2)$ terms of Eqs.~(\ref{eq:MGC_SH}) to the half-sum of Eqs.~(\ref{BBC}), we relate the irreducible components of QT and SA decompositions:%  
\begin{subequations}\label{QT_SA_relation}
\begin{align}
	\label{Q0} \Qz   &= \tfrac{5}{9} \,\Sz  + \tfrac{2}{9} \, \Az \,,
	\\ \label{Q2}
	\Qt &= \tfrac79 \,\St  - \tfrac29 \, \At \,,
	\\ \label{T0} \Tz  &= \tfrac{4}{9} \,\Sz  - \tfrac{2}{9} \, \Az \,,
	\\ \label{T2}
	\Tt &= \tfrac29 \,\St  + \tfrac29 \, \At \,,
	\\ \label{T4}
	\Tf &= \Sf \,.
	%	       \label{S0} \Sz &= \Tz  +  \Qz \,,
	%		\\ \label{A0} \Az  &= -\tfrac52 \,\Tz  + 2 \, \Qz \,,
	%		\\ \label{S2} \St  &=  \Tt  +  \Qt \,,
	%		\\ \label{A2} \At   &= \tfrac72 \,\Tt  - \, \Qt \,,
	%		\\ \label{S4} \Sf  &= \Tf  \,,
\end{align}
\end{subequations}
These relations respect 
\begin{equation} \label{S=Q+T}
\Sl = \Ql + \Tl \,,  %\quad \ell = 0, \, 2,\,4,
\end{equation}
evident by comparing Eqs.~(\ref{DKI}) and (\ref{eq:MGC_SH}) for $\beta=1$, and formally setting $\mathsf{Q}^{(4)}\equiv0$.
Upon SO(3) rotations, components $\Sl$, $\Al$, or $\Tl$, $\Ql$, do not mix with each other.

Which decomposition is more ``fundamental"? 
Since each representation with $\ell =0$ and $\ell = 2$ enters Eq.~(\ref{DxD}) twice, {\it any} decomposition of $\C$ with two independent linear combinations of representations with $\ell=0$, and separately for $\ell=2$, is legitimate. 
The  decompositions (\ref{QT}) and (\ref{SA}) are selected by their distinct physical meaning: 
The SA decomposition breaks symmetry between the LTE and non-LTE acquisitions, 
while the QT decomposition is natural to describe tissue properties.

%%%%%%%%%%%%%%%%%%%%%%%%%%%%%%%%
\begin{figure*}[th!!]
	\centering
		\includegraphics[width=0.95\textwidth]{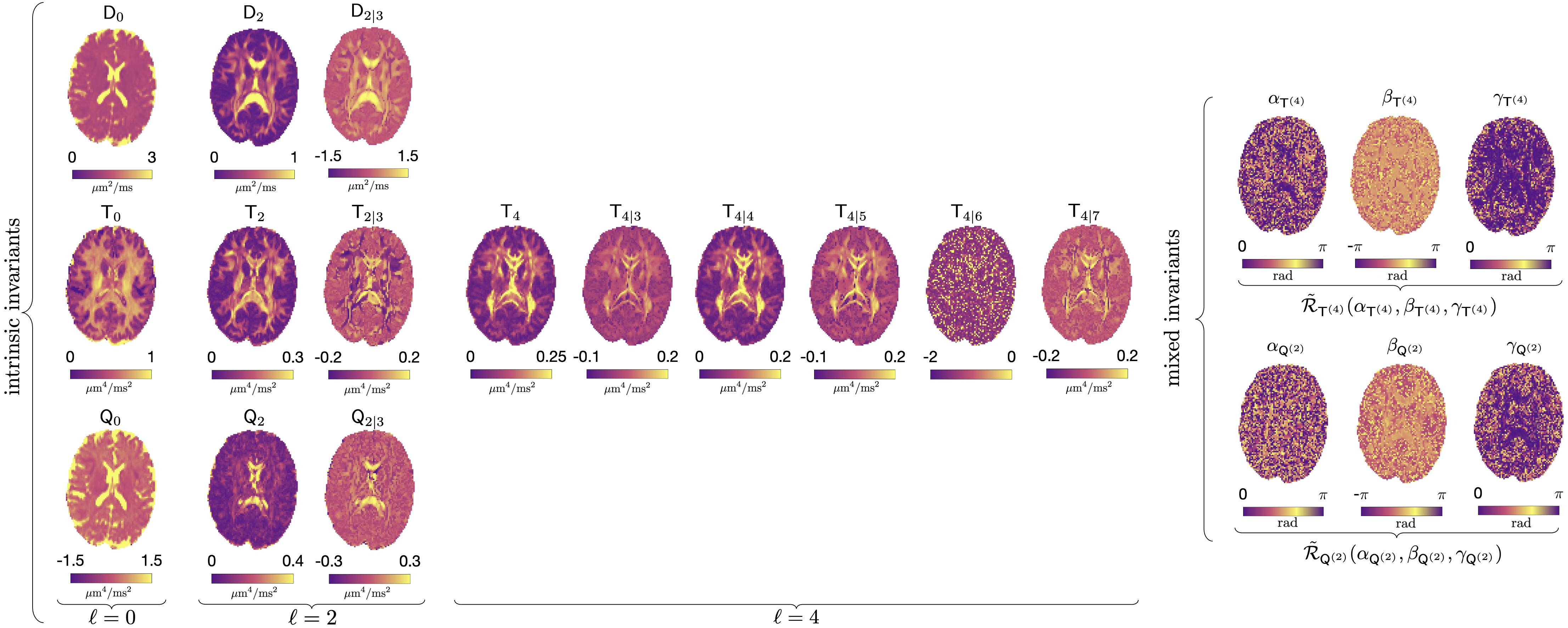}
	\caption[caption FIG]{RICE maps for a normal brain (33 y.o. male). 
		Intrinsic invariants for each irreducible decomposition of $\D$, $\T$ and $\Q$ 
		are shown as powers of corresponding traces, to match units of $\D$ and $\C$. 
		The 6 mixed invariants correspond to Euler angles of eigenframes of $\Tf$ and $\Qt$ relative to that of $\Tt$ (see text). 
		The underlying tissue microstructure introduces correlations between invariants: e.g. small relative angles $\tilde\beta$ in white matter tracts exemplify the alignment of eigenframes of different representations of SO(3) with the tract. 
	}\label{fig:maps_QT}
\end{figure*}
%%%%%%%%%%%%%%%%%%%%%%%%%%%%%%%%

The geometric meaning of the irreducible decompositions such as (\ref{D=irrep}), (\ref{QT}), and (\ref{SA}), comes from the correspondence between spherical tensors and spherical harmonics (cf. Eq.~(\ref{STF=SH}) in {\it Methods}), Fig.~\ref{fig:glyphsQTSA}. 
%For example, in the tensor glyphs of Fig.~\ref{fig:glyphsQTSA}, $\D(\n)=\D_{ij}\, n_i n_j$, $\Q(\n)=\Q_{ij}\, n_i n_j$ and 
%$\T(\n) = \T_{ijkl} \, n_i n_j n_k n_l$, where $\n$ is a unit vector, 
The $\ell=0$ parts give directional averages;  the $\ell=2$ parts are responsible for glyphs parametrized by five $Y^{2m}(\n)$, turning the corresponding ball into an ellipsoid, such as $\D(\n) = \D_{ij} n_i n_j$ in DTI \cite{JONES2010}.  
The $\ell=0$ and $\ell=2$ parts can capture a single fiber tract (a single pair of opposite lobes on a sphere, both at the ${\cal O}(b)$ and  ${\cal O}(b^2)$ level), but cannot represent geometries with multiple pairs of lobes, such as in fiber crossings. The unique $\Tf=\Sf$ part, parametrized by nine $Y^{4m}(\n)$, is the only part of $\C$ that captures multiple pairs of lobes coming from fiber crossings, see, e.g., Fig.~\ref{fig:glyphsQTSA}c.  
Thereby, the beyond-LTE $\A$-tensor part of $\C$ can only accommodate an ellipsoidal glyph shape  --- albeit physically distinct from that of the diffusion tensor.

As for the parameter estimation, one can either directly use Eqs.~(\ref{eq:MGC_SH}), or go the SA route (\ref{W}) and (\ref{A}) to compute the irreducible components $\Sl$ and $\Al$ [Eqs.~(\ref{eq_Slm}) in {\it Methods}],  and then transform to QT via Eqs.~(\ref{QT_SA_relation}). 
Once the irreducible components of $\Q$ and $\T$ tensors are known, tissue properties (compartmental spherical tensor covariances) can be found by inverting Eqs.~(\ref{QTLM}), see Supplementary Eqs.~(\ref{eq:DDsolved}). 
 % Eqs.~(\ref{eq:DCcart_Dsph})--(\ref{eq:ellSYSTEM_QT}) and (\ref{eq:TQ_DD_system})  relate all  irreducible components to compartment tensor covariances. 

\update{Above, the \rvleft[R2]{} source of kurtosis in Eq.~(\ref{DKI}) was solely the heterogeneity of compartment  tensors $\Dc$, also responsible for the covariance  (\ref{eq:DCdefinitions}). 
For non-Gaussian compartments, each characterized by its own microscopic kurtosis tensor $S(t)$, 
the narrow-pulse LTE\citep{STEJSKAL&TANNER1965} DKI signal, compared to  Eq.~(\ref{DKI}), reads 
\begin{equation} \label{DKI_t}
\begin{aligned}
	\ln \la \Sig\ra &= \ln \la e^{- bD(t,\g) + \frac12 b^2 S(t,\g) + \O(b^3)}\ra \\
	&=- b \D(t, \g) +   {b^2\over 2}  \lb \S(t,\g) + \Smu(t,\g)\rb + {\cal O}(b^3) \,, 
\end{aligned}
\end{equation}
where $\D(t,\g) = \D_{ij}(t) g_i g_j$, 
$\S(t)$ is the symmetric part of $\C(t)$, cf. Eq.~(\ref{W}), and the overall microscopic kurtosis tensor
\begin{equation} \label{Smu}
	\Smu(t) = \langle S(t) \rangle  
\end{equation}
contributes to the overall kurtosis tensor $\S(t)+\Smu(t)$. Hence, LTE cannot separate intrinsic and heterogeneity-based kurtosis contributions. In other words, while one can still  ``rotate" from $\Q(t)$ and $\T(t)$ via Eqs.~(\ref{QT_SA_relation}) to $\S(t)$ and $\A(t)$, the resulting $\S$ tensor will not yield the measured overall kurtosis via Eq.~(\ref{DKI_t}). 
To determine the $\Q(t)$, $\T(t)$ and $\Smu(t)$ tensors one has to go beyond Eq.~(\ref{eq:CumExpinB}), as discussed in Eqs.~(\ref{eq:DDE_cumExp})--(\ref{eq:SH_DDE}).
}

\iffalse 

\update{Above, the  source of kurtosis in Eq.~(\ref{DKI}) was solely the heterogeneity of compartment diffusion tensors $\Dc$, also responsible for the covariance  (\ref{eq:DCdefinitions}). 
	For non-Gaussian compartments, each characterized by its own microscopic kurtosis tensor $S(t)$, 
	$\Sig = \exp[- bD(t,\g) + \frac12 b^2 S(t,\g) + \O(b^3)]$ 
	for the narrow-pulse LTE \citep{STEJSKAL&TANNER1965}, 
	the DKI signal, compared to  Eq.~(\ref{DKI}), reads 
	\begin{equation} \label{DKI_t}
		\ln \la \Sig\ra = - b \D(t, \g) +   {b^2\over 2}  \lb \S(t,\g) + \Smu(t,\g)\rb + {\cal O}(b^3) \,, 
	\end{equation}
	where $\D(t,\g) = \D_{ij}(t) g_i g_j$, 
	$\S(t)$ is the symmetric part of $\C(t)$, cf. Eq.~(\ref{W}), and the overall microscopic kurtosis tensor
	\begin{equation} \label{Smu}
		\Smu(t) = \langle S(t) \rangle  
		%\quad \Smu(t,\g) =  {\Smu}_{ijkl}(t) g_i g_j g_k g_l \,, 
	\end{equation}
	contributes to the overall kurtosis tensor $\S(t)+\Smu(t)$. Hence, LTE cannot separate intrinsic and heterogeneity-based kurtosis contributions. In other words, while one can still  ``rotate" from $\Q(t)$ and $\T(t)$ via Eqs.~(\ref{QT_SA_relation}) to $\S(t)$ and $\A(t)$, the resulting $\S$ tensor will not yield the measured overall kurtosis via Eq.~(\ref{DKI_t}). 
	To determine the $\Q(t)$, $\T(t)$ and $\Smu(t)$ tensors one has to go beyond Eq.~(\ref{eq:CumExpinB}), as discussed in Eqs.~(\ref{eq:DDE_cumExp})--(\ref{eq:SH_DDE}).
	%Microscopic kurtosis also biases the estimation of $\C(t)$-tensor from Eq.~(\ref{eq:CumExpinB}). 
}

\fi

\subsection*{Invariants: intrinsic and mixed}\label{ss:invs}
\noindent
%Irreducible components, such as $\Sl, \Al$, or $\Tl, \Ql$, have some DOF whose values do not change upon rotations of the basis, i.e., they are rotationally invariant. 
%that are independent of the coordinate basis and also their orientation with respect to a given basis. 
%The following subsection discusses how many rotational invariants we can find in $\D$ and $\C$ and how to compute them.
Tensor invariants are the combinations of parameters that do not change upon rotations of the basis. 
We now construct all invariants of $\C$ within either QT or SA decomposition, by splitting them into the ones intrinsic to a given irreducible representation, and the ones mixing representations. 
Let us define {\it intrinsic invariants} as those that belong purely to a single irreducible component, such as $\Q^{(\ell)}$ and $\T^{(\ell)}$ in Eq.~(\ref{QT}), or $\S^{(\ell)}$ and $\A^{(\ell)}$ in Eq.~(\ref{SA}).  
Conversely, {\it mixed invariants} determine the relative orientations between different irreducible components with $\ell>0$, such as between $\Tt$ and $\Tf$. 
In what follows, we will focus on the invariants of the QT decomposition, Fig.~\ref{fig:RotInvsALL}, with all 
maps for a human brain shown in Fig.~\ref{fig:maps_QT}. 
A completely analogous treatment yields the corresponding $\S$ and $\A$ invariants, cf. %Supplementary Section \ref{SM:SA_invariants} and 
Supplementary Fig.~\ref{fig:maps_SA}.

How many intrinsic and mixed invariants are there? 
First note that the total number of  invariants for any tensor equals its number of DOF minus $3$ angles defining its overall orientation \citep{GHOSH2012}, yielding 3 for $\D$ (DTI), 12 for $\S$ or $\W$ (DKI), and 18 for $\C$. %the $\C$ tensor. 
Applying this argument to each irreducible representation, % --- either for Eq.~(\ref{QT}) or for Eq.~(\ref{SA}) --- 
the number of intrinsic invariants is 1 for $\ell=0$, and $(2\ell+1) - 3 = 2(\ell-1) = 2, \ 6,\ \dots$ for $\ell=2,\ 4, \ \dots\,$. 
This yields $(1+2+6)+(1+2)=12$ intrinsic invariants of $\C$, Fig.~\ref{fig:RotInvsALL}. 

%\dn{edits; changed Eq 20 to involve averages of $\S$ and $\A$ not $\T$ and $\Q$}
The isotropic, $\ell=0$ component of a symmetric tensor is an invariant, 
normalized here to its angular average as in Eq.~(\ref{MD}): 
\begin{equation}\label{ell0invs}
\begin{aligned}
\A_0  & \equiv \A^{00} =\overline{\A(\n)}= \tfrac13 \A_{ii} \,, \\  
\S_0  & \equiv \S^{00} = \overline{\S(\n)} = \tfrac15 \S_{iijj} \,, 
%	 \Q_0 &\equiv \Q^{00}  = \tfrac{5}{9} \,\S_0  + \tfrac2{9} \, \A_0\,, \\
%	 \T_0 & \equiv \T^{00}  = \tfrac{4}{9} \,\S_0  - \tfrac2{9} \, \A_0\,,
%	 \Q_0 &\equiv \Q^{00} = \overline{\Q(\n)} = \tfrac{5}{9} \,\S_0  + \tfrac2{9} \, \A_0\,, \\
%	 \T_0 & \equiv \T^{00} = \overline{\T(\n)} = \tfrac4{9} \,\S_0  - \tfrac2{9} \, \A_0\,,
\end{aligned}
\end{equation}
cf. Eq.~(\ref{eq_Slm}). 
%and the 
The respective invariants $\Q_0\equiv \Q^{00}$ and $\T_0\equiv \T^{00}$ are given in terms of $\S_0$ and $\A_0$ via Eqs.~(\ref{Q0}) and (\ref{T0}).

Consider now the irreducible components with $\ell=2$. 
We take $\Dt$ as an example, with the analogous definitions for $\Tt$, $\Qt$, $\St$, and $\At$. 
Among the 5 DOF, 3 angles define the orientation of the glyph $\Dt(\n)$ and are not invariants.
The remaining 2 DOF parametrize the 3 eigenvalues that sum up to zero trace. 
The corresponding 2 invariants can be written as traces of the powers of $\Dt$, where we introduce the index notation $\ell | n$ useful for higher $\ell$ in what follows: 
\begin{equation}\label{ell2invs}
\begin{aligned}
	\iDt{n} &=\left (  \tfrac23\, \mathrm{tr\,} (\Dt)^n \right)^{1/n} ,\quad \,\, n=2,3\,.
\end{aligned}
\end{equation}
%Since the $n=2$ ($L_2$-norm) invariants 
Since $n=2$ invariants ($L_2$-norm) play special role\cite{FRANK2002,MIRZAALIAN2016,REISERT2017}, we will henceforth drop the $n=2$ index for % all
degrees $\ell \geq 2$, Fig.~\ref{fig:RotInvsALL}. 

Based on the definition (\ref{ell2invs}), the square of $\D_2 \equiv \D_{2|2}$, 
\begin{equation} \label{VarD}
{\D_2}^2 
%=  \tfrac23\, \mathrm{tr\,} (\Dt)^2 
= 2\, \mathbb{V}_{\lambda}(\D)\,, \quad 
	\mathbb{V}_{\lambda}(\D) \equiv \frac13 \sum_{i=1}^3 (\lambda_i- \bar{\D})^2 , 
\end{equation}
is related to the variance $\mathbb{V}_{\lambda}(\D)$ of the eigenvalues of $\D$. 
Geometrically, ${\D_2}^2$ characterizes the {\it directional variance} 
\begin{equation} \label{VarD_dir}
	\mbox{var}\, \D(\n) = \overline{\D^2(\n)} - \D_0^2 = \tfrac15 {\D_2}^2 \,, \, {\D_2}^2 \equiv  \sum_{m} \D^{2m*} \,\D^{2m} , 
\end{equation}
cf. the expansion (\ref{D=Cart}) and spherical harmonics orthogonality. 
The equivalence of  the directional and eigenvalue variances  \citep{NOVIKOV2018} is readily seen in the eigenbasis:  
for $\Dt = \mathrm{diag\,}\{ \eta_1, \eta_2, -\eta_+\}$, where $\eta_\pm = \eta_1\pm\eta_2$, 
using $\D^{2 m} = \tfrac23\, {\Y^{2 m\, *}_{ij}} \,\D_{ij}$, cf. Eq.~(\ref{eq_Slm}) from {\it Methods}, 
we get $\D^{2,\pm2} = \eta_-/\sqrt{6}$, $\D^{2,\pm1} = 0$, and $\D^{20} = -\eta_+$, proving 
$\sum_m |\D^{2m}|^2 = 2\cdot\frac13 \, (\eta_1^2 + \eta_2^2  + \eta_+^2 )$, cf. Eq.~(\ref{VarD}). 

%\dn{$=2\det \Dt$, not $\det \D$ ? }
The invariants $\D_0$, $\D_2$, and  ${\D_{2|3}}^3 =2\det \Dt$ (the volume of the ellipsoid $\D(\n)$), 
%the 3 invariants of $\D$:  $\D_0=\iDz{1}$, $\D_2\equiv \iDt{2}$, and $\iDt{3}$ 
provide 3 symmetrized combinations of the eigenvalues of $\Dz\oplus\Dt$ which fully determine its shape. 

%\dn{added Eq 24 and moved other invs to Methods}
For the $\ell=4$ components $\Tf=\Sf$, the picture is more complex. The 3  angles determine the orientation of the glyph 
$\T^{(4)}(\n)=\sum_{m=-4}^4 \T^{4m}Y^{4m}(\n)$, while the remaining 6 DOF determine its shape (Fig.~\ref{fig:glyphsQTSA}a,b). 
Here we construct the principal invariant $\T_4\equiv \T_{4|2}$, the glyph variance analogously to Eq.~(\ref{VarD_dir}): 
\begin{equation} \label{VarTf_dir}
	\mbox{var}\, \Tf(\n) = \tfrac19 {\T_4}^2 \,, \,\,
	{\T_4}^2 \equiv  \sum_{m} \T^{4m*} \,\T^{4m} = \tfrac8{35}\, \Tf_{ijkl} \Tf_{ijkl} \,, 
\end{equation}
where the normalization factor in the last equality follows from Eq.~(\ref{eq:ThorneSTFprod}) in {\it Methods}.  
The remaining 5 intrinsic invariants are constructed in Eq.~(\ref{ell4invs}) of {\it Methods}. 

The relative angles between irreducible components of a given tensor do not change upon rotations (the tensor transforms as a whole). Hence, $18-12=6$ DOF define the two sets of mixed invariants of $\C$ that parametrize the relative rotations  between the frames of $\Tf$, $\Qt$, and $\Tt$. 
%\dn{removed tildes on ${\mathcal{R}}$ as well}
Without the loss of generality, we take them as the Euler angles that define active rotations 
$\tilde{\mathcal{R}}_\Tf(\alpha_\Tf,\beta_\Tf,\gamma_\Tf)$ and 
$\tilde{\mathcal{R}}_\Qt(\alpha_\Qt,\beta_\Qt,\gamma_\Qt)$ 
 of the $\Tt$ frame along $z$ by $\alpha$, then along new $x'$ by $\beta$, and along new $z''$ by $\gamma$ to obtain the $\Tf$ and $\Qt$ frames (Fig.~\ref{fig:RotInvsALL}).
 These are mapped for the human brain in the right panel of Fig.~\ref{fig:maps_QT} (cf. Supplementary Fig.~\ref{fig:maps_SA}  for $\S$ and $\A$).
%(This is equivalent to the product of extrinsic active rotations $\tilde{\mathcal{R}}(\tilde\alpha,\tilde\beta,\tilde\gamma)=X(\tilde\alpha) Y(\tilde\beta) Z(\tilde\gamma)$ in the fixed $\St$ frame to obtain the $\Sf$ and $\At$ frames). 

The major and minor symmetries of $\C$ mimic those of the elasticity tensor in continuous media. 
The proposed construction of tensor invariants via representation theory has the benefit of symmetry and geometric meaning, as compared to approaches within the elasticity theory \cite{BACKUS1970,BETTEN1987,BONA2004,MOAKHER2008}, and later in dMRI \cite{BASSER2007,QI2009,PAPADOPOULO2014}. Previous works used a Cartesian representation of $\C$ as a symmetric $6\times6$ matrix %with ``super"-indices $11, 22, 33, 23, 31, 12$ 
(Kelvin/Voigt notation,  Eq.~(\ref{eq:C_6x6}) in {\it Methods}). The invariants (generally, not a complete set) were introduced as coefficients of the  characteristic polynomial of two variables in the $6\times6$ matrix representation of $\C$  \cite{BETTEN1987,BASSER2007} or of $\S$ \cite{QI2009}, or via Hilbert's theorem on non-negative ternary quartics\cite{PAPADOPOULO2014} for $\S$. 
%While yielding a number of invariants (often an incomplete set), these approaches provided limited intuition for their geometric interpretation.  

%%%%%%%%%%%%%%%%%%%%%%%%%%%%%%%%
\begin{figure*}[htbp]
	\centering
	\includegraphics[width=0.95\textwidth]{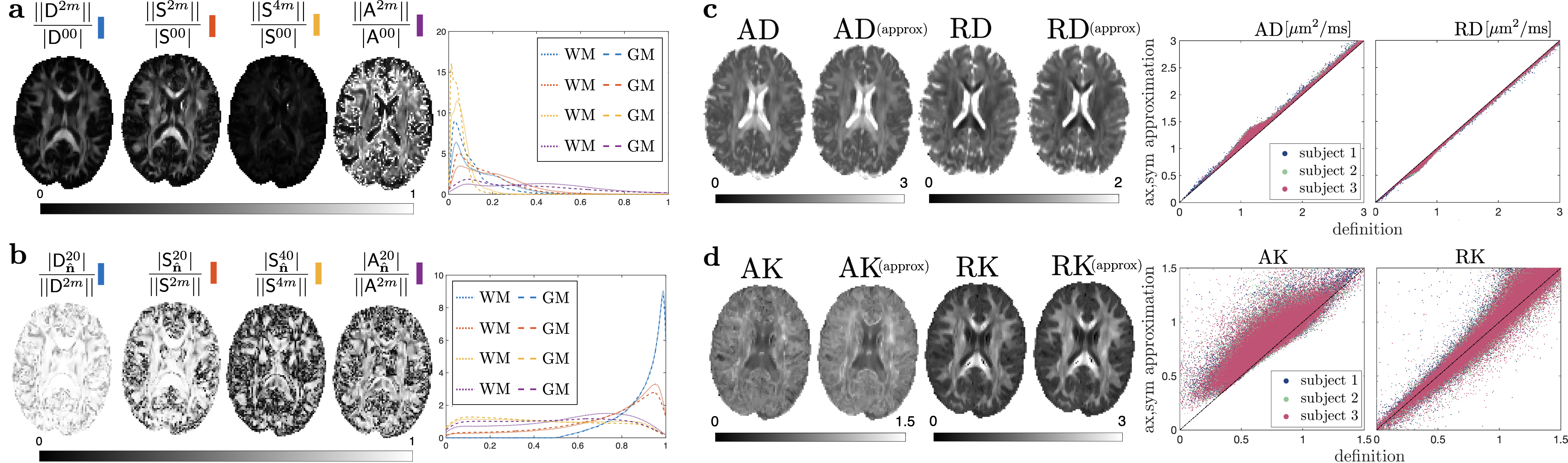}
	\caption[caption FIG]{Significance of $\ell>0$ and axial symmetry in healthy brains. 
		(a) Normalized maps of the $L_2$-norms for degree-$\ell$ components of $\D$, $\S\propto\W$, and $\A$, and their histograms for white and gray matter (WM, GM) voxels. 
		$\S^{4m}$ elements are $5-10\times$ smaller than $\S^{00}$. 
		(b) Relative contribution of the $m=0$ components 
		in the principal fiber coordinate frame, 
		such ratio $=1$ for perfect axial symmetry. 
		(c,d): Axial and radial projections of the diffusion (c) and kurtosis (d) tensors, 
		calculated both exactly and via Eqs.~(\ref{axsym}) relying on the axial symmetry approximation.}
	\label{fig:AxSymRICE}
\end{figure*}
%%%%%%%%%%%%%%%%%%%%%%%%%%%%%%%%

\subsection*{Previously used dMRI contrasts from RICE}
\noindent
The present unifying group theory approach allows us to express all previously used ${\cal O}(b^2)$ dMRI contrasts, sometimes related to specific acquisition schemes, via the smallest possible subset of the full system of invariants. 
Below we express the most popular contrasts  in terms of only 4 invariants: 2 from $\D$ ($\D_0$ and $\D_2$), 
and  2 from $\C$ ($\Q_0$  and $\T_0$). 

The ${\cal O}(b)$ contrasts, MD (\ref{MD}) and fractional anisotropy
\begin{equation}\label{FA}
		%\text{MD}  =  \D_0\,, \quad 
		\text{FA}  \equiv \sqrt{\frac32  \frac{{\mathbb{V}_{\lambda}(\D)}}{{\mathbb{V}_{\lambda}(\D)}+ \D_0^{2}}}	
		=\sqrt{ \frac{3 {\D_2}^2}{2 {\D_2}^2 + 4 {\D_0}^2 }} 
\end{equation}
[cf. Eq.~(\ref{VarD})], involve only $\D_0$ and $\D_2$. 
Although the dimensionless ratio $\D_2/\D_0$ is a more natural way to quantify the anisotropy of $\D$ tensor given Eq.~(\ref{VarD_dir}), 
the function (\ref{FA}) of $\D_2/\D_0$, bounded by $0\leq \mathrm{FA}\leq 1$, has been widely used; the bounding leads to a less transparent noise propagation. 
%Placing $\mathbb{V}_{\lambda}(\D) $ in the denominator makes its noise propagation properties less transparent. 

Including $\Q_0$  and $\T_0$ yields a number of ${\cal O}(b^2)$ contrasts. We begin with mean kurtosis (MK),  cf. Eqs.~(\ref{W}) and (\ref{S=Q+T}): 
\begin{equation} \label{MK}
	\mbox{MK}\equiv
%	 \W_0 = \frac{3\S_0}{{\D_0}^2} \,, \quad \S_0 = \Q_0+\T_0\,,
	 \overline{\W(\n)} = \W_0 = \frac{3\overline{\S(\n)}}{{\D_0}^2}= \frac{3\S_0}{{\D_0}^2} \,, \quad \S_0 = \Q_0+\T_0\,,
\end{equation}
%where any pair of $\{\T_0,\,\Q_0,\,\S_0,\,\A_0\}$ suffices, see Eqs.~(\ref{QT_SA_relation}).
%here defined via the directional average $\S_0 = \overline{\S(\n)}$ of the $\S$ tensor \cite{HANZHANG2006,HANSEN2013,JESPERSEN2017}, equivalent to its full trace, which  can be computed fast and precisely. 
defined via the directional average of $\W$ \cite{HANZHANG2006,HANSEN2013}. Eq.~(\ref{MK}) is sometimes referred to as mean kurtosis tensor (MKT) \cite{HANSEN2016}, as it is equivalent to the tensor trace $\W_0=\frac15 \W_{iijj}$ and, thus, can be computed fast and precisely. 
%This definition is  since it is equivalent to the tensor trace $\S_{iijj}$.} %and, thus, can be computed fast and precisely.}  
%\$\tfrac15 \W_{iijj} = \tfrac15 \tr \W =\W_0$, which  can be computed fast and precisely from the estimated $\W$ tensor. 
%The definition of mean kurtosis is ambiguous in the dMRI literature. In this work we define MK (\ref{MK}) 
The original DKI paper \cite{JENSEN2005} defined mean kurtosis as the directional average of 
$K({\bf \hat{n}})= 3 \S(\hat{\mathbf{n}})/ \D^2(\hat{\mathbf{n}})$. 
%Note that both $\bar{\W}$ and $\bar{K}$ are dimensionless and have qualitatively similar contrasts. 
While perhaps more intuitive, that definition suffers from two drawbacks. First, it cannot be compactly represented as a trace of a tensor [akin to Eqs.~(\ref{ell2invs}) and (\ref{ell4invs})] due to a nontrivial directional dependence of the denominator. Indeed, 
as $K(\hat{\mathbf{n}})$ involves an {\it infinite series} in the powers of $\hat{\mathbf{n}}$ due to the expansion of  $1/\D^2(\n)$ in the powers of $\Dt(\n)$, it cannot be written as a convolution of a product ${n}_i {n}_j\dots$  with a tensor or even a few tensors \cite{ADESARON2024}. Second, $\overline{K({\bf \hat{n}})}$ has lower precision than Eq.~(\ref{MK}), and is strongly affected by outliers coming from low-diffusivity directions, for which 
$\D(\hat{\mathbf{n}})\ll \D_0$.

The curvature $\mathbb{V}_\mathrm{I} = \Q_0$ of the STE cumulant series $\L_{\rm STE}$ [item (iii) below Eqs.~(\ref{eq:MGC_SH})] is known as the {\it isotropic variance} \cite{TOPGAARD2017,WESTIN2016}. 
Speaking precisely, it is the variance of isotropic components of compartment tensors. 
The directional average of the LTE signal, represented in the cumulant form 
$\ln \langle \bar S_{\rm LTE}\rangle = -b\D_0 +  \frac12 b^2 \mathbb{V}_\mathrm{A} + \O(b^3)$, 
defines the so-called {\it anisotropic variance} \cite{SZCZEPANKIEWICZ2016,LASIC2014}   
\begin{equation} \label{VA}
	\mathbb{V}_\mathrm{A} \equiv \la \overline{D^2(\n)} \ra - {\D_0}^2 = \tfrac25 \la \mathbb{V}_\lambda (D) \ra = \tfrac15 \la {D_2}^2\ra =  \T_0 +  \tfrac15 {\D_2}^2 ,
\end{equation} 
proportional to the variance (\ref{VarD}) of the eigenvalues of compartmental $\Dc$ averaged over $\P(\Dc)$. 
Note that taking spherical mean of LTE signal before the logarithm makes $\mathbb{V}_\mathrm{A}$ a moment, rather than a cumulant: 
$\la {D_2}^2\ra = \sum_m \la D^{2m*} \,D^{2m} \ra = 5\T_0 + \sum_m \D^{2m*} \D^{2m}$, cf. Eq.~(\ref{eq:T00}).  
The above variances define the isotropic and anisotropic kurtoses  $K_\mathrm{I,A} = 3 \mathbb{V}_\mathrm{I,A}/\D_0^2$, cf. Eq.~(\ref{MK}) \cite{TOPGAARD2017,WESTIN2016,LASIC2014,SZCZEPANKIEWICZ2016,NETOHENRIQUES2020}. 
Note that this somewhat misleading nomenclature refers to  iso/anisotropic components of compartmental diffusion tensors, while all the above quantities correspond to the {\it fully isotropic} ($\ell=0$) parts of  $\C$,  Eq.~(\ref{DxD}). 

Finally, the analogy with Eq.~(\ref{FA}) has motivated the definition of {\it microscopic fractional anisotropy} \cite{JESPERSEN2013,LASIC2014,WESTIN2016,SZCZEPANKIEWICZ2016}
\begin{equation}
	\label{muFA}
	\upmu\mbox{FA}  
	 \equiv  \sqrt{\frac32  \frac{\la \mathbb{V}_{\lambda}(D)\ra}{\la \mathbb{V}_{\lambda}(D)\ra+ {\D_0}^{2}}}
	= \sqrt{\frac{15 \T_0  + 3 {\D_{2}}^2}{10 \T_0  + 2 {\D_{2}}^2 + 4 {\D_0}^2}} \,. \quad
\end{equation}
%Representing \mFA in the invariant form (the second equality above) enables its evaluation without the need to assume axial symmetry for $\Dc$ and/or a specific functional form for $\P(D)$, as in Refs.~\cite{LASIC2014,TOPGAARD2017}. 
Writing \mFA using RICE invariants %(the second equality above) 
enables its evaluation without the need to assume axial symmetry for $\Dc$ and/or a specific functional form for $\P(D)$, as in Refs.~\cite{LASIC2014,TOPGAARD2017}. 
Note that the definition of \mFA is inconsistent, in a sense that the averages over $\P(\Dc)$ are taken in the numerator and denominator separately, i.e., \mFA
is {\it not} the compartmental $\mbox{FA}(\Dc)$ averaged over $\P(D)$ \cite{NETOHENRIQUES2019}. Unfortunately, information for calculating voxel-averaged $\la \mbox{FA}(\Dc)\ra$ is not contained in the $\mathcal{O}(b^2)$ signal, as it involves all higher-order cumulants.

All the above 6 popular contrasts --- $\W_0$, $\mathbb{V}_\mathrm{I,A}$, $K_\mathrm{I,A}$, \mFA --- rely just on two $\ell=0$ invariants of $\C$, and are thereby redundant. Contrasts involving higher-degree 
%\dn{higher-degree?}
invariants have been largely unexplored. Empirically, one can rationalize this as the invariants decrease with $\ell$, shown in Fig.~\ref{fig:AxSymRICE}a for the $\ell=2$ and $\ell=4$ principal invariants relative to their $\ell=0$ counterparts. 
For example, the 9 elements of $\Sf$ are $\sim 5-10$ times smaller than $\S_0$. 
Such decrease with $\ell$ is more pronounced in gray matter, and less so in crossing fibers or highly aligned white matter regions, e.g., the corpus callosum. 

Invariants with $\ell>0$ quantify the anisotropies of the irreducible components of $\C$  
(while \mFA does not characterize the anisotropy of any specific tensor). Is there a natural definition for an overall anisotropy measure of a higher-order tensor, analogous to $\D_2$ or FA for the 2nd-order tensor? Our approach suggests that each anisotropic irreducible component is characterized by its own independent invariants, such as $\S_2$ and $\S_4$ for the $\S$ tensor. Combining such invariants into a single metric seems arbitrary and mixes different symmetry sources. An example is the {\it kurtosis fractional anisotropy} (KFA) \cite{HANSEN2013,GLENN2015}:
%\dn{removed $\delta_{(ij}\delta_{kl)}$ from numerator since we subtract tensors not components}
\begin{equation}\label{eq:KFA}
\begin{aligned}
\mathrm{KFA} &\equiv \frac{\| \S - \Sz   \|_\mathrm{F}}{\| \S \|_\mathrm{F}} = %\sqrt{ \frac{(\S_{ijkl} - \S_0  \delta_{(ij}\delta_{kl)})(\S_{ijkl} - \S_0  \delta_{(ij}\delta_{kl)})}{\S_{ijkl}\S_{ijkl}}} \\
%&= 
\sqrt{ \frac{14 {\S_2}^2 + 35 {\S_4}^2}{40 {\S_0}^2 + 14 {\S_2}^2 + 35 {\S_4}^2}},
\end{aligned}
\end{equation}
where we express the Frobenius norm 
$\| \S \|_\mathrm{F}^2 \! \equiv \! \S_{ijkl} \S_{ijkl} \! =\! \sum_{\ell} \zeta(4,\ell) \, {\S_\ell}^2$  in terms of the principal invariants $\S_\ell$ using Eq.~(\ref{eq:alpha_STF}). 
Akin to FA, KFA is bounded between 0 and 1. However, it seems equally reasonable to replace the Frobenius norm with the directional average 
$\overline{\S^2(\n)} = \sum_{\ell}  \S_\ell^2/(2\ell+1)$, yielding different relative weights with which the invariants $\S_\ell$ would enter. As there is no ``best'' combination of invariants with different $\ell$, it is generally logical to consider them separately. 

There is a case where combining invariants with different $\ell$ is natural. 
The {\it axial and radial kurtoses}  \cite{HUI2008,JENSEN2010} $\W_\|$ and $\W_\perp$ are projections of the kurtosis tensor $\W = 3\S/{\D_0}^2$ [Eq.~(\ref{W})] onto the principal fiber axis $\hat{\V}$ (the eigenvector for the largest eigenvalue of $\D$), and transverse to it. 
In general, they are not rotational invariants %\dn{added ``inherent"}
inherent to $\W$, as $\W\propto \S$ does not ``know'' about the tensor $\D$ and its eigenvectors. 
However, in voxels with a dominant fiber direction  $\hat{\V}$, microstructure aligns the $\S$ and $\D$ tensors, Fig.~\ref{fig:maps_QT}, 
and axial/radial nomenclature becomes meaningful. 
Under an extra assumption of axial symmetry of $\S$ around $\hat{\V}$, 
we express its projections  via the principal invariants for $\ell=0, 2, 4$:% 
\begin{subequations} \label{axsym}
\begin{align}
\D_\|^\text{ax,sym}    &= \D_0 +  \D_2 \,, \\
\D_\perp^\text{ax,sym} &= \D_0 - \tfrac{1}{2} \D_2 \,,  \\
%\W_\|^\text{ax,sym}    &= \W_0 +   \W_2 + \W_4 , \\ 
%\W_\perp^\text{ax,sym} &= \W_0 - \tfrac{1}{2} \W_2 + \tfrac{3}{8} \W_4 
\W_\|^\text{ax,sym}    &= \tfrac{3}{{\D_0}^2} \left(\S_0 +   \S_2 + \S_4 \right) , \\ 
\W_\perp^\text{ax,sym} &= \tfrac{3}{{\D_0}^2} \left(\S_0 - \tfrac{1}{2} \S_2 + \tfrac{3}{8} \S_4  \right) ,
\end{align}%\label{eq:AxSymApprox}
\end{subequations}
without the need to transform to the eigenbasis of $\D$. 
Indeed, in the basis with  $z$-axis along $\hat{\V}$, only $m=0$ components are nonzero for each $\ell$. 
This implies that $\D_\ell = |\D^{\ell 0}|$ and $\S_\ell =  |\S^{\ell 0}|$. 
In this basis, $Y^{\ell 0}(\hat{\mathbf{z}})=1$ for $\|$ invariants; the average of $Y^{\ell 0}(\n) = P_\ell(\n\cdot {\bf \hat z})$ around the equator for $\perp$ invariants is given by 
 Legendre polynomial values $P_2(0) = -\frac12$ and $P_4(0)=\frac38$.
%Of course, the ``axial'' and ``radial'' nomenclature is meaningful only when there is a dominant fiber direction in a voxel. 

How valid is the axial symmetry assumption?
In Fig.~\ref{fig:AxSymRICE}b,  we rotate all voxels' principal fiber axes $\n$ to $\z$, and quantify the  axial symmetry in each irreducible component $\F^{(\ell)}$ via the ratios $|\F^{\ell 0}_{\n}|/\F_\ell$ of the $m=0$ component to its principal invariant, the 2-norm $\F_\ell = \| \F^{\ell m}\|$ over all $m$. 
For the $\D$ tensor, the ratio $|\D^{20}_{\n}|/\D_2 = (1 + \eta_-^2/3\eta_+^2)^{-1/2}$ in the notations after Eq.~(\ref{VarD_dir}). 
Perfect axial symmetry would correspond to the unit ratios. 
%(Note that these ratios characterize how axially symmetric $\F^{(\ell)}$ is relative to its 3d glyph; to quantify strictly in-plane axial symmetry transverse to $\n$, do not measure a strictly in-plane, 2d degree of axial symmetry, as they also depend on .) 
Fig.~\ref{fig:AxSymRICE}b shows that $\Dt$ and $\St$ have a high axial symmetry, while the opposite holds for $\Sf$ and $\At$. 

Figures \ref{fig:AxSymRICE}c and \ref{fig:AxSymRICE}d show axial and radial projections for %of the $\D$ and $\S$ tensors: 
$\D$ and $\S$:
the exact ones calculated by projecting onto principal fiber direction $\hat{\V}$ and transverse to it, and the ones via assuming axial symmetry, Eqs.~(\ref{axsym}) . 
%A good agreement between the exact and approximate expressions in found in the whole brain, although $\Sf$ may not be always axially symmetric, because its components are much smaller than $\Sz$ and $\St$. 
Although $\Sf$ is generally not axially symmetric, a good agreement between exact and approximate expressions in found in the whole brain because $\Sf$  are much smaller than $\Sz$ and $\St$.

\subsection*{New invariants}
\noindent
The QT decomposition is motivated by the separation of different sources of covariances of compartmental diffusivities. While the $\ell = 0$ parts of $\T$ and $\Q$ have been studied in the dMRI literature under different names, their anisotropic complements (with $\ell>0$) have remained largely unexplored. 
They provide access to novel rotationally invariant dMRI contrasts.

Overall, just 6 independent invariants: $\D_0$ and $\D_2$;  $\S_0$ and $\A_0$ (equivalently, $\Q_0$ and $\T_0$); and $\S_2$ and $\S_4$,  are enough to synthesize all previously used model-independent dMRI contrasts up to ${\cal O}(b^2)$. 
Since individual DTI eigenvalues may be used as contrasts, the total number of previously studied  ${\cal O}(b^2)$ invariants (explicitly or implicitly) is at most 3 from $\D$ ($\D_0$, $\D_{2}$, $\iDt{3}$), and at most 4 from $\C$ ($\A_0$, $\S_0$, $\S_{2}$, $\S_{4}$).

The remaining 14 invariants of the $\C$ tensor contain essentially unexplored information. Their definition, symmetries, and geometric meaning constitute the main results of the present comprehensive group theory-based approach.  Just as an example, one can think of a novel physically motivated contrast --- the size-shape correlation coefficient 
\begin{equation} \label{SSC}
	\mathrm{SSC} \equiv \frac{\left \|\lla \Dc^{00} \, \Dc^{2m} \rra\right \|}{\sqrt{\lla (\Dc^{00})^2\rra \cdot \textstyle \sum_m \lla \Dc^{2m} {\Dc^{2m}}^* \rra}} 
	= \frac{1}{2} \frac{{\Q_2}}{\sqrt{5\,\Q_0 \T_0}}
\end{equation}
that involves $\Q_2$ and is normalized between $0$ and $1$, where SSC=0 for independent shapes and sizes and SSC=1 for a linear relationship. 
%the dependence of compartmental shapes and sizes. 
The $\ell = 2$ and $\ell =4$ sectors of $\T$ have not been looked upon at all. 
The six mixed invariants relate to the underlying fiber tract geometry, quantifying correlations  between eigenframes of different irreducible components. 
In particular, small relative angles $\tilde\beta$ in Fig.~\ref{fig:maps_QT} within major white matter tracts are set by the underlying aligned fiber geometry.

Such a large set of complementary tissue contrasts is  well suited for machine learning algorithms to study human development, aging and disease. Much like RGB pictures contain $N \!=\!3$ colors, $N \!=\!21$ invariant contrasts can be viewed as a large-$N$ generalization of computer vision data, prompting the development of large-$N$ classifiers.

A comprehensive $\B$-tensor encoding human dataset has been recently made public \cite{SZCZEPANKIEWICZ2019b}, from which all SA and QT invariants can be determined and studied. 
For invariants not involving the $\A$ tensor, one can explore hundreds of thousands human data sets from imaging consortia such as the Alzheimer’s Disease Neuroimaging Initiative (ADNI) \cite{CLIFFORD2008}, Human Connectome Project \cite{JAHANSHAD2013}, UK Biobank \cite{MILLER2016}, Adolescent Brain Cognitive Development (ABCD) \cite{VOLKOW2018}, and Cambridge Centre for Ageing and Neuroscience 
%Cam-CAN 
data repository \cite{TAYLOR2017}, 
which are all compatible with the DKI  representation (\ref{DKI}). 
Below we  demonstrate the clinical relevance  of the added information content from novel invariants in detecting neurodegeneration.

%\dn{different thoughts}

%We also note that such a large set of complementary tissue contrasts is  well suited for machine learning algorithms to study human development, aging and disease. Much like RGB pictures contain $N \!=\!3$ colors, $N \!=\!21$ invariant contrasts can be viewed as a large-$N$ generalization of computer vision data, prompting the development of large-$N$ classifiers.  

%\cite{CLIFFORD2008,JAHANSHAD2013,GLASSER2016,MILLER2016,VOLKOW2018,TAYLOR2017}
\subsection*{Multiple sclerosis classification based on clinical dMRI}
\noindent 
We evaluated the clinical relevance of RICE invariants in a cohort of 1189 
subjects that received a clinically dedicated brain MRI. This included 627 multiple sclerosis patients and 562 age- and sex-matched healthy controls. 
%	participants, 
%including 562 healthy controls (aged $42.9 \pm 14.1$ years, 386 females) and 627 MS patients (aged $42.7 \pm 13.6$ years, 443 females). 
%including 562 healthy controls and 627 MS patients (age and sex matched). 
Two-shell LTE dMRI was acquired \cite{LIAO2024,CHEN2024}, see {\it Methods} for details. 
All clinical scans were retrospectively analyzed, and invariants were computed using DTI, DKI, and $\mathrm{RICE}_\mathrm{LTE}$. 
Logistic regression models were trained using all DTI–DKI–$\mathrm{RICE}_\mathrm{LTE}$ invariants as predictors, where $\mathrm{RICE}_\mathrm{LTE}$ contains all invariants of the kurtosis tensor $\W\propto \S$ in the SA decomposition of $\C$. Due to LTE-only data, the $\A$-tensor invariants were inaccesible.

\begin{figure}[H]
	\centering
	\includegraphics[width=0.90\columnwidth]{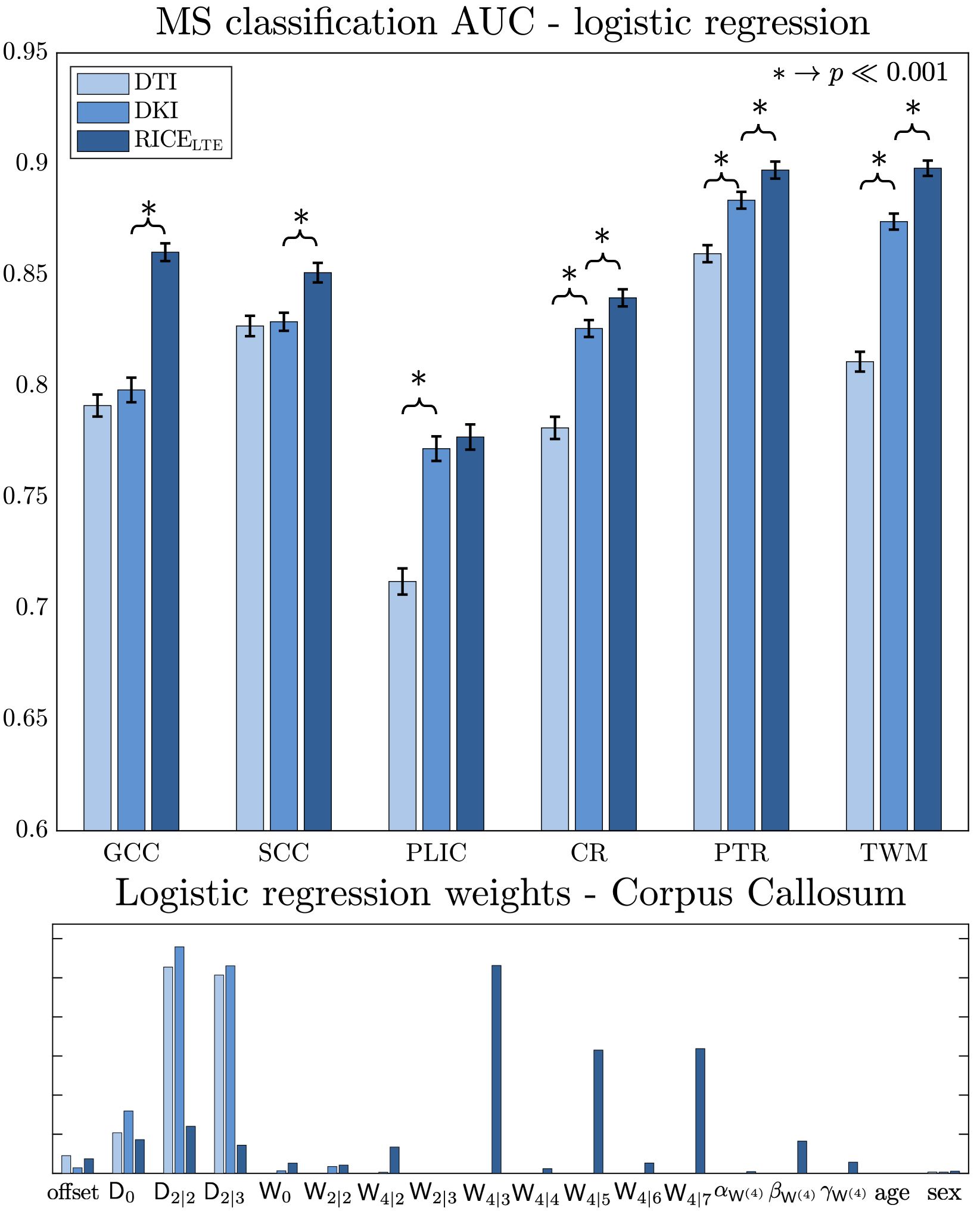}
	\caption[caption FIG]{Regression performance for classifying multiple sclerosis from controls based on median values across white matter regions (genu of corpus callosum, GCC; splenium of corpus callosum, SCC; Posterior limb of internal capsule, PLIC; Corona Radiata, CR; Posterior thalamic radiation, PTR; Total white matter, TWM). AUCs were averaged over 100 stratified bootstrap iterations for logistic regressions based on DTI, DKI, and $\mathrm{RICE}_\mathrm{LTE}$ invariants on 238 test subjects ($p\ll0.001$, denoted by $*$, error bars denote $95\%$ confidence interval of the means). Regression weights averaged for the corpus callosum are displayed in the bottom plot.
	}\label{fig:MSAUC}
\end{figure}

%To assess the practical utility of the complete set of RICE invariants, we trained a 
Classification was evaluated with stratified bootstrap resampling ($80\%$ training, $20\%$ testing), preserving group proportions and sex balance. This yielded a distribution of area under the ROC curve (AUC) values across iterations, providing a robust measure of model performance.

Figure~\ref{fig:MSAUC} shows AUC values in six different white matter regions involved in multiple sclerosis, extracted from the Johns Hopkins University white matter atlas \cite{MORI2005}. DKI outperforms DTI, as expected due to a more informative dMRI acquisition (an extra LTE $b$-shell). 
Remarkably, \rvleft[R4]{} \update{up to a $30\%$ reduction in pairwise ranking error (ranking loss, $1-\mathrm{AUC}$) is achieved} via $\mathrm{RICE}_\mathrm{LTE}$ invariants based on the same DKI acquisition, essentially for free, underscoring the nontrivial information content of the present approach. Regression weights highlight the importance of incorporating invariants beyond the conventional six, see \textit{Previously used dMRI contrasts from RICE}.

%This analysis demonstrates that RICE invariants are not only mathematically independent, but also biologically and clinically meaningful and informative. Incorporating non-LTE data yielding  the asymmetric part $\A$ of the covariance tensor, is likely to further enhance RICE's classification performance.

%\update{In regions such as the genu of the corpus callosum, the improvement corresponds to a $\sim30\%$ reduction in ranking error, with more moderate gains elsewhere. These increases are noteworthy given that the invariants are extracted retrospectively from standard LTE acquisitions without additional data or model-based fitting. Because our analysis was intentionally conservative in both anatomical scope and model complexity, the reported gains represent a lower bound on their potential clinical utility.}
This analysis demonstrates RICE invariants are not only mathematically independent but also biologically and clinically informative. 
%\dn{see changes}
\rvleft[R4]{} 
\update{Since these invariants were extracted retrospectively from standard 2-shell LTE acquisitions, with no extra data involved, the classification improvement represents pure gain in performance based on our symmetry-based approach. As our analysis was intentionally restricted to individual anatomical regions and summary statistics (median values within a small number of large ROIs), and employed simple logistic regression,  the reported gains should be interpreted as a lower bound on potential information gain.} 
Incorporating non-LTE data, yielding the asymmetric part $\A$ of the covariance tensor, is likely to further enhance RICE's classification performance.

%%%%%%%%%%%%%%%%%%%%%%%%%%%%%%%%%%%%%%%%%%%%%%%%%%%%%%%%
\begin{table*}[th!!]%[th!!]
	\centering
	\caption{%Comparison of existing and proposed fast protocols. 
		Theoretically minimal protocols  contain the minimum unique number of directions and distinct $b$-values (specific %exemplary 
		$b$-values can be altered). For STE, more than 1 direction implies  rotation of the waveform for accuracy. All protocols include a %non-diffusion weighted image ($b=0$)
		$b=0$ image to estimate $S|_{b=0}$.}\vspace{0.05cm}
	\begin{tabular}{@{}l@{}c@{}c@{}}
		\hline
		\multicolumn{3}{c}{Fast protocols comparison, $b$-values are in $\unit{ms}/\unit{\upmu m^2}$} \\
		\hline
		Output maps &  MD+MK & MD+FA+MK \\
		\hline
		\rowcolor{lightgray}
		& $6\!\times\!\mathrm{LTE}_{b=2}$   & \\
		\rowcolor{lightgray} 	\multirow{-2}[0]{*}{Theoretical minimum}  & $1\!\times\!\mathrm{STE}_{b=1}$   &  \multirow{-2}[0]{*}{$\,\,6\!\times\!\mathrm{LTE}_{b=1};\,6\!\times\!\mathrm{LTE}_{b=2}\,\,$}  \\
		%		\hline
		\vspace{0.01cm}\bcicosaedre  iRICE (MD+FA+MK) &       & \multirow{-2}[0]{*}{$6\!\times\!\mathrm{LTE}_{b=1};\,6\!\times\!\mathrm{LTE}_{b=2}$} \\
		%		\hline
		\rowcolor{lightgray}
		&     & \\
		\rowcolor{lightgray} 	\multirow{-2}[0]{*}{Ref.~\cite{HANSEN2013,HANSEN2016b}}  & \multirow{-2}[0]{*}{$3\!\times\!\mathrm{LTE}_{b=1};\,9\!\times\!\mathrm{LTE}_{b=2}$}   & 
		\multirow{-2}[0]{*}{$9\!\times\!\mathrm{LTE}_{b=1};\,9\!\times\!\mathrm{LTE}_{b=2}$}   \\

		%		\vspace{0.22cm} Ref.~\cite{HANSEN2013} & $3\!\times\!\mathrm{LTE}_{b=1};\,9\!\times\!\mathrm{LTE}_{b=2}$ &  \\
		\hline
		Output maps &  MD+MK+${\upmu}$FA & MD+FA+MK+${\upmu}$FA \\ 
		%		\textbf{Output maps} &  \textbf{MD+MK+$\boldsymbol{\upmu}$FA} & \textbf{MD+FA+MK+$\boldsymbol{\upmu}$FA} \\ 
		\hline
		\rowcolor{lightgray}
		& $6\!\times\!\mathrm{LTE}_{b=2}$   & $6\!\times\!\mathrm{LTE}_{b=1};\,6\!\times\!\mathrm{LTE}_{b=2}$ \\
		\rowcolor{lightgray}
		\multirow{-2}[0]{*}{Theoretical minimum} & $1\!\times\!\mathrm{STE}_{b=1};\,1\!\times\!\mathrm{STE}_{b=1.5}$ & $1\!\times\!\mathrm{STE}_{b=1.5}$ \\
		%		\hline
		\multirow{2}[0]{*}{\bcicosaedre  iRICE (MD+FA+MK+\mFA)} &       & $6\!\times\!\mathrm{LTE}_{b=1};\,6\!\times\!\mathrm{LTE}_{b=2}$ \\
		&       & $3\!\times\!\mathrm{STE}_{b=1.5}$ \\
		%		\hline
		\rowcolor{lightgray}
		& $3\!\times\!\mathrm{LTE}_{b=0.1};\,3\!\times\!\mathrm{LTE}_{b=0.7};\,6\!\times\!\mathrm{LTE}_{b=1.4};\,6\!\times\!\mathrm{LTE}_{b=2}$ &  \\
		\rowcolor{lightgray}
		\multirow{-2}[0]{*}{Ref.~\cite{NILSSON2019}}& $\,\,6\!\times\!\mathrm{STE}_{b=0.1};\,6\!\times\!\mathrm{STE}_{b=0.7};\,10\!\times\!\mathrm{STE}_{b=1.4};\,16\!\times\!\mathrm{STE}_{b=2}\,\,$ &  \\
		\hline
	\end{tabular}%
	\label{tab:TableProtocolsComparison}
\end{table*}
%%%%%%%%%%%%%%%%%%%%%%%%%%%%%%%%%%%%%%%%%%%%%%%%%%%%%%%%%%%%

\subsection*{Minimal protocols: iRICE}
\noindent
%Clinical dMRI is constrained by scan time. A practical compromise is to use fewer measurements (much fewer than $1+6+21$) to estimate only a few $\ell=0$ invariants, which are the most pronounced (Fig.~\ref{fig:AxSymRICE}a). The orthogonality of the spherical-tensor (spherical-harmonic) basis ensures unbiased estimation of any $\ell$ sector without computing the full set. Moreover, arranging $\B$-tensors into spherical designs yields the minimal number of directions needed to preserve this orthogonality. These ideas lead to the minimal iRICE (“instant RICE”) protocols described below.
Clinical dMRI is constrained by scan time. A compromise could be to use a small number of measurements (much fewer than $1+6+21$) to estimate just a few $\ell=0$ invariants, given that they are most pronounced, Fig.~\ref{fig:AxSymRICE}a. The orthogonality of %the %spherical tensor / 
spherical harmonics %basis 
ensures an unbiased estimation of any sector $\ell$ of spherical tensor components without having to estimate the entire set. Furthermore, arranging $\B$-tensors into {\it spherical designs} provides the minimum number of directions that guarantee such orthogonality. These ideas result in the minimal iRICE (``instant RICE") protocols, as described below. 

We first note that since the axially symmetric family (\ref{eq:AxSymB}) couples to all components of $\C$, without the loss of generality we can consider spherical designs on $\mathbb{S}^2$ rather than on the SO(3) manifold $\mathbb{S}^3/\mathbb{Z}_2$ of generic $\B$-tensors. 
%, and cancel all $\ell>0$ components in Eq.~(\ref{eq:MGC_SH}). 
Next, we remind that a spherical $L-$design on $\mathbb{S}^2$  is a set of $N_L$ points $\{\g^\alpha \in \mathbb{S}^2 \}_{\alpha=1}^{N_L}$  on a unit sphere, that  satisfies \citep{SEYMOUR1984}
\begin{equation}\label{eq:SphDesign}
	\frac1{N_L}  \sum_{\alpha=1}^{N_L} f^{(L)}(\g^{\alpha}) \equiv \int_{\mathbb{S}^2}  f^{(L)}(\g)\, {\d\Omega_{\g} \over 4\pi}  = f^{00}
\end{equation}
for any degree-$L$ function $f^{(L)}(\g)$, i.e., expressible as a spherical harmonics series up to degree $L$. 
%a function that can be expressed by a degree $L$ truncated spherical harmonics series. 
In other words, any integration over $\mathbb{S}^2$ is exactly represented by a finite sum (\ref{eq:SphDesign}) for any rotation of the set $\{\g^{i}\}$, provided that the integrand is degree-$L$, and yields its angular average $f^{00}$. 
In particular, taking $f^{(L)}(\g) = Y^{\ell m \, *}(\g) \, Y^{\ell' m'}(\g)$, the orthogonality of $Y^{\ell m}(\g)$ and $Y^{\ell' m'}(\g)$ is ensured exactly in the finite-$N_L$ sums (\ref{eq:SphDesign}) for all $ \ell+\ell' \leq L$. 
%Thus, SH terms of degree $\ell$ and $\ell'$ can be isolated with these discrete measurements if $\ell+\ell'\leq L$.
The smallest 2- and 4-designs are provided by tetrahedron and icosahedron vertices, $N_2\!=\!4$ and $N_4\!=\!12$, respectively. However, for antipodal-symmetric $f^{(L)}(\n) = f^{(L)}(-\n)$ relevant for dMRI, $N_2\!=\!3$ reduces to half of the octahedron vertices, and $N_4\!=\!6$ to half of the icosahedron vertices, cf. Supplementary Section \ref{SM:SphDesigns}. 
%$N_L$ reduces to half of the octahedron vertices for $L=2$ [the $N_2=3$ cyclic permutations of $\n = (1,0,0)$], and to half of the icosahedron vertices for $L=4$ (the $N_4=6$ cyclic permutations of $\n = \tfrac{1}{\sqrt{1+\varphi^2}} (1, \pm \varphi,0)$, where $\varphi = (1+\sqrt{5})/2$), cf. Supplementary Section \ref{SM:SphDesigns}. 

\begin{figure*}[th!!]
	\centering
	\includegraphics[width=0.84\textwidth]{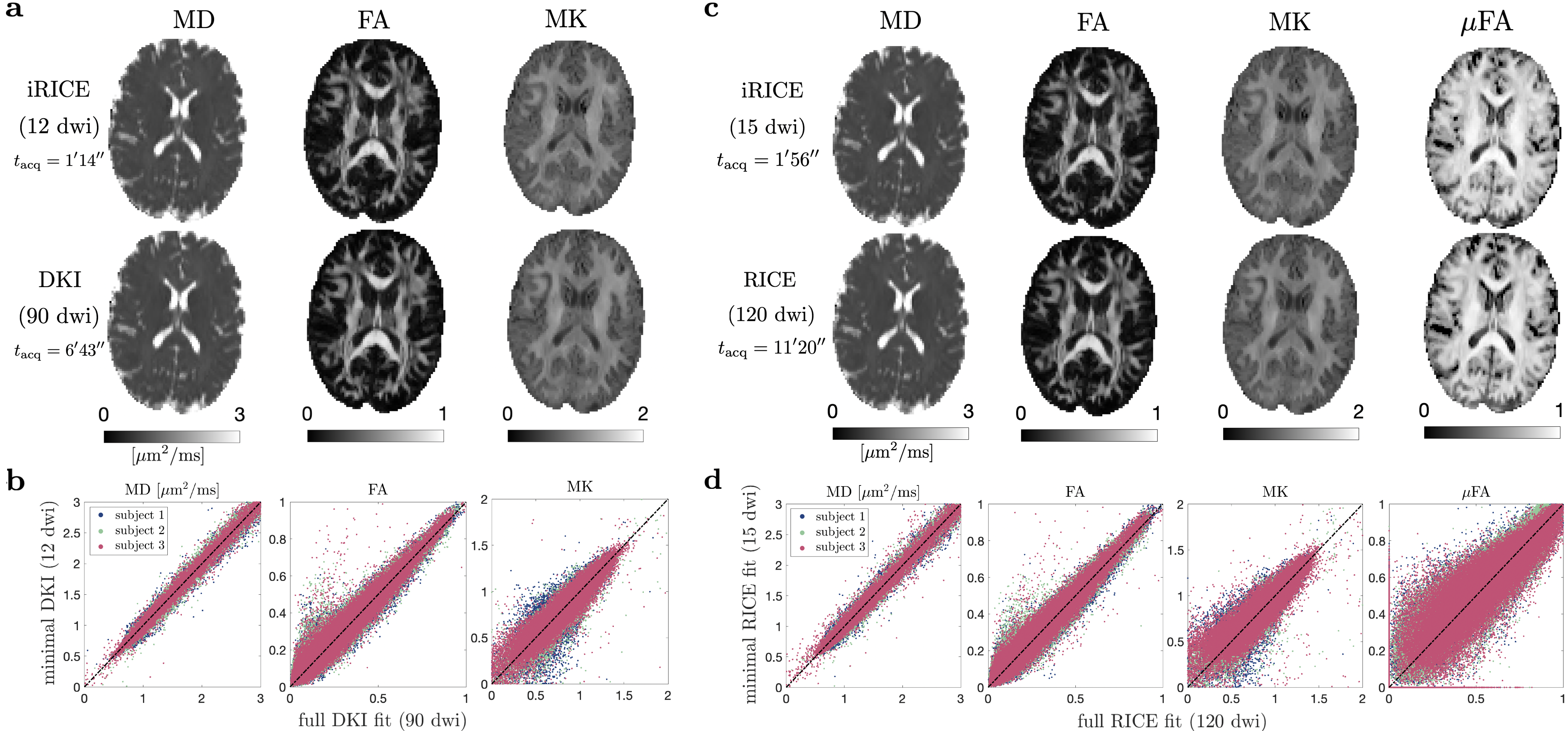}
	\caption[caption FIG]{Comparison of iRICE (1-2 minutes) with fully sampled acquisitions (6-11 minutes). 
		(a,c): iRICE maps (top) vs fully sampled DKI maps (bottom) for a healthy volunteer. Panels (b) and (d) show scatter plots for all brain voxels in 3 normal volunteers (3 colors).
	}\label{fig:bothfastRICEcomparison}
\end{figure*}

%\dn{$\g\to \g^\alpha$ }
Constructing fixed-$b$ shells (\ref{eq:AxSymB}) from 4-designs $\{\g^\alpha\}$ allows us to become insensitive to  $\ell=4$ terms in Eq.~(\ref{eq:MGC_SH}) and fit 
\begin{align} \nonumber
	&\ln {\Sig(b,\g^\alpha) \over \Sig|_{b=0}} = 
%	&\ln S(b,\g^\alpha) = 	
	- b \left(\D_0 + \beta \sum_{m=-2}^2 \D^{2m} \, Y^{2m}(\g^\alpha)  \right) \\
	& + \frac{b^2}{2} \left ( \Q_0 + \beta^2 \T_0 
	+  \sum_{m=-2}^2  \left( \beta \Q^{2m} + \beta^2 \T^{2m} \right)  Y^{2m}(\g^\alpha) \right ).
%	\ln {S(b,\g^\alpha) \over S|_{b=0}} &= 
%- b \left(\D_0 + \beta \sum_{m=-2}^2 \D^{2m} \, Y^{2m}(\g^\alpha)  \right) 
%+ \frac{b^2}{2} \left(\beta^2 \T_0 + \Q_0 \right)
%\\
%&+ \frac{b^2}{2} \sum_{m=-2}^2  \left( \beta^2 \T^{2m} + \beta \Q^{2m} \right)  Y^{2m}(\g^\alpha)\,.
\label{eq:iRICE}
\end{align}
We kept the $\ell=2$ terms in Eq.~(\ref{eq:iRICE}), not just $\ell=0$, because $N_4=6$ directions are enough to obtain unbiased $\D^{2m}$ estimates (and hence, DTI) at no extra cost. Unfortunately, the $\O(b^2)$ estimates of $\beta\Q^{2m}+\beta^2 \T^{2m}$ terms will be biased by the presence of $\T^{4m}$, as the 4-design cannot detect these two as orthogonal. 

Minimal 4-designs combined with required $b$ and $\beta$ contain fewer measurements than the total DOF in $\D$ and $\C$. 
Thus, Eq.~(\ref{eq:iRICE}) provides a method for the fastest unbiased estimation of $\Dz$, $\Dt$, $\Tz$, and $\Qz$  components, and the invariants (\ref{FA})--(\ref{muFA}).  
Table~\ref{tab:TableProtocolsComparison} shows  the proposed minimal iRICE protocols for MD, FA, MK, based on measurements with 
12 $\B$ tensors; and MD, FA, MK, \mFA with 15 $\B$ tensors. These numbers are notably smaller than $21=6+15$ or $27=6+21$ that were thought needed to simultaneously estimate all DOF (besides $S|_{b=0}$).  
%While short acquisitions focused on just a few invariants have been developed\cite{HANSEN2013,HANSEN2016},  \update{the ones proposed here are more optimal. Ref. \cite{HANSEN2013} relies on two-step estimations which may lead to spatially varying biases, and Ref. \cite{HANSEN2016b} involves $50\%$ additional measurements.}  The present comprehensive approach based on spherical designs finds the protocols that involve fewer measurements (Table~\ref{tab:TableProtocolsComparison})  than was previously anticipated, as well as guarantees that these protocols are in fact the shortest possible.
%
%\dn{why ref 78 is not cited together with 69-70?}
While short acquisitions focused on just a few invariants have been developed\cite{HANSEN2013,HANSEN2016b}, \rvleft[R4]{} \update{the present formulation allows for shorter and fully optimal protocols. In the 1-3-9 scheme of Ref.~\cite{HANSEN2013}, 
%MK and MD are estimated independently, 
%\dn{independent is not bad, need to explain}
MD is estimated from the lower-$b$ shell assuming no $\sim b^2$ contributions, 
which may lead to spatially varying biases, while the 1-9-9 scheme of Ref.~\cite{HANSEN2016b} uses approximately $50\%$ more measurements than the minimum 1-6-6 required, Table~\ref{tab:TableProtocolsComparison}.}
Our comprehensive approach based on spherical designs identifies protocols that require fewer measurements (Table~\ref{tab:TableProtocolsComparison}) while guaranteeing that these are, in fact, the shortest possible for a given cumulant order.

Minimal and extensive acquisitions are compared in Figs.~\ref{fig:bothfastRICEcomparison}a,c. iRICE maps show a near-identical contrast, even in regions with large $\Sf$ such as WM fiber crossings, despite having 8-fold fewer measurements and not estimating full tensors. 
Scatter plots, Figs.~\ref{fig:bothfastRICEcomparison}b,d, do not show biases.

\subsection*{\update{Generalization to time-dependent cumulants}}\label{ss:DDE_tdep}
\rv[R2]{new subsection}
\noindent
Consider  now the general case of time-dependent non-Gaussian diffusion in at least some of the compartments, characterized by the time-dependent diffusion and kurtosis tensors $D(t)$ and $S(t)$, as discussed around Eq.~(\ref{DKI_t}), with a voxel  represented by the probability distribution functional $\P[D(t), S(t)]$. 
To  isolate physically distinct time-dependent $\O(b^2)$ contributions, one can use double diffusion encoding\cite{JESPERSEN2013,SHEMESH2015,NETOHENRIQUES2020}. DDE consists of two diffusion encoding blocks with diffusion times $t_{1,2}$, weightings $b_{1,2}$, and directions $\g_{1,2}$, separated by the mixing time $t_m$. In the limit $t_m\to\infty$, with $t_1=t_2\equiv t$, analogously to  Eqs.~(\ref{eq:MGC_SH})--(\ref{DKI}), the logarithm $\mathcal{L}_{\rm DDE}$ of a voxel-wise DDE signal  up to $\O(b^2)$ reads \cite{JESPERSEN2013} 
\begin{equation}\label{eq:DDE_cumExp}
	\begin{aligned}
		\mathcal{L}_{\rm DDE} &= \ln  \la\! e^{-b_1\Dc(t,\g_1) -b_2\Dc(t,\g_2) + \tfrac12 {b_1^2}\, S(t,\g_1) + \tfrac12 {b_2^2}\, S(t,\g_2)} \!\ra  \\
		&\simeq -b_1 \D(t,\g_1) - b_2 \D(t,\g_2) +\tfrac12{b_1^2}\,\Smu(t,\g_1) \\
		&+ \tfrac12{b_2^2}\, \Smu(t,\g_2) + \tfrac12{b_1^2} \lla D^2(t,\g_1)\rra \\
		&+ \tfrac12{b_2^2}  \lla D^2(t,\g_2)\rra + b_1 b_2 \lla D(t,\g_1)D(t,\g_2)\rra  
%		\mathcal{L}_{\rm DDE} &= \ln  \la e^{-b_1\Dc(\g_1) -b_2\Dc(\g_2) + \tfrac12 {b_1^2}\, S(\g_1)+ \tfrac12 {b_2^2}\, S(\g_2)} \ra  \\
%&\simeq -b_1 \D(\g_1) - b_2 \D(\g_2) +\tfrac12{b_1^2}\,\Smu(\g_1) + \tfrac12{b_2^2}\, \Smu(\g_2) \\
%&+ \tfrac{b_1^2}{2} \lla D^2(\g_1)\rra+ \tfrac{b_2^2}{2}  \lla D^2(\g_2)\rra + b_1 b_2 \lla D(\g_1)D(\g_2)\rra  
	\end{aligned}
\end{equation}
where  $\lla D(t,\g_1)D(t,\g_2)\rra=\C_{ijkl}(t) \,g_{1i} g_{1j} g_{2k} g_{2l}$, and 
the overall microscopic kurtosis $\Smu(t)$, Eq.~(\ref{Smu}), can now be disentangled\cite{JESPERSEN2013,NETOHENRIQUES2020} from the $\C(t)$-tensor, in contrast to Eq.~(\ref{DKI_t}).

To equally probe all orientations, the signal (\ref{eq:DDE_cumExp}) for a given DDE pair  $(b_1, \g_1; \ b_2, \g_2)$ 
%$( \g_1, \g_2)$ 
should be isotropically sampled across all rotations of this pair, keeping the angle $\psi$ between $\g_1$ and $\g_2$ fixed.\cite{JESPERSEN2025}  
Akin to the discussion before Eq.~(\ref{eq:MGC_SH}), the corresponding DDE ``shell" is the SO(3) group manifold $\mathbb{S}^3/\mathbb{Z}_2$. 
The extra dimension relative to the  2-dimensional LTE shell  $\mathbb{S}^2$
%$=\mathrm{SO(3)}/\mathrm{SO(2)}$ for LTE 
enables the estimation of all irreducible $t$-dependent components of Eq.~(\ref{eq:DDE_cumExp}). One way to do so is consider its spherical harmonic expansion with respect to $\g_1$ while sampling the DDE shell, equivalent to the integral over $\mathrm{SO(3)}\cong\mathbb{S}^3/\mathbb{Z}_2$
\begin{equation}\label{eq:DDE_g1proj}
	\mathcal{L}_\mathrm{DDE}^{\ell m}=(2\ell+1)\!\int_\mathrm{SO(3)}\!\!\!\d\R\,Y^{\ell m*}(\R\g_1)\,\mathcal{L}_\mathrm{DDE}(\R\g_1,\R\g_2)  \ \
%		\mathcal{L}_\mathrm{DDE}^{\ell m}=(2\ell+1)\!\int_\mathrm{SO(3)}\!\!\d\R\,Y^{\ell m*}(\R^{-1}\g_1)\,\mathcal{L}_\mathrm{DDE}(\R^{-1}\g_1,\R^{-1}\g_2)
\end{equation}
with canonical measure normalized to $\int_\mathrm{SO(3)}\!\d\R\equiv1$. 
In Supplementary Section \ref{SM:DDE_WignerD} we derive SO(3) Fourier coefficients of $\mathcal{L}_\mathrm{DDE}(\R\g_1,\R\g_2)$, 
and the spherical harmonics   (\ref{eq:DDE_g1proj}): 
\begin{equation}\label{eq:SH_DDE}
	\begin{aligned}
	\mathcal{L}_{\mathrm{DDE}}^{00}&=-\left(b_1+b_2\right)\mathsf{D}^{00}(t) +\tfrac12 \left(b_1^2+b_2^2\right)\Smu^{00}(t)  \\
	&+ \tfrac{1}{2}\left(b_1+b_2\right)^2\mathsf{Q}^{00}(t) +\tfrac12 \left(b_1^2+b_2^2+2p_2\, b_1b_2\right)\mathsf{T}^{00}(t)\,, \\
	\mathcal{L}_{\mathrm{DDE}}^{2m}&=-\left(b_1+p_2\, b_2\right)\mathsf{D}^{2m}(t) +\tfrac12 \left(b_1^2+p_2\, b_2^2\right)\Smu^{2m}(t) \\
	&+\tfrac12\left(b_1^2+p_2\, b_2 ^2+\left(1+p_2\right)b_1b_2 \right)\mathsf{Q}^{2m}(t)\\
	&+\tfrac12 \left( b_1^2+p_2\, b_2^2+2p_2\, b_1b_2\right)\mathsf{T}^{2m}(t)\,, \\
	\mathcal{L}_{\mathrm{DDE}}^{4m}&=\tfrac12 \left(b_1^2+p_4\, b_2^2\right)\Smu^{4m}(t) \\
	& + \tfrac12 \left(b_1^2+p_4\,b_2^2+2p_2\, b_1b_2\right)\mathsf{T}^{4m}(t)\,,
	\end{aligned}
%	\begin{aligned}
%	\mathcal{L}_{\mathrm{DDE}}^{00}&=-\left(b_1+b_2\right)\mathsf{D}^{00}+\tfrac{1}{2}\left[\left(b_1+b_2\right)^2\mathsf{Q}^{00}+\left({b_1}^2+{b_2}^2+2b_1b_2p_2\right)\mathsf{T}^{00}+\left({b_1}^2+{b_2}^2\right)\Smu^{00}\right],\\\
%	\mathcal{L}_{\mathrm{DDE}}^{2m}&=-\left(b_1+p_2b_2\right)\mathsf{D}^{2m}+\tfrac{1}{2}\left[\left({b_1}^2+p_2{b_2}^2+b_1b_2\left(1+p_2\right)\right)\mathsf{Q}^{2m}+\left({b_1}^2+p_2{b_2}^2+2b_1b_2p_2\right)\mathsf{T}^{2m}+\left({b_1}^2+p_2{b_2}^2\right)\Smu^{2m}\right],\\
%	\mathcal{L}_{\mathrm{DDE}}^{4m}&=\tfrac{1}{2}\left[\left({b_1}^2+p_4{b_2}^2+2b_1b_2p_2\right)\mathsf{T}^{4m}+\left({b_1}^2+p_4{b_2}^2\right)\Smu^{4m}\right]
%\end{aligned}
\end{equation}
where $p_\ell(\psi)\equiv\,\!P_\ell(\g_1\!\cdot\!\g_2)$ are Legendre polynomials. 
The  components $\Q^{\ell m}(t)$ and $\T^{\ell m}(t)$ of $\Q(t)$ and $\T(t)$ are defined via the covariances 
$\lla D^{\ell_1 m_1}(t) D^{\ell_2 m_2}(t)\rra$ in Eqs.~(\ref{QTLM}), while 
$\Smu(t)=\Smu^{(0)}(t)\oplus\Smu^{(2)}(t)\oplus\Smu^{(4)}(t)$ is the  decomposition of the overall microscopic $t$-dependent kurtosis, akin to Eq.~(\ref{S=S0+S2+S4}). 
%DDE measurements are not available in the current datasets but are the scope of future work.

$\mathcal{L}_{\mathrm{DDE}}^{00}$ is related to the spherical mean of the DDE signal\cite{NETOHENRIQUES2020}; it serves as the basis of correlation tensor imaging yielding the mean microscopic kurtosis $\Smu^{00}(t)$. Remarkably, the anisotropic components $\mathcal{L}_{\mathrm{DDE}}^{2m}$ and $\mathcal{L}_{\mathrm{DDE}}^{4m}$ enable the estimation of all $t$-dependent spherical tensor components of the diffusion, covariance and microscopic kurtosis tensors. 
While the $\cos2\psi$ DDE modulation coming from $p_2(\psi)$ has been observed \cite{SHEMESH2011}, the unexpected 4-fold $\cos4\psi$ modulation emerges from the $\ell =4$ sector via $p_4(\psi)$.\cite{COELHO2025}

The microscopic kurtosis tensor $\Smu(t)$ has its own 15 DOF and 12 invariants:  
$1+2+6=9$ intrinsic and 3 mixed (angles between the $\Smu^{(2)}$ and $\Smu^{(4)}$ frames), 
similar to the invariants of $\S$ tensor in {\it Invariants: intrinsic and mixed} section.

\section*{Discussion}\label{s:Discussion}
\noindent
\update{It \rv[EDITOR]{} may be only a slight exaggeration to say that modern physics is the study of symmetry. Here we uncovered how the fundamental SO(3) symmetry of our 3-dimensional world constrains the geometry and the degrees of freedom in the diffusion MRI cumulant series. The representation theory approach is so general that it applies for both Gaussian and non-Gaussian diffusion in the individual compartments. While the coarse-graining by diffusion \citep{NOVIKOV2019,NOVIKOV2021} within each compartment results in a particular time-dependence of its cumulants $D(t)$ and $S(t)$, the SO(3) symmetry defines the structure of the irreducible components of the voxel-wise cumulant tensors. We can think of the symmetry providing a ``non-negotiable'' tensor structure within which the plethora of time-dependent diffusion effects occurs based on the coarse-graining over the compartment-specific microstructure. }
Based only on symmetry considerations, the formalism readily extends onto all 4th-order tensors with minor and major symmetries such as the elasticity tensor in continuous media \citep{BACKUS1970,BETTEN1987,BONA2004,MOAKHER2008}, yielding applications in mechanics, geology, materials science, and soft condensed matter physics. 

%{A central distinction underlying the present framework is between symmetry and scale. The irreducible decompositions introduced here are dictated entirely by rotational symmetry, and therefore apply independently of how diffusion processes are coarse-grained in time. In contrast, diffusion-time dependence reflects a choice of temporal resolution, rather than a modification of the admissible tensorial structure. Whether diffusion encoding is sensitive to time-dependent microstructural effects or not is therefore a separate experimental question, which does not alter the symmetry-constrained form of the cumulant tensors or their irreducible components. This distinction allows time-dependent effects to be incorporated naturally into RICE.}

\update{Can a non-Gaussian compartment be made of ``more elementary" Gaussian ones in the spirit of Eqs.~(\ref{eq:CumExpinB})--(\ref{eq:DCdefinitions})? 
If this were true, we could without the loss of generality build our formalism based on $\P(\Dc)$ for the Gaussian compartments. 
In general, the answer is no: Coarse-graining of the microstructure by diffusion generally leads to %the 
{\it time-dependent} cumulants in a given %compartment, with $\Dc(t)$ monotonically decreasing towards a constant tensor, and $S(t)$ and higher-order ones vanishing as $t\to\infty$.\cite{NOVIKOV2010}  
compartment. $\Dc(t)$ monotonically decreases towards a constant tensor, and $S(t)$ and higher-order cumulants vanish as $t\to\infty$.\cite{NOVIKOV2010}  
Hence, the picture of Eqs.~(\ref{eq:CumExpinB})--(\ref{eq:DCdefinitions}) is valid at sufficiently long times, when the transient coarse-graining processes become negligible, $\C(t)\to\C = \mathrm{const}$, and $\Smu\to 0$. 
%The e
Exchanging compartments\cite{JENSEN2005,NETOHENRIQUES2020,CHAKWIZIRA2025} can be viewed as a single non-Gaussian one with time-dependent $S(t)$ that vanishes as $t\to\infty$, contributing to the overall $\Smu(t)$ at finite $t$.    
}

\update{While the present framework is  model-independent and general, our main limitation has been working up to $\O(b^2)$. However, this is not a limitation of the  symmetry-based approach, which can be extended  to higher-order cumulant tensors.} 
For example, the next, 6th-order cumulant at $\O(b^3)$ maps onto the addition of 3 angular momenta of $\ell\!=\!0$ or $2$ (for the Gaussian compartments).  
The generalization of QT decomposition would yield the covariances  $\lla D^{\ell m} D^{\ell' m'}D^{\ell'' m''}\rra$, which split into the irreducible decomposition 
\begin{equation}\label{DDD}
\mathrm{Sym\,}  \left[ (V_0 \oplus V_2)^{\otimes 3}\right]
\;\cong\; V_6 \;\oplus\; 2V_4 \;\oplus\; 4V_2 \;\oplus\; 5V_0 
\end{equation}
with the multiplicities as indicated, totaling 56 DOF, of which 35 are intrinsic and 18 mixed invariants. 
The corresponding SA decomposition gives $V_6 \oplus V_4 \oplus V_2 \oplus V_0$ for the fully symmetric part, and $V_4 \oplus 3V_2 \oplus 4V_0$ for the remaining asymmetric part, each carrying 28 DOF. 
Probing all 56 DOF  of $\lla D\otimes D\otimes D\rra$ requires non-axially symmetric $\B$ tensors \cite{NING2021}. In other words, this and higher-order cumulants involve waveform rotations over the SO(3) manifold $\mathbb{S}^3/\mathbb{Z}_2$ rather than $\mathbb{S}^2$, \update{increasing scan time substantially}. 

%\dn{edits}
\update{The cumulant expansion provides a powerful model-independent description of the dMRI signal within its convergence radius $b_*$ that depends on the microstructure in a given sample (voxel).\cite{KISELEV2007}
When the acquisition range $0\leq b \leq b_{\rm max}$ is within 
 the convergence radius, $b_{\rm max}< b_*$, adding the successive terms increases the accuracy of the cumulant tensors' estimation (at a cost of the loss of precision); 
the truncation at a given order is justified when the neglected contributions remain below the noise floor. 
When valid, symmetry-based irreducible representations of the cumulant tensors provide maximally informative and mutually orthogonal decompositions at each  order. 
%Its limitation is that the accuracy deteriorates at strong diffusion weightings, as the higher-order terms become non-negligible, and the truncation at a given order is only justified when the neglected contributions remain below the noise floor. In practice, this condition depends jointly on the encoding scheme, microstructural heterogeneity, and signal-to-noise ratio, rather than on the $b$-value alone. Within a valid regime, symmetry-based irreducible representations provide a maximally informative and internally consistent description at each truncation order.
%Acquisitions sensitive to terms $\sim b^3$ and beyond may come close or exceed the convergence radius of the cumulant expansion \cite{KISELEV2007}, in which case their estimation may be challenging, 
For $b \geq b_*$, the series diverges; hence, for protocols with $b_{\rm max}\geq b_*$, adding extra terms can uncontrollably modify the estimates for the lower-order ones\cite{FROHLICH2006,MOUTAL2022}.   The signal's description in terms of the power series then ceases to be useful since the estimates depend on the range $b_{\rm max}$ rather than on tissue properties, and a full functional form of $\P(\Dc,\dots)$ should rather be employed.} 
The latter has been performed using a multi-dimensional inverse Laplace transform  \cite{DEALMEIDA2016,TOPGAARD2017,MAGDOOM2021} assuming Gaussian compartments, which requires regularization or constraints, and uses extensive, so-far clinically unfeasible  acquisitions.

Practically, for diffusion NMR/MRI, we uncovered 14 invariants in addition to 7 previously used in the life sciences context of diagnostic MRI.  
Publicly available consortia dMRI datasets provide a venue to study the novel contrasts at a population level. 
Our results show that for conventional multi-shell dMRI (LTE-only), the information added by the RICE invariants for the kurtosis tensor improves multiple sclerosis detection. 
RICE maps belong to distinct irreducible representations of rotations, and thereby represent  ``orthogonal" contrasts up to $b^2$. Such independence may improve sensitivity and specificity to detect specific tissue microstructure changes in disease, aging and development. 
Furthermore, the proposed minimal iRICE acquisition protocols 
for estimating MD, FA, MK, and \mFA will enable clinical translation of beyond-DTI diffusion metrics  hampered by long scan times.

\clearpage
\newpage

%%%%%%%%%%%%%%%%%%%%%%%%%%%%%%%%%%%%%%%%
\section*{Methods}\label{s:Methods}
%\section{Theory}\label{s:Theory}

\subsection*{Notations}
\noindent
Throughout this work, 
sans serif font refers to  voxel-wise tensors, such as $\D$, $\C$, $\A$, $\S$, $\Q$, $\T$, whereas italic font  refers to tensors $\Dc$ of the microscopic compartments. 
These may have either Cartesian components such as $\D_{ij}$ and $\Dc_{ij}$, or spherical tensor components such as $\D^{\ell m}$ and $\Dc^{\ell m}$.  
We refer to irreducible components, such as $\Sl$, both as a collection of $\S^{\ell m}$ spherical tensor elements, or Cartesian ones $\Sl_{i_1 \hdots i_\ell}$. Unless specified otherwise, the order $\ell$ of the latter will coincide with the number of Cartesian subindices since this is the most natural representation. 
We assume Einstein's convention of summation over repeated Cartesian subindices $ijkl...$.%, as well as over repeated spherical tensor indices $m$ from $-\ell$ to $\ell$ for a given degree-$\ell$ representation. 

Brackets $\la\dots\ra$ denote averages over the compartmental diffusion tensor distribution $\P(\Dc)$,  
and double brackets $\lla \dots \rra$ denote the cumulants of $\P(\Dc)$.  
Finally, for readability, given the plethora of conventional superscript indices $^{(\ell)}$ for irreducible tensor components, 
we omit outer parentheses when referring to tensor traces, e.g., 
$\tr \B = \tr (\B)$, $\tr \B\D = \tr (\B\D)$, $\tr (\Dt)^3 =\tr \big( (\Dt)^3 \big)$, 
and $\tr^\nu \B = \big(\tr (\B)\big)^\nu$.

%Rotational invariants of $\Tl$ are denoted $\iTl{n}$, where $n$ indicates the power of the tensor before taking the trace. We drop $n=1$ index for $\ell=0$ and $n=2$ index (Euclidean norm) for $\ell > 2$. 

\subsection*{Multiple Gaussian compartments }
\noindent
The dMRI signal from a given voxel 
\begin{equation} \label{S-gen}
\Sig = \langle e^{i \phi} \rangle_\mathrm{paths+spins}\,, \quad \phi =- \int_0^t \! \d\tau\, \Gg(\tau)\!\cdot\! \mathbf{r}(\tau)
\end{equation}
is the transverse magnetization $e^{i\phi}$ in the rotating frame, averaged over all spins traveling along all possible Brownian paths $\mathbf{r}(\tau)$ between $\tau=0$ and measurement time $t$.  Experiments are performed under the condition 
$\int_0^t \Gg(\tau)\,\d \tau = 0$ on the applied Larmor frequency gradient $\Gg(\tau)$ \cite{JONES2010,KISELEV2017,NOVIKOV2019}. 
In general, the dMRI signal is a functional of $\Gg(t)$, or, equivalently, of the encoding function $\q(t)=\int_0^t \Gg(\tau)\,\d \tau$: 
%(the anti-derivative of $\Gg(t)$): 
$\Sig=\Sig[\q(t)]$. 
Even for a conventional pulsed-gradient (LTE) diffusion sequence \citep{STEJSKAL&TANNER1965}, one obtains a multi-dimensional phase diagram in the space of sequence parameters \citep{HURLIMANN1995,GREBENKOV2007,NOVIKOV2021}. 

However, after coarse-graining the dynamics within a given tissue compartment over sufficiently long $t$, the microstructure-induced temporal velocity correlations become forgotten, such that the distribution of spin phase 
$\phi=\int_0^t\! \d\tau\, \q(\tau) \!\cdot\! \mathbf{v} (\tau)$ becomes asymptotically Gaussian, 
defined by its velocity autocorrelation function 
$\lla v_i(\tau)v_j(\tau') \rra \simeq 2D_{ij} \, \delta(\tau-\tau')$.\citep{KISELEV2017,NOVIKOV2019,NOVIKOV2021}
The averaging in Eq.~(\ref{S-gen}) over a compartment  
is then represented in terms of the second cumulant of the phase, 
$\Sig = e^{-\B_{ij} D_{ij}}$, and the measurement is determined by specifying the $\B$-tensor 
\begin{equation}\label{eq:Btensor}
\B_{ij} = \int_0^t\! \d\tau\, q_i(\tau) \, q_j(\tau)
%\quad b=\tr \B \,,
\end{equation}
calculated based on $\q(t)$.  
%The trace of $\B$ is the conventional  $b$-value. 
The distribution $\P(D)$ of compartment tensors in a voxel 
gives rise to the overall  signal \citep{BASSER2003,JIAN2007,RUSSEL2015}
\begin{equation} \label{eq:signal_DTD}
    \Sig(\B) = \int\! \d D\,\P(D)\, e^{-\tr \B D} = \sum_\alpha f_\alpha\, e^{-\tr \B D^\alpha} \,. 
\end{equation}
Normalization 
$\int\!\d D\,\P(D) \!= \! 1$ 
implies $\sum_\alpha f_\alpha \!= \!1$.% For readability, we omit outer parentheses when referring to tensor traces, e.g., $\tr \B = \tr (\B)$, $\tr \B\D = \tr (\B\D)$, and $\tr^\nu \B = \big(\tr (\B)\big)^\nu$. 

Equation~(\ref{eq:signal_DTD}) is the most general form of a signal from multiple Gaussian compartments. It is valid when the transient processes have played out, such that compartmental diffusion tensors $D$ have all become time-independent, and thereby higher-order cumulants in each compartment are negligible \citep{NOVIKOV2019}. 
In this case, 
%instead of being a functional of $\q(t)$, 
the signal (\ref{eq:signal_DTD}) is a function of the $\B$-tensor: $\Sig[\q(t)] \to \Sig(\B)$, while  tissue is fully represented by the voxel-wise distribution $\P(D)$.  
This long-$t$ picture of multiple Gaussian compartments (anisotropic and non-exchanging) underpins a large number of dMRI modeling approaches, in particular, the Standard Model (SM) of diffusion  \citep{NOVIKOV2019} and its variants \citep{JESPERSEN2007,JESPERSEN2010,FIEREMANS2011,ZHANG2012,SOTIROPOULOS2012,JENSEN2016}. Furthermore, this picture contains the SM extension onto different fiber populations in a voxel, lifting the key SM assumption of a single-fascicle ``kernel" (response).

Given the forward model (\ref{eq:signal_DTD}), an inverse problem is to restore $\P(D)$ from measurements with different $\B$. This problem is a matrix version of the inverse Laplace transform and is therefore ill-conditioned. Since in clinical settings, typical encodings are moderate ($\tr \B \D \sim 1$), the  inverse problem can be formulated term-by-term for the 
cumulant expansion (\ref{eq:CumExpinB}) of the signal (\ref{eq:signal_DTD}).  
% --- or, more conventionally, for expansion of $\ln S(\B)$, the cumulant expansion \citep{VANKAMPEN1981,KISELEV2010}:

\subsection*{The cumulant series for multiple Gaussian compartments}
\noindent
The higher-order signal terms in Eq.~(\ref{eq:CumExpinB}) couple to successive cumulants of $\P(D)$. 
The inverse problem maps onto finding the  cumulants $\lla D_{i_1j_1}\dots D_{i_nj_n}\rra$
(tensors of even order $2n$) from a set of measurements.
%Here we expand on the statement that the higher-order signal terms in Eq.~(\ref{eq:CumExpinB}) couple to successive cumulants of $\P(\D^\alpha)$.
This becomes obvious by noting the analogy $\B\to i\lambda$ with the standard cumulant series \citep{VANKAMPEN1981} 
$\ln p(\lambda) = \sum_{n=1}^\infty 
{(-i\lambda)^n\over n!} \lla x^n \rra$
for the characteristic function
$p(\lambda)= \int\!\d x\, e^{-i\lambda x} p(x) $
of a probability distribution $p(x)$, such that the $n$-th term in Eq.~(\ref{eq:CumExpinB}) is 
$\frac{(-1)^n}{n!} \B_{i_1j_1}\dots \B_{i_nj_n} 
\lla D_{i_1j_1}\dots D_{i_nj_n}\rra$. 

Hence, the introduction of the $\B$-tensor, enabled by the Gaussian diffusion assumption in every compartment, 
lowers the order  by half: The $2n$-th cumulant of the phase (\ref{S-gen}) 
[i.e., the $2n$-th order term of expanding $\ln \Sig [{\bf q}(t)]$ in ${\bf q}(t)$] 
maps onto the $n$-th cumulant $\lla D_{i_1j_1}\dots D_{i_nj_n}\rra$ of $\P(D)$ corresponding to
the $n$-th order of expansion of $\ln \Sig(\B)$ in $\B$.   
The number of DOF for this cumulant equals to that for a fully symmetric order-$n$ tensor of dimension $d=6$, which is a number of assignments of $n$ indistinguishable objects into $d$ distinguishable bins: $\binom{n+d-1}{n}= (n+5)!/(n! \cdot 5!) = 6, \ 21, \ 56, \ 126,\ \dots $ for $n=1,\ 2,\ 3,\ 4, \ \dots $ .

The range of diffusion weightings  used in this work ensures the sensitivity to $\mathcal{O}(b^2)$ signal components, necessary for estimating $\C$ in the living human brain. While it is not required to replicate our exact diffusion weightings to estimate the cumulants, the maximal $b$ should be high enough to capture  $\mathcal{O}(b^2)$ contributions, yet low enough to keep higher-order terms small. Extending $b$ beyond the convergence radius of the cumulant expansion\cite{KISELEV2007} will diminish the accuracy \cite{FROHLICH2006,MOUTAL2022}. 
For \textit{in vivo} brain dMRI, we recommend $b\lesssim2-3\,\mathrm{ms}/\mu\mathrm{m}^2$.
%\dn{maybe 2-3?}
%Beyond this limit, high-order cumulant expansions may lead to inaccurate signal representations, compromising the interpretation and utility of the extracted coefficients \cite{MOUTAL2022}. 

\subsection*{Irreducible representations of SO(3)}
\label{ss:rep}
\noindent
Here we remind key aspects from representation theory of SO(3) \citep{Tinkham,HALL2015} used throughout this work. 
A $d$-dimensional representation of a group is a mapping of each element (rotation) onto a $d\times d$ matrix that acts on a $d$-dimensional vector space. Representation theory provides a way to split a complex object (such as tensor $\D$ or $\C$) into a set of independent simpler ones with certain symmetries, on which a group acts. 
In particular, the elements of an {\it irreducible representation} transform among themselves, and hence can be studied separately.

Every irreducible representation of SO(3) is labeled by an integer {\it degree} $\ell=0,\ 1,\ 2,\ \dots$, and acts on a $2\ell+1$-dimensional space of 
STF tensors \cite{THORNE1980} --- fully symmetric and trace-free (for $\ell>0$) tensors of {\it order} $\ell$ (i.e., with $\ell$ indices). 
Any order-$\ell$ STF tensor (STF-$\ell$ tensor) can be represented by its $2\ell+1$ {\it spherical tensor} components  $\F^{\ell m}$, $m=-\ell \dots \ell$:  
\begin{equation} \label{sph-STF}
	\F^{(\ell)}_{i_1\hdots i_\ell} = \sum_{m=-\ell}^\ell  \F^{\ell m} \, \Y^{\ell m}_{i_1\hdots i_\ell}\, ,
\end{equation}
where $\Y^{\ell m}_{i_1\hdots i_\ell}$ form the standard complex-valued STF basis \cite{THORNE1980}. 
STF property means that a trace over any pair of indices vanishes for $\ell\geq 2$: 
$\Y^{\ell m}_{i_1\hdots i_\ell} \delta_{i_n i_{n'}} \equiv 0$, $1\leq n, n' \leq \ell$. The case $\ell=0$ corresponds to a (generally nonzero) scalar. 
For each $\ell$, basis STF tensors generate $2\ell+1$ spherical harmonics on $\mathbb{S}^2$: 
\begin{equation}\label{Eq:SHandSTF}
    Y^{\ell m}(\n) = \Y^{\ell m}_{i_1\hdots i_\ell} \, n_{i_1} \hdots n_{i_\ell} \,, \quad |\n | = 1 \,.
    %m = -\ell \dots \ell  \,.
    %= \Y^{\ell m}_{i_1\hdots i_L} n_{i_1} \hdots n_{i_L} \,, 
\end{equation}
%for $m=-\ell \dots \ell$, the spherical tensor components transform upon SO(3) rotations as the SH components \citep{THORNE1980}. 
As a result, a spherical tensor is uniquely mapped onto a  ``glyph" (e.g., Fig.~\ref{fig:glyphsQTSA}): 
\begin{equation} \label{STF=SH}
\F^{(\ell)}(\n) = \F^{(\ell)}_{i_1\hdots i_\ell} \, n_{i_1} \hdots n_{i_\ell} =  \sum_{m=-\ell}^\ell \F^{\ell m}\, Y^{\ell m}(\n) \,,
\end{equation} 
with its components $ \F^{\ell m} $ being  spherical harmonics coefficients of the glyph.

Throughout this work, we use {\it Racah normalization} \cite{RACAH1942}  of spherical harmonics and spherical tensors ($^*$ denotes complex conjugation): 
\begin{equation} \label{STF-SH-orthogonality}
\int_{\mathbb{S}^2}\d\Omega_{\n}\, {Y^{\ell m\, *}}(\n)\, Y^{\ell' m'}(\n) = \frac{4\pi}{2\ell+1} \,\delta_{\ell \ell'}\, \delta_{mm'}\,.
\end{equation} 
To obtain Racah-normalized spherical harmonics and basis tensors $ \Y^{\ell m}_{i_1 \hdots i_\ell}$, 
one needs to multiply the orthonormal $Y^{\ell m}(\n)$ and $ \Y^{\ell m}_{i_1 \hdots i_\ell}$ (found, e.g., in Thorne~\cite{THORNE1980}) by $\sqrt{4\pi/(2\ell+1)}$. 
Thus, Racah-normalized spherical tensor coefficients $\F^{\ell m}$ do not carry the $\sqrt{(2\ell+1)/4\pi}$ factors ubiquitous for the orthonormal spherical harmonics. 
This has a benefit of the $\ell=0$ component being identically equal to the angular-averaged glyph since $\Y^{00}\equiv 1$ (see, e.g., Eq.~(\ref{MD}) for mean diffusivity), 
as well as of simplifying the relations between the spherical harmonics products and Clebsch-Gordan coefficients, see Supplementary Section \ref{SM:CGcoeffsSTF}. We also use Condon-Shortley convention \citep{CONDON1964} 
\begin{equation} \label{condon-shortley}
\Y^{\ell m \, *}_{i_1 \hdots i_\ell} = (-1)^m \, \Y^{\ell, -m}_{i_1 \hdots i_\ell}(\n) \,, \quad 
Y^{\ell m\, *}(\n) = (-1)^m \, Y^{\ell, -m}(\n) \,,
\end{equation}
and thus $\F^{\ell m\,*} = (-1)^m \,  \F^{\ell,-m}$ for real-valued $\F^{(\ell)}_{i_1\hdots i_\ell}$.

To go from Cartesian to spherical tensor components, we invert Eq.~(\ref{sph-STF}) via Eq.~(2.5) of Ref.~\cite{THORNE1980}, Eqs.~(\ref{Eq:SHandSTF}) and (\ref{STF-SH-orthogonality}):  
\begin{equation}\label{eq:ThorneSTFprod}
{\Y^{\ell m\, *}_{i_1 \hdots i_\ell}} \, \Y^{\ell m'}_{i_1 \hdots i_\ell} =\frac{(2\ell-1)!!}{\ell!}\, \delta_{mm'} \,,
\end{equation}
where $\ell!!\equiv \ell (\ell-2)(\ell-4)\cdot \cdot \cdot (2 \text{ or } 1)$, such that 
\begin{equation}\label{Flm}
\F^{\ell m} =  \frac{\ell!}{(2\ell-1)!!} \,  {\Y^{\ell m \,*}_{i_1\hdots i_\ell}} \, \F_{i_1\hdots i_\ell}^{} \,.
\end{equation}

\subsection*{Decomposition into irreducible representations}\label{subsection:STF}
\noindent
Any fully symmetric order-$L$ tensor $\F_{i_1\hdots i_L}$ (generally, with nonzero traces), such as the kurtosis tensor (\ref{S=S0+S2+S4}), can be decomposed into a direct sum of STF tensors $\F=\F^{(L)}\oplus\F^{(L-2)}\oplus\dots$ of degrees $L$, $L-2$, $\dots$, (0 or 1),  effectively by subtracting traces \cite{THORNE1980}. 
To convert this direct sum into an ordinary sum in the space of order-$L$ tensor components 
\begin{align}\label{Eq:STFdecomposition}
	\F_{i_1\hdots i_L} &= \sum_{\ell = L, L-2, \dots}\F^{(\ell)}_{i_1\hdots i_L} \,, \quad 
	\F^{(\ell)}_{i_1\hdots i_L} =  \sum_{m=-\ell}^\ell \F^{\ell m}\, \Y^{\ell m}_{i_1\hdots i_L} \,, 
%	= \sum_{\ell = L, L-2, \dots} \sum_{m=-\ell}^\ell \F^{\ell m}\, \Y^{\ell m}_{i_1\hdots i_L} \,, 
\end{align}
we introduce the basis elements that are  {\it non-STF} for $L>\ell$: 
\begin{equation}\label{Eq:Ylm_deltas}
	\Y^{\ell m}_{i_1\hdots i_L} = \Y^{\ell m}_{(i_1\hdots i_\ell}\delta_{i_{\ell+1}i_{\ell+2}}^{}\hdots \delta_{i_{L-1}i_{L)}}^{}\,,\quad \ell \leq L\,, 
\end{equation}
through symmetrization of the STF-$\ell$ basis elements with the remaining $(L-\ell)/2$ Kronecker symbols, 
%fully determines the spherical tensor components $\F^{\ell m}$ for $\ell = L, L-2, \dots$ in Eq.~(\ref{Eq:STFdecomposition}) and 
naturally generalizing Eqs.~(\ref{Eq:SHandSTF})--(\ref{STF=SH}): 
\begin{equation} \label{STF=SH-L}
\F^{(\ell)}(\n) = \F^{(\ell)}_{i_1\hdots i_L} \, n_{i_1} \hdots n_{i_L} 
=  \sum_{m=-\ell}^\ell \F^{\ell m}\, Y^{\ell m}(\n) \,.
\end{equation} 
The basis (\ref{Eq:Ylm_deltas}) of order-$L$ symmetric tensors is orthogonal with respect to taking the trace over all $L$ indices. 

To find all spherical tensor components $\F^{\ell m}$ for $\ell = L, L-2, \dots$ of a fully symmetric order-$L$ Cartesian tensor according to Eq.~(\ref{Eq:STFdecomposition}), in Supplementary Section~\ref{SM:STF_generalized} we generalize the normalization (\ref{eq:ThorneSTFprod}) for $L>\ell$: 
\begin{equation}
\begin{aligned}\label{eq:alpha_STF}
\Y^{\ell m\, *}_{i_1\hdots i_L} \Y^{\ell' m'}_{i_1\hdots i_L}
 & = \zeta(L,\ell)  \, \delta_{\ell \ell'}^{} \, \delta_{m m'}^{} \,, \\
\zeta(L,\ell) & = \frac{(L+\ell+1)!!(L-\ell)!!}{L!\,(2\ell+1)} \,,
\end{aligned}
\end{equation}
such that $\zeta(\ell,\ell) = (2\ell-1)!! / \ell!$ matches Eq.~(\ref{eq:ThorneSTFprod}), yielding 
\begin{equation}\label{Flm-gen}
\F^{\ell m} = \frac1{\zeta(L,\ell)}  \, 
\Y^{\ell m\, *}_{i_1\hdots i_\ell}\,  
\delta_{i_{\ell+1}i_{\ell+2}}^{} \hdots \delta_{i_{L-1}i_{L}}^{} \,\F_{i_1\hdots i_L}^{}\,.
\end{equation}
In other words, the degree-$\ell$ spherical tensor components $\F^{\ell m}$ in the decomposition (\ref{Eq:STFdecomposition}) are obtained by taking the remaining $(L-\ell)/2$ traces and convolving with the standard STF-$\ell$ basis element $\Y^{\ell m \, *}_{i_1\hdots i_\ell}$, albeit using the normalization coefficient $1/\zeta(L,\ell)$ that ``remembers'' the original $L>\ell$.

So far, our discussion of SO(3) representations has been completely general. 
In dMRI context, every degree $\ell$ is even due to time-reversal invariance of the Brownian motion, dictating even parity $Y^{\ell m}(-\n)=Y^{\ell m}(\n)$. 
Hence, each cumulant or moment tensor, as in Eq.~(\ref{eq:CumExpinB}), can be split into a direct sum of irreducible representations with even $\ell$ in Eq.~(\ref{Eq:STFdecomposition}), connecting it with the orientation dispersion in the spherical harmonics basis \citep{NOVIKOV2018,POZO2019}. 

Applying the above methodology to the $\C$ tensor, we first split it into $\S$ and $\A$,  
Eq.~(\ref{W})--(\ref{A}). 
%where $\A_{ijkl} = \C_{ijkl}-\S_{ijkl} = \tfrac13 (2\C_{ijkl} - \C_{iljk} - \C_{ikjl})$. 
To decompose $\A$, we use Eqs.~(\ref{A=A}), where 
Eq.~(\ref{Apq}) is $\A_{pq}=4\Delta_{pq}$ from Eqs.~(45)--(46) in \citep{ITIN2013}; here we use the same symbol $\A$ for 2nd and 4th order tensors, as they are isomorphic (and represent the same geometric object).   
Eq.~(\ref{Apq=}) is obtained by using the identity $\epsilon_{i j k} \epsilon_{l m n}=\delta_{i l} \delta_{j m} \delta_{k n}+\delta_{i m} \delta_{j n} \delta_{k l}+\delta_{i n} \delta_{j l} \delta_{k m}-\delta_{i l} \delta_{j n} \delta_{k m}-\delta_{i n} \delta_{j m} \delta_{k l}-\delta_{i m} \delta_{j l} \delta_{k n}$.
Eq.~(\ref{Aijkl}) (the inverse relation) follows from  $\epsilon_{ijn}\epsilon_{kln}=\delta_{ik}\delta_{jl}-\delta_{il}\delta_{jk}$ and from the property $\A_{i(jkl)}\equiv 0$ that is a consequence of Eq.~(\ref{A}).

Now that all Cartesian tensors $\S_{ijkl}$ and $\A_{pq}$ derived from $\C$ are fully symmetric, we can represent them as combinations of spherical tensors, cf. Eq.~(\ref{Eq:STFdecomposition}).  
Applying Eq.~(\ref{Flm-gen}), we get% 
\begin{equation}
	\begin{aligned}
		\D^{00}  &= \tfrac13\,\D_{ii}\,  \quad
		&&
		\D^{2 m} = \tfrac23\, {\Y^{2 m}_{ij}}^* \,\D_{ij}^{}\,;  
		\\
		\A^{00}  &= \tfrac13 \,\A_{ii}\,,   \quad
		&& 
		\A^{2 m} = \tfrac23\, {\Y^{2 m}_{ij}}^*\, \A_{ij}^{}\,;
		\\
		\S^{00}  &= \tfrac15 \,\S_{iijj}\,, 
		\quad
		&&
		\S^{2 m} = \tfrac47\, {\Y^{2 m}_{ij}}^* \,\S_{ijkk}^{}\,,    
		\\ 
		& \quad && \S^{4m}  = \tfrac{8}{35}\,  {\Y^{4 m}_{ijkl}}^* \,\S_{ijkl}^{}\,.
	\end{aligned}\label{eq_Slm}
\end{equation}
Eqs.~(\ref{QT_SA_relation}) then relate the components (\ref{eq_Slm}) to those of $\Q$ and $\T$. 
%From the above, one can compute the spherical tensor components of the QT decomposition following .

\subsection*{From tissue diffusivities to QT decomposition}\label{ss:CG}
\noindent
We derived Eqs.~(\ref{QT_SA_relation}) using a  shortcut that happens to work for the $\C$ tensor: the special $\B$-tensor family (\ref{eq:AxSymB}) allowed us to match the spherical harmonics components (\ref{eq:MGC_SH}) with those of Eqs.~(\ref{BBC}). It turns out that for higher-order tensors (arising at the $\O(b^3)$ level and beyond), the axially-symmetric family (\ref{eq:AxSymB}) does not probe all their DOF. Hence, below we provide a more generalizable way to derive Eqs.~(\ref{QT_SA_relation}) by explicitly following the components  (\ref{eq_Slm}) through triple products of spherical harmonics and the corresponding Clebsch-Gordan coefficients. This highlights the connection between the compartmental diffusion tensor covariances and the addition of angular momenta (cf. Supplementary Section \ref{SM:CGcoeffsSTF}). 

We begin from the irreducible decomposition of compartmental diffusion tensors (\ref{D=Cart}), this can be used to explicitly write $\D$ and $\C$,  Eqs.~(\ref{eq:DCdefinitions}):
%of $\C$ (\ref{eq:DCdefinitions}):
\begin{equation} \label{eq:DCcart_Dsph}
	\begin{aligned}
		\D_{ij}    &=  \la D^{00} \ra  \delta_{ij} + \sum_m \la D^{2m} \ra \Y_{ij}^{2m},\\
		\C_{ijkl} &=  \lla (\Dc^{00} )^2 \rra \, \delta_{ij} \delta_{kl}  + \sum_{m,m'} \lla D^{2m} D^{2m'} \rra \, \Y_{ij}^{2m} \Y_{kl}^{2m'}\\
		&+ \sum_{m}  \lla D^{00} D^{2m} \rra \Big ( \delta^{}_{ij} \Y^{2m}_{kl} + \Y_{ij}^{2m} \delta^{}_{kl}\Big )\, .\\
%		&+ \sum_{m,m'} \lla D^{2m} D^{2m'} \rra \, \Y_{ij}^{2m} \Y_{kl}^{2m'} \,. 
%		\D_{ij} %&= \lla D_{ij} \rra \\
%&=  \la D^{00} \ra  \Y_{ij}^{00} + \sum_m \la D^{2m} \ra \Y_{ij}^{2m},\\
%\C_{ijkl} %&= \lla D_{ij}^\alpha \, D_{kl}^\alpha \rra \\
%&=  \lla (\Dc^{00} )^2 \rra \, \Y^{00}_{ij} \Y^{00}_{kl}  \\
%&+ \sum_{m}  \lla D^{00} D^{2m} \rra \Big ( \Y_{ij}^{00} \Y^{2m}_{kl} + \Y_{ij}^{2m} \Y^{00}_{kl}\Big )\, \\
%&+ \sum_{m,m'} \lla D^{2m} D^{2m'} \rra \, \Y_{ij}^{2m} \Y_{kl}^{2m'} \,. 
	\end{aligned}
\end{equation}
The SA decomposition of $\C$, Eqs.~(\ref{W})-(\ref{A}), yields   
\begin{equation}
\begin{aligned}\nonumber
	\S_{ijkl} & =  \lla (\Dc^{00} )^2 \rra \, \delta_{(ij} \delta_{kl)} + 2 \textstyle \sum_{m}  \lla D^{00} D^{2m} \rra \delta_{(ij}^{} \Y^{2m}_{kl)} \, \\
	&+\textstyle \sum_{m,m'} \lla D^{2m} D^{2m'} \rra \, \Y_{(ij}^{2m} \Y_{kl)}^{2m'}   \,,\\
	\A_{ijkl} &= \C_{ijkl} - \S_{ijkl} =  \lla (\Dc^{00} )^2 \rra \, \Big( \delta_{ij} \delta_{kl}  -  \delta_{(ij} \delta_{kl)} \Big) \\
	&+ \textstyle \sum_{m}  \lla D^{00} D^{2m} \rra \Big ( \delta_{ij}^{} \Y^{2m}_{kl} + \Y_{ij}^{2m} \delta^{}_{kl} - 2 \delta_{(ij}^{} \Y_{kl)}^{2m} \Big )\, \\
	&+ \textstyle \sum_{m,m'} \lla D^{2m} D^{2m'} \rra \, \Big( \Y_{ij}^{2m} \Y_{kl}^{2m'} -  \Y_{(ij}^{2m} \Y_{kl)}^{2m'} \Big)   \, .
%	\S_{ijkl} & =  \lla (\Dc^{00} )^2 \rra \, \Y^{00}_{(ij} \Y^{00}_{kl)} + 2 \sum_{m}  \lla D^{00} D^{2m} \rra \Y_{(ij}^{00} \Y^{2m}_{kl)} \, \\
%&+ \sum_{m,m'} \lla D^{2m} D^{2m'} \rra \, \Y_{(ij}^{2m} \Y_{kl)}^{2m'}   \,,\\
%\A_{ijkl} &= \C_{ijkl} - \S_{ijkl} =  \lla (\Dc^{00} )^2 \rra \, \Big( \Y^{00}_{ij} \Y^{00}_{kl}  -  \Y_{(ij}^{00} \Y^{00}_{kl)} \Big) \\
%&+ \sum_{m}  \lla D^{00} D^{2m} \rra \Big ( \Y_{ij}^{00} \Y^{2m}_{kl} + \Y_{ij}^{2m} \Y^{00}_{kl} - 2 \Y_{(ij}^{00} \Y_{kl)}^{2m} \Big )\, \\
%&+ \sum_{m,m'} \lla D^{2m} D^{2m'} \rra \, \Big( \Y_{ij}^{2m} \Y_{kl}^{2m'} -  \Y_{(ij}^{2m} \Y_{kl)}^{2m'} \Big)   \, .
\end{aligned}
\end{equation}
%This leaves us with symmetric tensors $\D_{ij}$ and $\S_{ijkl}$, and with 
Although $\A_{ijkl}$ does not seem to have any particular symmetry at first glance, it can be mapped to a second order symmetric tensor using Eq.~(\ref{Apq=}):
\begin{equation} \nonumber
\begin{aligned}
		\A_{pq}   &= \delta_{pq} (\A_{iikk} - \A_{ikik}) + 2 (\A_{pkqk} -  \A_{pqkk}) \\
		& = 2   \lla (\Dc^{00} )^2 \rra \delta_{pq}^{}  - 2 \textstyle \sum_{m}  \lla D^{00} D^{2m} \rra  \Y^{2m}_{pq}   \\
		&- \textstyle \sum_{m,m'} \lla D^{2m} D^{2m'} \rra \Big(  \Y^{2m}_{ij} \Y^{2m'}_{ij}  \delta_{pq}^{} - 2 \Y^{2m}_{pk} \Y^{2m'}_{qk} \Big)  .
\end{aligned}
\end{equation}
This leaves us with symmetric tensors $\D_{ij}$, $\S_{ijkl}$, and $\A_{pq}$ written as a function of compartmental diffusion tensor covariances, for which we can compute irreducible decompositions following Eqs.~(\ref{eq_Slm}). Below we group their spherical tensor elements according to their degree $\ell$:
\begin{equation}\label{eq:ellSYSTEM_SA}
	\begin{aligned}
		\D^{00} & =\lla \Dc^{00}\rra\,, \\
		\S^{00} & = \lla (\Dc^{00} )^2 \rra \! %\\
		%&
		+\!\tfrac{1}{5 } \!\! \sum_{m^\prime ,m^{\prime\prime}} \!\! \lla D^{2 m^{\prime}} \! D^{2 m^{\prime\prime}}\rra\! \mathcal{Y}_{(i j}^{2 	m^{\prime}} \mathcal{Y}_{kl)}^{2 m^{\prime\prime}} \! \mathcal{Y}_{ijkl}^{00^*} \,, \\
		\A^{00} & =2 \lla (\Dc^{00} )^2 \rra-\tfrac12 \!\!  \sum_{m^\prime ,m^{\prime\prime}} \!\! \lla D^{2 m^{\prime}}\! D^{2 m^{\prime\prime}}\rra\! \mathcal{Y}_{(i j}^{2 	m^{\prime}} \mathcal{Y}_{kl)}^{2 m^{\prime\prime}} \! \mathcal{Y}_{ijkl}^{00^*} \,, \\[5pt]
%		\S^{00} & = \lla (\Dc^{00} )^2 \rra + \tfrac15 \textstyle \sum_{m} \lla {\Dc^{2 m}}^* \Dc^{2 m} \rra \,, \\
%		\A^{00} & =2 \lla (\Dc^{00} )^2 \rra-\tfrac12 \, \textstyle \sum_{m} \lla {\Dc^{2 m}}^* \Dc^{2 m} \rra \,, \\[5pt]
		\D^{2 m} & =\lla D^{2 m}\rra \,, \\
		\S^{2 m} & =2 \lla D^{00} D^{2 m}\rra\!+\!\tfrac{4}{7} \! \!\sum_{m^\prime ,m^{\prime\prime}} \!\!\lla D^{2 m^{\prime}}\! D^{2 m^{\prime\prime}}\rra\! \mathcal{Y}_{(i j}^{2 	m^{\prime}} \mathcal{Y}_{kl)}^{2 m^{\prime\prime}} \! \mathcal{Y}_{ijkl}^{2 m^*} \,, \\
		\A^{2 m} & =\shortminus 2 \lla D^{00} D^{2 m}\rra\! +\!2\!\!  \sum_{m^\prime ,m^{\prime\prime}}\!\! \lla D^{2 m^{\prime}} \! D^{2 m^{\prime\prime}}\rra\! \mathcal{Y}_{(i j}^{2 	m^{\prime}} \mathcal{Y}_{kl)}^{2 m^{\prime\prime}} \! \mathcal{Y}_{ijkl}^{2 m^*} \,, \\[5pt]	
%		\S^{2 m} & =2 \lla D^{00} D^{2 m}\rra\!\\
%		&+\!\tfrac{8}{21 } \textstyle \sum_{m^\prime ,m^{\prime\prime}} \lla D^{2 m^{\prime}} D^{2 m^{\prime\prime}}\rra\! \mathcal{Y}_{i j}^{2 m^{\prime}} \mathcal{Y}_{j k}^{2 m^{\prime\prime}} \mathcal{Y}_{k i}^{2 m^*} \,, \\
%		\A^{2 m} & =\shortminus 2 \lla D^{00} D^{2 m}\rra\! \\
%		&+\!\tfrac{4}{3} \textstyle \sum_{m^\prime ,m^{\prime\prime}}\lla D^{2 m^{\prime}} D^{2 m^{\prime\prime}}\rra\! \mathcal{Y}_{i j}^{2 m^{\prime}} \mathcal{Y}_{j k}^{2 m^{\prime\prime}} \mathcal{Y}_{k i}^{2 m^*} \,, \\[5pt]
		\S^{4 m} & =\tfrac{8}{35 }   \sum_{m^\prime ,m^{\prime\prime}} \lla D^{2 m^{\prime}} D^{2 m^{\prime\prime}}\rra \mathcal{Y}_{(i j}^{2 m^{\prime}} \mathcal{Y}_{k l)}^{2 m^{\prime\prime}} \! \mathcal{Y}_{i j k l}^{4 m^*} \,,
	\end{aligned}
\end{equation}
%using the relations $\Y^{2m}_{ij}\Y^{2m'}_{ij} %= \Y^{2m}_{ij}\Y^{2m'}_{jk} \Y^{00}_{ki} 
%= \tfrac32 \Y^{2m}_{(ij}\Y^{2m'}_{kl)} \Y^{00}_{ijkl}$ and $ \Y^{2m'}_{ij}\Y^{2m''}_{jk} \Y^{2m^*}_{ki} = \tfrac32 \Y^{2m'}_{(ij}\Y^{2m''}_{kl)} \Y^{2m^*}_{ijkl}$.
using $\Y^{2m}_{ij}\Y^{2m'}_{ij} %= \Y^{2m}_{ij}\Y^{2m'}_{jk} \Y^{00}_{ki} 
= \tfrac32 \Y^{2m}_{(ij}\Y^{2m'}_{kl)} \!\Y^{00}_{ijkl}$ and $ \Y^{2m'}_{ij}\Y^{2m''}_{jk} \Y^{2m^*}_{ki} = \tfrac32 \Y^{2m'}_{(ij}\Y^{2m''}_{kl)}\! \Y^{2m^*}_{ijkl}$.

%Although not obvious at first sight, Eqs.~(\ref{eq:ellSYSTEM_SA}) are related with Eqs.~(\ref{QTLM}) via  Eqs.~(\ref{QT_SA_relation}).  
The connection (\ref{QT_SA_relation}) to the QT decomposition (\ref{QTLM}) is established  via the relations of the triple products of spherical basis tensors to the products of Clebsch-Gordan coefficients:
\begin{equation}\label{eq:YYY_CGcoeffs}
\begin{aligned}
\mathcal{Y}_{(i j}^{2 m^{\prime}}  \mathcal{Y}_{k l)}^{2 m^{\prime\prime}} \!\mathcal{Y}_{i j k l}^{0 0^*} & = 5 \braket{2020 | 00} \braket{2m' 2m'' | 00}, \\
\mathcal{Y}_{(i j}^{2 m^{\prime}}  \mathcal{Y}_{k l)}^{2 m^{\prime\prime}} \!\mathcal{Y}_{i j k l}^{2 m^*} & = \tfrac74 \braket{2020 | 20}\braket{2m' 2m'' | 2m},\\
\mathcal{Y}_{(i j}^{2 m^{\prime}}  \mathcal{Y}_{k l)}^{2 m^{\prime\prime}} \! \mathcal{Y}_{i j k l}^{4 m^*} & = \tfrac{35}{8}  \braket{2020 | 40} \braket{2m' 2m'' | 4m},\\
\end{aligned}
\end{equation}
%$\mathcal{Y}_{(i j}^{2 m^{\prime}}  \mathcal{Y}_{k l)}^{2 m^{\prime\prime}} \mathcal{Y}_{i j k l}^{2 m^*} $ and $\Y_{(i j}^{2 m^{\prime}} \Y_{k l)}^{2 m^{\prime\prime}} {\Y_{i j k l}^{4 m}}^*$ become proportional to the Clebsch-Gordan coefficients $\la 2, m^\prime, 2, m^{\prime\prime} | 2, m \ra$ and $\la 2, m^\prime, 2, m^{\prime\prime} | 4, m \ra$, 
see derivation of the Supplementary Eq.~(\ref{eq:YYY_CGcoeffs}). 
We then use Eq.~(\ref{TLM}) to express the sums in Eqs.~(\ref{eq:ellSYSTEM_SA}) via $\T^{\ell m}$: 
\begin{equation}\label{eq:ellSYSTEM_QT}
	\begin{aligned}
%		\D_{00} & =\lla D_{00}\rra\,, \\
%		\D_{2 m} & =\lla D_{2 m}\rra \,, \\[5pt]
%		\Q^{00} & = \lla (\Dc^{00} )^2\rra \,, \\
%		\Q^{2 m} & =  2 \lla \Dc^{00} \Dc^{2 m}\rra\! \,, \\[5pt]
		 \textstyle \sum_{m^\prime ,m^{\prime\prime}} \lla \Dc^{2 m^{\prime}} \Dc^{2 m^{\prime\prime}}\rra\! \mathcal{Y}_{(i j}^{2 m^{\prime}}  \mathcal{Y}_{k l)}^{2 m^{\prime\prime}} \!\mathcal{Y}_{i j k l}^{00^*}
		& = 5\T^{00} \,, \\
		 \textstyle \sum_{m^\prime ,m^{\prime\prime}} \lla \Dc^{2 m^{\prime}} \Dc^{2 m^{\prime\prime}}\rra\! \mathcal{Y}_{(i j}^{2 m^{\prime}}  \mathcal{Y}_{k l)}^{2 m^{\prime\prime}} \!\mathcal{Y}_{i j k l}^{2 m^*}
		 & = \tfrac74 \T^{2 m} \,, \\
		 \textstyle \sum_{m^\prime ,m^{\prime\prime}}\lla \Dc^{2 m^{\prime}} \Dc^{2 m^{\prime\prime}}\rra \mathcal{Y}_{(i j}^{2 m^{\prime}} \mathcal{Y}_{k l)}^{2 m^{\prime\prime}} \!\mathcal{Y}_{i j k l}^{4 m^*}
		 & = \tfrac{35}8 \T^{4 m} \,,
	\end{aligned}
\end{equation}
which after simple algebra proves the system (\ref{QT_SA_relation}). 

At first glance, Clebsch-Gordan coefficients may seem a non-intuitive 
way to combine the components $\lla D^{2m'} D^{2m''}\rra$. However, they can yield elegant simplifications. Consider $\T^{00}$ as defined in Eq.~(\ref{TLM}). 
Using $\braket{ 2 0 2 0 | 0 0}=1/\sqrt{5}$ and $\braket{2 m 2 m' | 0 0} = (-1)^m \delta_{m,-m'}/\sqrt{5}$, as well as Eq.~(\ref{condon-shortley}), we get 
\begin{equation}\label{eq:T00}
\begin{aligned}
	\T^{00} 
	%&=  \tfrac15  \textstyle \sum_{m ,m'} \lla \Dc^{2 m} \Dc^{2 m'}\rra\! \mathcal{Y}_{(i j}^{2 m}  \mathcal{Y}_{k l)}^{2 m'} \mathcal{Y}_{i j k l}^{00^*} \\
	&=\braket{ 2 0 2 0 | 0 0} \textstyle \sum_{m,m'} \braket{2 m 2 m' | 0 0} \lla \Dc^{2m} \Dc^{2m'} \rra \\
	&= \tfrac15 \textstyle\sum_{m} (-1)^{m} \lla \Dc^{2m} \Dc^{2 -m} \rra
	= \tfrac15 \textstyle\sum_{m}  \lla \Dc^{2m} \Dc^{2 m \,*} \rra .
\end{aligned}
\end{equation}

\subsection*{Intrinsic invariants of $\Tf = \Sf$}\label{ss:RotInvs}
%\dn{should we change $\S$ to $\T$ everywhere here? or change Eq 60}
\noindent
The 3 independent invariants for a 2nd-order tensor, such as $\D$, can be constructed from the coefficients of its characteristic polynomial 
$p(\lambda)=\det(\D-\lambda \mathsf{I}) =-\lambda^3+c_2\lambda^2+c_1\lambda+c_0$:   $c_0=\det \D $, $c_1=\tfrac12 (\tr \D^2 - \tr^2\D)$, and $c_2=\tr \D$. In the main text, we mapped them onto $\D_0$, $\D_{2}$, and $\D_{2|3}$, Eqs.~(\ref{MD}) and (\ref{ell2invs}), by first isolating the scalar and the STF-2 parts, Eq.~(\ref{D=Cart}). 
%Constructing invariants for 2nd-order tensors, such as $\St$, involves the characteristic polynomial $p(\lambda)=\det(\St-\lambda \mathsf{I})$. Roots of $p(\lambda)$ are the eigenvalues of $\St$, and the coefficients of the terms $1,\lambda,\lambda^2$ yield algebraically independent invariants.  

For 4th-order tensors, the characteristic polynomial needs to be defined. 
Fortunately, 4th-order 3d tensors with minor symmetry, like the elasticity or covariance tensors, can be mapped to a 2nd-order 6d tensor $\C \rightarrow \C_{6\times 6}$ using the Kelvin/Mandel notation \cite{BETTEN1987,BASSER2007}. Explicitly \citep{BASSER2007}:
\begin{equation}\label{eq:C_6x6}
	\C_{6\times 6}=
	\left(\begin{array}{cccccc}\!\!
		{\scriptstyle \C_{1111}}            & {\scriptstyle \C_{1122} }           & {\scriptstyle \C_{1133} }           & {\scriptstyle \sqrt{2}\, \C_{1112} }& {\scriptstyle \sqrt{2}\, \C_{1113} }& {\scriptstyle \sqrt{2}\, \C_{1123}  }\vspace{0.1cm} \\ \vspace{0.1cm}
		{\scriptstyle \C_{1122}}            & {\scriptstyle \C_{2222} }           & {\scriptstyle \C_{2233} }           & {\scriptstyle \sqrt{2}\, \C_{2212} }& {\scriptstyle \sqrt{2}\, \C_{2213} }& {\scriptstyle \sqrt{2}\, \C_{2223} }\\ \vspace{0.1cm}
		{\scriptstyle \C_{1133}}            & {\scriptstyle \C_{2233} }           & {\scriptstyle \C_{3333} }           & {\scriptstyle \sqrt{2}\, \C_{3312} }& {\scriptstyle \sqrt{2}\, \C_{3313} }& {\scriptstyle \sqrt{2}\, \C_{3323} }\\ \vspace{0.1cm}
		{\scriptstyle \sqrt{2}\, \C_{1112} }& {\scriptstyle \sqrt{2}\, \C_{2212} }& {\scriptstyle \sqrt{2}\, \C_{3312} }& {\scriptstyle 2\, \C_{1212} }       & {\scriptstyle 2\, \C_{1213} }       & {\scriptstyle 2\, \C_{1223} }\\ \vspace{0.1cm}
		{\scriptstyle \sqrt{2}\, \C_{1113} }& {\scriptstyle \sqrt{2}\, \C_{2213} }& {\scriptstyle \sqrt{2}\, \C_{3313} }& {\scriptstyle 2\, \C_{1312} }       & {\scriptstyle 2\, \C_{1313} }       & {\scriptstyle 2\, \C_{1323}} \\ \vspace{0.1cm}
		{\scriptstyle \sqrt{2}\, \C_{1123} }& {\scriptstyle \sqrt{2}\, \C_{2223} }& {\scriptstyle \sqrt{2}\, \C_{3323} }& {\scriptstyle 2\, \C_{2312} }       & {\scriptstyle 2\, \C_{2313} }       & {\scriptstyle 2\, \C_{2323} }
	\end{array}\!\!\right),
\end{equation}
where the major symmetry of $\C$ makes $\C_{6\times 6}$ symmetric, and factors $\sqrt{2}$ and $2$ account for repeated elements, such that we can define traces of the powers of 4th-order tensors as 
\begin{equation} \label{tr_Sn}
	\tr\, \C^n \equiv \C_{i_1 j_1 i_2 j_2}  \hdots \C_{i_n j_n i_1 j_1}= \tr\, (\C_{6\times 6})^n \,.
%	\tr\, (\Sf)^n \equiv \S^{(4)}_{i_1 j_1 i_2 j_2}  \hdots \S^{(4)}_{i_n j_n i_1 j_1}= \tr\, (\Sf_{6\times 6})^n.
\end{equation}

The mapping  $\C \rightarrow \C_{6\times 6}$ allows one to extract 6 invariants from the characteristic polynomial of a square matrix $\C_{6\times 6}$.   
This approach was extended \citep{BETTEN1987} by considering the coefficients of the polynomial 
$\tilde p(\nu,\mu)=\det\big( \C_{6\times 6}- \mathsf{I^{(4)}}(\nu,\mu)\big)$ with $\mathsf{I}_{ijkl}^{(4)}(\nu,\mu) = \tfrac{\nu}{2} (\delta_{ik}\delta_{jl} + \delta_{il}\delta_{jk}) + \mu \,\delta_{ij}\delta_{kl}$. 
However, such invariants do not form a complete set, and besides, they mix the irreducible representations and symmetries.

The irreducible decomposition (\ref{DxD}) allows us to keep track of  symmetries, and consider a more constrained problem: find the $6$  intrinsic invariants of the only component $\Sf=\Tf$ of $\C$ that is neither a scalar nor an STF-2 tensor. For that, we use the mapping (\ref{eq:C_6x6}) to define a $6\times 6$ matrix $\Sf_{6\times 6}$, cf. Supplementary Eq.~(\ref{S4_6x6}). We consider standard characteristic polynomial $p(\lambda)=\det\big(\Sf_{6\times 6}-\lambda \mathsf{I}\big)$ because, as it turns out, no extra invariants arise from the coefficients of $\tilde p(\lambda,\mu)$ for $\Sf_{6\times 6}$.  

From direct inspection of $\Sf_{6\times 6}$ (Supplementary Section~\ref{SM:C6x6}) we find that its trace and one of its eigenvalues are zero.  
The polynomial $p(\lambda) = \lambda^6 + p_5 \lambda^5 + \dots + p_1 \lambda + p_0$ has only 4 algebraically independent coefficients: 
\begin{equation} \label{p14}
\begin{aligned}
p_1 & = \tfrac16 \tr \big(\Sf\big)^2 \tr \big(\Sf\big)^3 - \tfrac15 \tr \big(\Sf\big)^5 \,,\\
p_2 & = \tfrac18 \tr \big(\Sf\big)^2 \tr \big(\Sf\big)^2 - \tfrac14 \tr \big(\Sf\big)^4 \,,\\
p_3 & = -\tfrac13 \tr \big(\Sf\big)^3 \,,\\
p_4 & = -\tfrac12 \tr \big(\Sf\big)^2 \,,
%p_1 & = \tfrac16 \tr (\Sf_{6\times 6})^2 \tr (\Sf_{6\times 6})^3 - \tfrac15 \tr (\Sf_{6\times 6})^5 \,,\\
%p_2 & = \tfrac18 \tr (\Sf_{6\times 6})^2 \tr (\Sf_{6\times 6})^2 - \tfrac14 \tr (\Sf_{6\times 6})^4 \,,\\
%p_3 & = -\tfrac13 \tr (\Sf_{6\times 6})^3 \,,\\
%p_4 & = -\tfrac12 \tr (\Sf_{6\times 6})^2 \,,
\end{aligned}
\end{equation}
since $p_0=  \det \Sf_{6\times 6} = 0$ and $p_5=-\tr \Sf = 0$
(traces are defined according to Eq.~(\ref{tr_Sn})). 
Thus $p(\lambda)$ has 4 nonzero independent roots (eigenvalues of $\Sf_{6\times 6}$). 
Since by Cayley–Hamilton theorem, each matrix satisfies its characteristic equation, 
$p\big(\Sf_{6\times 6}\big)=0$,  the 4 traces $\tr \big(\Sf\big)^2$, $\tr \big(\Sf\big)^3$, $\tr \big(\Sf\big)^4$, and $\tr \big(\Sf\big)^5$, are the 4 algebraically independent invariants that determine all traces of higher powers of $\Sf$. The latter can be iteratively obtained from relations 
$\tr \big[ \big(\Sf\big)^n p\big(\Sf\big)\big]=0$, $n=0,1,2,\dots$, by expressing $\tr \big(\Sf\big)^{6+n}$ via the above 4 independent traces that also define the characteristic polynomial coefficients, Eq.~(\ref{p14}).

% \vspace{-0.2cm}
\begin{table*}[th!!]%[!htbp]
	\centering
	\caption{Description of the four RICE protocols (rows) acquired for each volunteer. Imaging parameters were kept constant for all protocols and numbers denote the different directions sampled on each shell. All b-values are in microstructure units $\unit{ms}/\unit{\mu m^2}$.}\vspace{0.05cm}
	\begin{tabular}{@{}l@{}c@{}c@{}c@{}c@{}c@{}c@{}}
%	\begin{tabular}{lcccccc}
		\hline
		\multicolumn{7}{c}{$\text{Resolution}=2\times2\times2\,\unit{mm}^3$, $\text{PF}=6/8$, $R_\text{GRAPPA}=2$, $68$ slices, MB=2, TE=90ms, TR=4.2s} \\
		\hline
		Protocol & $\,\,b=0\,\,$ &  $\,\,b_\text{LTE}=1 \,\,$  & $\,\,b_\text{LTE}=2\,\,$ &  $\,\,b_\text{PTE}={1.5}\,\,$ &  $\,\,b_\text{STE}={1.5}\,\,$ & $\,\,t_\text{acq}$ [min' sec''] \\
		\hline
		DKI & 1     & 30    & 60    & -     & -     & 6' 43'' \\
		
		\rowcolor{lightgray}
		iRICE (MD+FA+MK) & 1     & 6     & 6     & -     & -     & 1' 14'' \\
		
		RICE & 2     & 30    & 60    & 30    & -     & 11' 20'' \\
		
		\rowcolor{lightgray}
		iRICE (MD+FA+MK+\mFA) & 2     & 6     & 6     & -     & 3     & 1' 56'' \\
		\hline
	\end{tabular}%
	\label{tab:TableProtocols}%
\end{table*}%
% \vspace{0.1cm}

The remaining 2 independent intrinsic invariants of $\Sf$ can be obtained from its eigentensor decomposition \citep{BASSER2007}
\begin{equation}\label{eq:eigtensordecomp}
    \S_{ijkl}^{(4)} = \sum_{a=1}^6 \lambda_a \E_{ij}^{(a)} \E_{kl}^{(a)},
\end{equation}
where $\lambda_a$ and $\E_{ij}^{(a)}$ are the eigenvalues and eigentensors of $\Sf$ according to the mapping (\ref{eq:C_6x6}). 
Namely, $\lambda_a$ are the eigenvalues of $\S^{(4)}_{6\times 6}$, 
and $\hat{\mathbf{v}}^{(a)} = \big[v^{(a)}_{xx},\, v^{(a)}_{yy},\, v^{(a)}_{zz},\,v^{(a)}_{xy},\, v^{(a)}_{xz},\, v^{(a)}_{yz} \big]'$ are its normalized 6d eigenvectors, yielding the eigentensors  
\begin{equation}
    \E_{ij}^{(a)}=\left(
    \begin{array}{ccc}
v_{x x}^{(a)} & \frac{1}{\sqrt{2}} v_{x y}^{(a)} & \frac{1}{\sqrt{2}} v_{x z}^{(a)} \\
\frac{1}{\sqrt{2}} v_{x y}^{(a)} & v_{y y}^{(a)} & \frac{1}{\sqrt{2}} v_{y z}^{(a)} \\
\frac{1}{\sqrt{2}} v_{x z}^{(a)} & \frac{1}{\sqrt{2}} v_{y z}^{(a)} & v_{z z}^{(a)}
\end{array}
\right) . 
\end{equation}
The orthogonality of $\hat{\mathbf{v}}^{(a)}$ induces the tensor orthogonality $\E^{(a)}_{ij} \E^{(b)}_{ij} = \delta_{ab}$. 
Based on Supplementary Section~\ref{SM:C6x6},  the eigenvector 
$\hat{\mathbf{v}}^{(a_0)}=\tfrac{1}{\sqrt{3}}\,(1,\,1,\,1,\,0,\,0,\,0)^t$, 
associated with a zero eigenvalue $\lambda_{a_0} = 0$, corresponds to the eigentensor 
\begin{equation}
\E^{(a_0)}_{ij} = \tfrac{1}{\sqrt{3}} \delta_{ij}\,; \quad \tr \mathsf{E}^{(a)} = 0 \,,  \ \  a\neq a_0 \,, 
\end{equation}
with the latter equality for all other 5 eigentensors stemming from their orthogonality to $\E^{(a_0)}$. 
Note that, although the tensor products $\E_{ij}^{(a)} \E_{kl}^{(a)}$ are not fully symmetric, when combined according to Eq.~(\ref{eq:eigtensordecomp}),  $\S_{ijkl}^{(4)}$ becomes fully symmetric.

We now define the remaining 2 independent invariants of $\Sf$ by arranging the above eigentensors into the following 2 combinations: 
\begin{equation} \label{def-E}
\E_{ij} = \sum_{a=1}^6  \lambda_a \E_{ij}^{(a)}, \quad \text{and}\quad \tilde{\E}_{ij} =  \sum_{a\neq a_0} \E_{ij}^{(a)}.
\end{equation}
Analogously to the $\D$-tensor example above, traces of the powers of 2nd-order tensors $\E$ and $\tilde{\mathsf{E}}$ are rotationally invariant; the traces of 1st, 2nd and 3rd powers of these matrices are algebraically independent. 
Further, by construction, $\tr \E =\tr \tilde{\E} =0$. 
One can also check that $\tr \E^2=\tr\big(\Sf\big)^2$ given by one of the previously found invariants, 
and $\tr \tilde{\E}^2=5$ independent of $\Sf$.  
Therefore, we identify the two remaining independent intrinsic invariants of $\Sf$ with $\tr \E^3$ and $\tr \tilde{\E}^3$. 
To summarize, the 6 intrinsic invariants of $\Sf$ are: 
\begin{equation}\label{ell4invs}
	\begin{aligned}
	\iSf{n}&= \left ( \tfrac8{35}\, \tr \big( \Sf \big)^n \right)^{1/n}, \quad n=2, \dots, 5 \,; \\
	\iSf{6}&={\tr}^{1/3}\, \E^3 \,, \qquad	\iSf{7}={\tr}^{1/3}\, \tilde{\E}^3,
\end{aligned}
\end{equation}
with $\E$ and $\tilde{\E}$  defined in Eq.~(\ref{def-E}). 
They are normalized such that $\S_4\equiv \S_{4|2}$ is the 2-norm of $\S^{4m}$, cf. Eq.~(\ref{VarTf_dir}). 
The independence of invariants from the same irreducible component can be checked with the Jacobian criterion for algebraic independence\cite{CARUYER2015}.

One could alternatively define $\Sf$ intrinsic invariants using the general formulation for $\ell=4$
\begin{equation} \label{RIF}
	\begin{aligned}
	\tilde\S_{\ell|n} & \!= \!\left((2\ell+1)\!\! \int_{\mathbb{S}^2} \!  \d\n \, \big(\Sl\!(\n)\big)^n \right)^{1/n} \!\!, \, n=2,\hdots,2\ell-1\,, \\
%			     &\!= \!\left((2\ell+1) \S^{\ell m_1}\hdots\S^{\ell m_n}\! \int_{\mathbb{S}^2}\!  \d\n\, Y^{\ell m_1}(\n)\hdots Y^{\ell m_n}(\n)\right)^{1/n}\,.
	\end{aligned}
\end{equation}
The integral over ${\mathbb{S}^2}$ acts as a projector of $n$-th order powers of $\Sf(\n)$ onto the isotropic subspace% ($\ell'=0$)
. 
This means that $\tilde\S_{4|n}$ are obtained by coupling the spherical harmonic components $\S^{4m_1}\hdots\S^{4m_n}$ to zero total angular momentum. 
For $n=2$, one can use Eq.~(\ref{STF=SH}) and Supplementary Eq.~(\ref{eq:tripleSHint})  to find 
$\tilde{\S}_{4|2}{}^2 = 9\sum_{m,m'}\S^{4m}\S^{4m'} \braket{4040| 00}\braket{4m4m'| 00} = \sum_m \S^{4m}\S^{4m*} = {\S_4}^2$, cf. Eq.~(\ref{VarTf_dir}).  
Here, Clebsch–Gordan coefficients enforce $m'=-m$, and the contraction reduces to a sum over conjugate pairs, as in Eq.~(\ref{eq:T00}). 
Both definitions (\ref{ell4invs}) and (\ref{RIF})  generate complete sets of invariants, and the first four invariants (\ref{RIF}), with $n=2,\dots,5$, correspond to the combinations of the traces $ \tr \big(\Sf \big)^n $, while the $n=6,7$ ones are independent, and therefore can serve as alternatives to $\S_{4|6}$ and $\S_{4|7}$. 
The construction (\ref{RIF}) is similar to the definition of spherical harmonic dMRI signal invariants in Ref.~\cite{ZUCCHELLI2020}, but is applied to a specific $\ell=4$ component $\Sf$, rather than to a mixture such as $\S=\Sz\oplus\St\oplus\Sf$. 
%Mixing irreducible components would obscure the group-theoretic meaning of the invariants.
Working with the pure $\ell=4$ component allows one to algebraically relate any $\tilde\S_{4|n}$ to the combination of the spherical tensor components $\S^{4m}$ via Clebsch-Gordan theory. 

\subsection*{MRI experiments}
\noindent
After providing inform consent, three healthy volunteers (23 yo and 25 yo females, 33 yo male) 
underwent MRI in a whole body 3T-system (Siemens, Prisma) using a 32-channel head coil. Maxwell-compensated free gradient diffusion waveforms \cite{SZCZEPANKIEWICZ2019} were used to yield linear, planar, and spherical $\B$-tensor encoding using a prototype spin echo sequence with EPI readout \cite{SZCZEPANKIEWICZ2019c}. Four RICE dMRI datasets were acquired according to Table \ref{tab:TableProtocols}. Imaging parameters:  voxel size$\,=2\times2\times2\,$mm$^3$, $T_R=4.2\,$s, $T_E=90\,$ms, bandwidth $=1818\,$Hz/Px, $R_\text{GRAPPA}=2$, partial Fourier$\,=6/8$, multiband$\,=2$. Scan time was 15 minutes per subject for all protocols (DKI was subsampled from RICE).

For the clinical validation\cite{LIAO2024,CHEN2024}, 627 subjects (aged $42.7 \pm 13.6$ years, 443 females) identified with a clinical diagnosis of multiple sclerosis %using the McDonald criteria\cite{POLMAN2011} 
who were referred for MRI of the head at NYU Langone Health between November 2014 and June 2020. 562 controls (aged $42.9 \pm 14.1$ years, 386 females) were selected from subjects with normal brain MRI and no history of neurological disorder, referred for imaging due to headache or dizziness. Groups were matched for age and sex. Subjects underwent MRI on whole body Siemens 3T-systems (Prisma: $47.1\%$, Skyra: $52.9\%$). The clinical dMRI protocol used a monopolar EPI sequence with linear $\B$-tensor encoding as follows: 4-5 $b=0$ images, $b_\mathrm{LTE}=1\mathrm{ms}/\mu\mathrm{m}^2$ along 20 directions and $b_\mathrm{LTE}=2\mathrm{ms}/\mu\mathrm{m}^2$ along 60 directions. Imaging parameters: voxel size$\,=1.7\times1.7\times3\,$mm$^3$, $T_R=3.2-4\,$s and $T_E=70-96\,$ms on Prisma, and $T_R=3.5-4.3\,$s and $T_E=70-96\,$ms on Skyra, $R_\text{GRAPPA}=2$, partial Fourier$\,=6/8$, multiband$\,=2$. Scan time was 6 minutes per subject.

\subsection*{Image pre-processing}
\noindent
All four protocols in Table \ref{tab:TableProtocols} were processed identically and independently for each subject. Magnitude and phase data were reconstructed. Then, a phase estimation and unwinding step preceded the denoising of the complex images \citep{LEMBERSKIY2019}. Denoising was done using the Marchenko-Pastur principal component analysis method \citep{VERAART2016} on the real part of the phase-unwinded data. An advantage of denoising before taking the magnitude of the data is that Rician bias is reduced significantly \cite{ADESARON2024}. 
We also processed this data using only magnitude data. Here, magnitude denoising and rician bias correction were applied, see results in Supplementary Section \ref{SM:MinimalDesigns}. 
Only magnitude denoising was applied to the clinical LTE dataset. 
Data was subsequently processed with the DESIGNER pipeline \citep{CHEN2024}. 
Denoised images were corrected for Gibbs ringing artifacts accounting for the partial Fourier acquisition \citep{LEE2021}, based on re-sampling images using local sub-voxel shifts. Images were then rigidly aligned and corrected for eddy current distortions and subject motion simultaneously \citep{SMITH2004}. A $b=0$ image with reverse phase encoding was included for correction of EPI-induced distortions \citep{ANDERSSON2003}. Finally, %to further boost SNR, 
MRI voxels were locally smoothed based on spatial and intensity similarity akin to Ref.~\cite{WIEST2007}.%based on similar spatial locations and signal intensities akin the method proposed by \cite{WIEST2007}.

\subsection*{Parameter estimation}
\noindent
Two variants of the cumulant expansion were fit to the RICE and clinical datasets. % described in Table \ref{tab:TableProtocols}. 
This depended on which parameters each protocol was sensitive to. 
For the full DKI, full RICE, and clinical LTE protocols,  Eq.~(\ref{eq:MGC_SH}) was used, while for the iRICE protocols,  parameters were estimated according to Eq.~(\ref{eq:iRICE}). 
Weighted linear least squares were used for fitting \citep{VERAART2013} to highlight the gain achieved purely by acquisition optimization. Including positivity constraints to improve parameters robustness \citep{HERBERTHSON2021} is straightforward for spherical tensors but is outside of the scope of this work. 

\section*{Data availability}
\noindent
Example anonymized and processed dMRI data used for the RICE analysis are available at \href{https://cai2r.net/resources/rotational-invariants-of-the-cumulant-expansion-the-rice-toolbox/}{https://cai2r.net/resources/rotational-invariants-of-the-cumulant-expansion-the-rice-toolbox/}. The large prospective clinical dMRI dataset contains protected health information and is therefore not publicly available; we provide access to the summary statistics computed using RICE at \href{https://github.com/NYU-DiffusionMRI/RICE}{https://github.com/NYU-DiffusionMRI/RICE}.
%The anonymized and processed RICE diffusion MRI data are available at \href{https://github.com/NYU-DiffusionMRI/RICE}{https://github.com/NYU-DiffusionMRI/RICE}. The large clinical dMRI prospective dataset is available upon request.
%The Anonymized and processed diffusion MRI data are available at \href{https://github.com/NYU-DiffusionMRI/RICE}{https://github.com/NYU-DiffusionMRI/RICE}. The raw diffusion MRI data are protected and are not publicly available due to the presence of protected health information (PHI).

%Anonymized data is publicly available as part of the RICE toolbox at \href{https://github.com/NYU-DiffusionMRI/RICE}{https://github.com/NYU-DiffusionMRI/RICE}.

\section*{Code availability}
\noindent
All codes for RICE parameter estimation were implemented in MATLAB (R2021a, MathWorks, Natick, Massachusetts). These are publicly available as part of the RICE toolbox at \href{https://github.com/NYU-DiffusionMRI/RICE}{https://github.com/NYU-DiffusionMRI/RICE}.

%All codes for RICE parameter estimation were implemented in MATLAB (R2021a, MathWorks, Natick, Massachusetts). These are publicly available together with an anonymized dataset as part of the RICE toolbox at \href{https://github.com/NYU-DiffusionMRI/RICE}{https://github.com/NYU-DiffusionMRI/RICE}.

% \nocite{*}
%
%\newpage
%\section*{References}
%\noindent
\bibliography{Coelho_bibliography_2025}% Produces the bibliography via BibTeX.

\begin{acknowledgments}
\noindent
This work was performed under the rubric of the Center for Advanced Imaging Innovation and Research ($\text{CAI}^2\text{R}$, \href{https://www.cai2r.net}{https://www.cai2r.net}), an NIBIB Biomedical Technology Resource Center NIH P41-EB017183 (DN, EF). This work has been supported by NIH under NINDS R01 NS088040 (DN, EF) and NIBIB R01 EB027075 (DN, EF) and K99 EB036080 (SC) awards, as well as by the Swedish Research Council 2021-04844 (FS) and the Swedish Cancer Society 22 0592 JIA (FS). 
Authors are grateful to Sune Jespersen, Valerij Kiselev and Jelle Veraart for fruitful discussions.
%Matlab processing code and example data (RICE toolbox) for the estimation of RICE maps from fully sampled data or the iRICE protocols is available at \href{https://github.com/NYU-DiffusionMRI/RICE}{https://github.com/NYU-DiffusionMRI/RICE}.
\end{acknowledgments}

\section*{Author contributions statement}
\noindent
SC and DSN developed theory and wrote the manuscript. SC performed data analysis. JC and EF curated, and JC preprocessed MRI data for multiple sclerosis classification. FS provided multidimensional dMRI sequence. SC, EF and DSN conceived and designed the study. EF and DSN supervised the project. All authors reviewed the manuscript and provided feedback.

\section*{Competing interests statement}
\noindent
SC, EF, DSN are co-inventors in technology related to this research; a PCT patent application has been filed and 
is pending. EF, DSN, and NYU School of Medicine are stock holders of Microstructure Imaging, Inc. --- post-processing tools for advanced MRI methods. FS is an inventor of technology related to this research and has financial interest in related patents. The remaining authors declare no competing interests.

%\section*{Tables}
%\noindent
% 
% 
% \section*{Figures}
% \noindent
% 

%====================================================
%========START OF SUPPLEMENTARY INFORMATION==========
%====================================================

\clearpage
\newpage

\onecolumngrid

\setcounter{section}{0}
\setcounter{figure}{0}
\setcounter{table}{0}
\setcounter{page}{1}
\setcounter{equation}{0}
\renewcommand{\thepage}{S\arabic{page}} 
\renewcommand{\thesection}{S\arabic{section}}  
\renewcommand{\thetable}{S\arabic{table}}  
\renewcommand{\thefigure}{S\arabic{figure}} 
\renewcommand{\theequation}{S\arabic{equation}}

\section*{{Supplementary Information}}
% --- \textit{Geometry of the cumulant series in neuroimaging}}}

Santiago Coelho, Jenny Chen, Filip Szczepankiewicz, Els Fieremans, and Dmitry S. Novikov

Corresponding authors: Santiago.Coelho@nyulangone.org (SC) and Dmitry.Novikov@nyulangone.org (DSN)

%\makeatletter

\section{Irreducible decomposition into a sum of tensors with the same order, Eqs.~(\ref{Eq:STFdecomposition})--(\ref{Flm-gen})}
\label{SM:STF_generalized}
\noindent
To find all the spherical tensor components (\ref{eq_Slm}) of the decomposition  (\ref{Eq:STFdecomposition}) used in this work we need to project on the generalized basis (\ref{Eq:Ylm_deltas}). The normalization coefficients (\ref{eq:alpha_STF}) can be found straightforwardly:
%To find all the spherical tensor components (\ref{eq_Slm}) of the decomposition  (\ref{Eq:STFdecomposition}) used in this work by projecting on the generalized basis (\ref{Eq:Ylm_deltas}), the normalization coefficients (\ref{eq:alpha_STF}) can be found straightforwardly:
\begin{equation}\label{eq:rank24_CART2STF}
	\begin{aligned}
		\Y^{00}_{ij}&=\delta^{}_{ij}\,, \quad \zeta(2,0) = {\Y^{00}_{ij}}^* \Y^{00}_{ij} = \delta_{ij} \delta_{ij} = 3 \,; 
		\\
		\Y^{00}_{ijkl} &=  \delta_{(ij}\delta_{kl)}  \,, \quad \zeta(4,0) = \Y^{00}_{ijkl} \delta^{}_{ij} \delta^{}_{kl} 
		%=  \delta_{(ij}\delta_{kl)}  \delta_{ij}\delta_{kl} 
		=  \tfrac13 \big(  \delta_{ij}\delta_{kl} + \delta_{ik}\delta_{jl} + \delta_{il}\delta_{jk} \big) \, \delta_{ij}		\delta_{kl} =   \tfrac13 \big( 3\cdot 3 + 3 + 3 \big) = 5 \,;
		\\	
		\Y^{2m}_{ijkl} \delta^{}_{kl}  &=  \Y^{2m}_{(ij}\delta^{}_{kl)}  \delta^{}_{kl} =
		\tfrac16 \big(\Y^{2m}_{ij}\delta^{}_{kl} + \Y^{2m}_{il}\delta^{}_{jk} + \Y^{2m}_{ik}\delta^{}_{lj} +
		\Y^{2m}_{kl}\delta^{}_{ij} + \Y^{2m}_{jk}\delta^{}_{il} + \Y^{2m}_{lj}\delta^{}_{ik}) \delta^{}_{kl}		
		= \tfrac76 \Y^{2m}_{ij} \,;
		\\
		{\Y^{2m}_{ijkl}}^* \Y^{2m'}_{ijkl} &=  {\Y^{2m}_{(ij}}^*\delta^{}_{kl)}  \delta^{}_{kl}  \Y^{2m'}_{ij} 
		= \tfrac76 {\Y^{2m}_{ij}}^*  \Y^{2m'}_{ij} = \tfrac76 \cdot \tfrac32\, \delta_{mm'} 
		\equiv \zeta(4,2) \, \delta_{mm'} \,, \quad \zeta(4,2)=\tfrac74 \,.
	\end{aligned}
\end{equation}
In the last equation we used the normalization $\zeta(\ell,\ell)$ (\ref{eq:ThorneSTFprod}) for $\ell=2$. 
%are sufficient to extract the zeroth-order component ($\D^{00}$) from a second-order tensor ($\D_{ij}$), and zeroth- ($\S^{00}$) and second-order ($\S^{2m}$) components from a fourth-order ($\S_{ijkl}$) tensor, completing Eq.~(\ref{eq_Slm}).

For completeness, in the remainder of this Section we motivate a general normalization formula (\ref{eq:alpha_STF}) for all $L$ and $\ell$. 
%\noindent {\bf Proof of Eq.~(\ref{eq:alpha_STF}).} 
%To extract degree-$\ell$ spherical components from a symmetric tensor $\F_{i_1\hdots i_L}$ of order $L>\ell$, according to Eq.~(\ref{Flm-gen}) we first build an $\ell$-th order tensor  $\F_{i_1\hdots i_L} \, \delta_{i_{\ell+1}i_{\ell+2}} \hdots \delta_{i_{L-1}i_{L}}$ by contracting $L-\ell$ indices, and then project the result onto the desired STF basis element $\Y^{\ell m\, *}_{i_1\hdots i_\ell}$. (Note that the choice of indices to contract in the first step is irrelevant since $\F_{i_1\hdots i_L}$ is fully symmetric.) This procedure keeps only $\F^{(\ell)}$ components since $\ell'>\ell$ information is removed when $L-\ell$ indices are contracted. Additionally, $\ell'<\ell$ is removed when projecting onto $\Y^{\ell m^*}_{i_1\hdots i_\ell}$, thus we can write:
%\begin{equation}\label{eq:rankL_projection}
%	\F_{i_1\hdots i_L} \delta_{i_{\ell+1}i_{\ell+2}} \hdots \delta_{i_{L-1}i_{L}} \, {\Y^{\ell m}_{i_1\hdots i_\ell}}^* =  \zeta(L,\ell)\,\F^{\ell m},\quad \text{where} \quad 
%		\zeta(L,\ell)= \Y^{\ell m}_{i_1\hdots i_L} \Y^{\ell' m'\,*}_{i_1\hdots i_L} \,\,\text{for }\ell=\ell',\,m=m'.
%\end{equation}
%To derive the proportionality coefficient $\zeta$ we note that to contract $L-\ell$ indices one can successively apply the following identity:
Equation (\ref{eq:alpha_STF}) relies on the following identity to be derived below: 
\begin{equation}\label{eq:YYextended_contracted2}
 \Y^{\ell m}_{i_1\hdots i_L} \delta_{i_{L-1} i_L} = \frac{(L+\ell+1)(L-\ell)}{L(L-1)} \;\Y^{\ell m}_{i_1\hdots i_{L-2}} 
\quad \text{for } L \geq \ell+2\,; \qquad \text{and} \quad 0 \quad \text{for } L=\ell.
\end{equation}
Assuming Eq.~(\ref{eq:YYextended_contracted2}) holds, we apply it recursively
\begin{equation}\label{eq:YYextended_contracted_arbitrary}
	\Y^{\ell m}_{i_1\hdots i_L} \delta_{i_{\ell+1} i_{\ell+2}} \hdots  \delta_{i_{L-1} i_L} =\frac{(L+\ell+1)(L+\ell-1)\hdots (2\ell+3)(L-\ell)(L-\ell-2)\hdots 2}{L(L-1)(L-2)\hdots (\ell+1)}\,\Y^{\ell m}_{i_1\hdots i_\ell} \,.
	%=   \frac{(L+\ell+1)!!(L-\ell)!!}{L!} \, \Y^{\ell m}_{i_1\hdots i_\ell}  .
\end{equation}
Combining Eqs.~(\ref{eq:YYextended_contracted_arbitrary}) and (\ref{eq:ThorneSTFprod}) allows us to obtain Eq.~(\ref{eq:alpha_STF}): 
\begin{equation}\nonumber %\label{eq:alpha_STF}
	\zeta(L,\ell) 
	= {\Y^{\ell m}_{i_1\hdots i_L}}^* \Y^{\ell m}_{i_1\hdots i_\ell} \delta_{i_{\ell+1} i_{\ell+2}} \hdots  \delta_{i_{L-1} i_L}
	= \frac{(L+\ell+1)!!(L-\ell)!!}{L!\,(2\ell+1)} \,.
\end{equation}
%which we move to the other side of Eq.~(\ref{eq:rankL_projection}) to obtain Eq.~(\ref{Flm-gen}).
To derive Eq.~(\ref{eq:YYextended_contracted2}), consider as an example the case for $L = \ell+2$:
\begin{equation}\label{eq:Ylm_iL_delta_contraction}
	\begin{aligned}
		\Y^{\ell m}_{i_1\hdots i_L}   \delta_{i_{L-1} i_L} 
		&= \Y^{\ell m}_{(i_1\hdots i_\ell}   \delta_{i_{L-1} i_L)}  \delta_{i_{L-1} i_{L}} \\
		%&= \Y^{\ell m}_{(i_1\hdots i_\ell}   \delta_{i_{\ell-1} i_\ell)}  \delta_{i_{\ell+1} i_{\ell+2}} \\
		& =	\frac{\ell! \,2!}{(\ell+2)!} \Big(
		 \Y^{\ell m}_{i_1\hdots i_\ell} \delta_{i_{L-1} i_L}
		 +\Y^{\ell m}_{i_{L-1} i_2 \hdots i_\ell} \delta_{i_{1} i_L}
		 +\Y^{\ell m}_{i_1 i_{L-1} i_3 \hdots i_\ell} \delta_{i_{2} i_L} + \hdots 
		 +\Y^{\ell m}_{i_1 \hdots i_{\ell-1} i_{L-1}} \delta_{i_{\ell} i_L} \\
		 & +\Y^{\ell m}_{i_{L} i_2 \hdots i_\ell} \delta_{i_{L-1}i_{1} }
		+\Y^{\ell m}_{i_1 i_{L} i_3 \hdots i_\ell} \delta_{i_{L-1}i_{2}} + \hdots 
		+\Y^{\ell m}_{i_1 \hdots i_{\ell-1} i_{L}} \delta_{i_{L-1}i_{\ell}} \\
		 & + \Y^{\ell m}_{i_{L-1} i_L i_3\hdots i_\ell} \delta_{i_1 i_2} + \Y^{\ell m}_{i_{L-1} i_2 i_L i_4\hdots i_\ell} \delta_{i_1 i_3} + \hdots +
		 \Y^{\ell m}_{i_1\hdots i_{\ell-2}i_{L-1} i_{L}} \delta_{i_{\ell-1} i_\ell} 
		  \Big)  \,\delta_{i_{L-1} i_L}  \,  \\
		  & = 	\frac{\ell! \,2!}{(\ell+2)!} \Big( 3  \Y^{\ell m}_{i_1\hdots i_\ell} + 2\ell \Y^{\ell m}_{i_1\hdots i_\ell}+  \binom{\ell}{2} \times 0  \Big)  
		   = \frac{2(2\ell+3)}{(\ell+1)(\ell+2)}  \Y^{\ell m}_{i_1\hdots i_\ell} \,,
	\end{aligned}
\end{equation}
which can be generalized for arbitrary $L>\ell$ to derive Eq.~(\ref{eq:YYextended_contracted2}). The rationale behind the derivation in Eq.~(\ref{eq:Ylm_iL_delta_contraction}) is that one must expand the symmetrized $\Y^{\ell m}_{i_1 \hdots i_L}$ into a sum of all permutations of subindices where $i_{L-1},i_L$ are either part of $\Y^{\ell m}$ or part of a product of $(L-\ell)/2$ Kronecker deltas (in this case, a single one). Then, we have to contract indices $i_{L-1},i_L$ on each term and add the contributions. Contracting any two indices of $\Y^{\ell m}_{i_1 \hdots i_\ell}$ yields 0 due to STF property, while all combinations of contractions of the products of the Kronecker symbols can be computed 
using %Eq.~(\ref{eq:THORNEintN}) defined in the following section.
\begin{equation}
%		I_{i_1 \hdots i_L} I_{i_{\ell+1} \hdots i_L}  = \frac{L+1}{\ell+1} \,  I_{i_1 i_L-2} \,.	
		\delta_{(i_1 i_2} \hdots \delta_{i_{L-1} i_L)} \delta_{(i_{\ell+1} i_{\ell+2}} \hdots \delta_{i_{L-1} i_L)}   = \frac{L+1}{\ell+1} \, \delta_{(i_1 i_2} \hdots \delta_{i_{L-3} i_{L-2})} \,.
\end{equation}

%\newpage
\section{Irreducible decomposition  of spherical tensor products}\label{SM:CGcoeffsSTF}
\subsection{Addition of angular momenta, Wigner rotation matrices, Clebsch-Gordan coefficients and spherical harmonics}
\noindent
A $d$-dimensional unitary representation ${\cal D}$ of SO(3) maps each group element $\R\in\mathrm{SO(3)}$ onto a $d\times d$ unitary matrix ${\cal D}(\R)$ realizing the rotation $\R$. 
All irreducible representations ${\cal D^\ell}(\R)$ of SO(3) are labeled by non-negative integers $\ell = 0, 1, 2, \dots$ and have dimensions $2\ell+1$.\cite{Tinkham,HALL2015} 
From the representation theory standpoint, the addition of  angular momenta $\vec{\ell_1}$ and $\vec{\ell_2}$ corresponds to the tensor product of two irreducible representations ${\cal D}^{\ell_1}$ and ${\cal D}^{\ell_2}$ of SO(3) with dimensions $2\ell_1+1$ and $2\ell_2+1$. 
This tensor product of dimension $(2\ell_1+1)(2\ell_2+1)$ is reducible, and splits into a sum of irreducible representations with $L=|\ell_1-\ell_2|, \ \dots , \ \ell_1+\ell_2$, each one entering exactly once \cite{Tinkham,HALL2015}: 
\begin{equation} \label{D1xD2}
{\cal D}^{\ell_1} \otimes {\cal D}^{\ell_2} = \bigoplus_{L=-|\ell_1-\ell_2|}^{\ell_1+\ell_2} {\cal D}^{L} \,.
%{\cal D}^{\ell_1} \otimes {\cal D}^{\ell_2} = \oplus\sum_{L=-|\ell_1-\ell_2|}^{\ell_1+\ell_2} {\cal D}^{L} \,.
\end{equation}
The orthonormal basis elements 
\begin{equation}
\ket{LM}=\sum_{m_1+m_2=M} \ket{\ell_1 m_1 \ell_2 m_2}  \braket{\ell_1 m_1 \ell_2 m_2 | LM} , \quad M = -L\dots L 
\end{equation}
of each such irreducible representation labeled by $L$, are expanded in terms of the basis elements of $\ket{\ell_1 m_1 \ell_2 m_2}=\ket{\ell_1 m_1}\otimes\ket{\ell_2 m_2}$ of the tensor product, 
where $m_1 = -\ell_1 \dots \ell_1$, $m_2 = -\ell_2\dots \ell_2$, and  $\braket{\ell_1 m_1 \ell_2 m_2 | LM}$ are the Clebsch–Gordan coefficients. 
In this work, we use complex spherical harmonics and spherical tensors, with the Condon-Shortley phase convention \citep{CONDON1964,THORNE1980}, for which the Clebsch-Gordan coefficients are purely real. 
Taking the  matrix elements of  the tensor product (\ref{D1xD2}),  we obtain 
\begin{equation}\label{eq:DD_CG}
 \mathcal{D}_{m_1 n_1}^{l_1}(\R) \mathcal{D}_{m_2 n_2}^{l_2}(\R) = \sum_{L,M,N} \braket{\ell_1 m_1 \ell_2 m_2 | LM} \braket{\ell_1 n_1 \ell_2 n_2 | LN}  \mathcal{D}_{MN}^{L}(\R) \,.
\end{equation}
By the Peter-Weyl theorem, elements of all unitary {\it Wigner matrices} ${\cal D}^\ell_{mm'}(\R)$  are dense in SO(3) and form an orthogonal basis, 
\begin{equation}\label{eq:wigner_ortho}
	\int_{\mathrm{SO}(3)}\! \d \R\;
	\mathcal{D}^{\ell}_{m n}(\R)\,
	\mathcal{D}^{\ell' *}_{m' n'}(\R)
	=
	\frac{1}{2\ell+1}\,
	\delta_{\ell\ell'}\delta_{mm'}\delta_{nn'}\,,
	\quad
	\mathcal{D}^{\ell}_{m n}(\R^{-1}) = \mathcal{D}^{\ell *}_{nm}(\R) \,, 
\end{equation}
under  the normalized Haar measure $ \int_{\mathrm{SO}(3)}\! \d \R =1$ on the group manifold. 
Following Eq.~(\ref{eq:DD_CG}), the integral of the triple product 
\begin{equation}\label{eq:DDD_int}
\int_{\mathrm{SO}(3)} \d\R\, \mathcal{D}^{\ell_1}_{m_1 n_1}(\R)  \mathcal{D}^{\ell_2}_{m_2 n_2}(\R)  \mathcal{D}^{L\,*}_{MN}(\R)  =  \frac{1}{2L+1}
\braket{\ell_1 m_1 \ell_2 m_2 | LM} \braket{\ell_1 n_1 \ell_2 n_2 | LN} \,.
\end{equation}
Spherical harmonics $Y^{\ell m}(\n)$ realize the $(2\ell+1)$-dimensional representations of SO(3), and transform {\it covariantly} (as basis elements) via the corresponding Wigner matrices $\mathcal{D}^{\ell}_{m' m}(\R)$: 
\begin{equation} \label{Y=DY}
\R Y^{\ell m} (\n) = Y^{\ell m}(\R^{-1}\n) = \sum_{m'=-\ell}^\ell Y^{\ell m'}(\n) \, \mathcal{D}^{\ell}_{m' m}(\R) \,.
\end{equation}
Importantly, $Y^{\ell m}(\n)$ can also be viewed as a subset of Wigner matrix elements defined on a 2-sphere $\mathbb{S}^2$ formed by the tips of all unit vectors $\n$, and viewed as a coset space $\mathbb{S}^2=\mathrm{SO(3)/SO(2)}$ of the original 3-dimensional SO(3) manifold $\mathbb{S}^3/\mathbb{Z}_2$ over the SO(2) rotations around $\z$. For that, consider rotating the North pole $\z$ of $\mathbb{S}^2$ by an SO(3) matrix $\R_{\n}$ that turns $\z$ into 
$\n \equiv \R_{\n} \z$.  
There is in fact a 1-dimensional family of such matrices $\R_{\n}$ that yield the same $\n \in \mathbb{S}^2$, with the same polar and azimuthal Euler angles, and an arbitrary third Euler angle of the SO(2) rotation leaving $\z$ unchanged (in the standard $ZYZ$ convention \cite{Tinkham}). 
Using Eq.~(\ref{Y=DY}) and $Y^{\ell m}(\z) = \delta^{m0}$ in the Racah normalization (\ref{STF-SH-orthogonality}),  $Y^{\ell m}(\n)$ are the ${\cal D}$-matrix elements independent of the third Euler angle (and up to complex conjugation): 
\begin{equation} \label{SH=Wigner}
Y^{\ell m}(\n) = Y^{\ell m}(\R_{\n} \z) = \sum_{m'=-\ell}^\ell Y^{\ell m'}(\z)\, {\cal D}_{m'm}^{\ell}(\R_{\n}^{-1}) 
= {\cal D}_{0m}^{\ell}(\R_{\n}^{-1}) 
\equiv {\cal D}_{m0}^{\ell *}(\R_{\n}) \,. 
\end{equation}
Using Eq.~(\ref{SH=Wigner}) in the product (\ref{eq:DD_CG}) and taking the $\bra{\ell_1 m_1 \ell_2 m_2} \dots \ket{\ell_1 0 \ell_2 0}$ matrix element yields 
\begin{equation} \label{YY=Y}
   Y^{\ell_1 m_1 }(\n) Y^{\ell_2 m_2 }(\n)   = \sum_{L,M}  \braket{\ell_1 0 \ell_2 0 | L 0} \braket{\ell_1 m_1 \ell_2 m_2 | L M} {Y^{LM}}(\n) \,, 
\end{equation}
equivalent to the well-known spherical harmonics identity in Racah normalization (assuming $\int_{\mathbb{S}^2} \d\n = \int\! \d\Omega_{\n}/4\pi \equiv 1$)
\begin{equation}\label{eq:tripleSHint}
    \int_{\mathbb{S}^2} \d\n\, Y^{\ell_1 m_1 }(\n) Y^{\ell_2 m_2 }(\n) {Y^{LM}}^*(\n)  =  \frac{1}{2L+1}
  \braket{\ell_1 0 \ell_2 0 | L 0} \braket{\ell_1 m_1 \ell_2 m_2 | L M} .
\end{equation}

\subsection{Derivation of Eq.~(\ref{eq:YYY_CGcoeffs})}
\noindent
The derivation will be based on the analog of Eq.~(\ref{eq:tripleSHint}) for the basis STF tensors, with the orthogonality over the unit sphere substituted by taking full trace. For that, we make use of the fully symmetric product of Kronecker symbols $I_{i_1\hdots i_\ell}$ 
and Eq.~(2.3) from Thorne \cite{THORNE1980}:
\begin{equation}\label{eq:THORNEintN}
	\begin{aligned}
		I_{i_{1} \hdots i_{\ell}} \equiv   \delta_{(i_{1}i_{2}}\hdots \delta_{i_{\ell-1}i_{\ell})} 
		=	(\ell+1) \int_{\mathbb{S}^2} \d\n\, n_{i_1} \hdots n_{i_\ell}  \,, 
		\quad \d\n \equiv \frac{\d\Omega_{\n}}{4\pi} \,.
	\end{aligned}
\end{equation}
A useful relation then comes from combining the orthogonality (\ref{STF-SH-orthogonality}) of spherical harmonics, and Eq.~(\ref{eq:THORNEintN}): 
%\dn{changed $\ell\to\ell'$ in $j_\ell$}
\begin{equation}
%\Y_{i_1 \hdots i_{\ell}}^{\ell m} \, \Y_{j_1 \hdots j_{\ell}}^{\ell' m' \,*} \, I_{i_1\hdots i_{\ell} j_1 \hdots j_{\ell}} 
\Y_{i_1 \hdots i_{\ell}}^{\ell m \,*} \, \Y_{j_1 \hdots j_{\ell'}}^{\ell' m'} \, I_{i_1\hdots i_{\ell} j_1 \hdots j_{\ell'}} 
= \delta_{\ell \ell'} \delta_{mm'} \,,
\end{equation}
which can be extended to a more general scenario applying Eq.~(\ref{STF=SH-L}):
\begin{equation}\label{eq:STForthogonalityEXT}
	\Y_{i_1 \hdots i_{L}}^{\ell m \,*} \, \Y_{j_1 \hdots j_{L'}}^{\ell' m' } \, I_{i_1\hdots i_{L} j_1 \hdots j_{L'}} \!=\! \frac{L+L'+1}{2\ell+1}\, \delta_{\ell \ell'} \delta_{mm'}
	\quad \text{for the case where}\quad L\geq \ell,\,\, L'\geq\ell'\,.
\end{equation}

Likewise, we now can rewrite Eq.~(\ref{eq:tripleSHint}) by replacing the integrals over $n_{i_1} n_{i_2}\hdots n_{i_k}$ with Eq.~(\ref{eq:THORNEintN}):
\begin{equation}\label{Eq:CG_w_STF_general_ell12_wI}
	\Y_{(i_1 \hdots i_{\ell_1}}^{\ell_1 m_1}\Y_{j_1\hdots j_{\ell_2})}^{\ell_2 m_2 }\Y^{L M\, *}_{k_1 \hdots k_{\ell_1+\ell_2}}\, I_{i_1\hdots i_{\ell_1}j_1 \hdots j_{\ell_2}k_1 \hdots k_{\ell_1+\ell_2}} 
	=   \frac{2\ell_1+2\ell_2  +1}{2L+1}   \braket{\ell_1 0 \ell_2 0 | L 0} \braket{\ell_1 m_1 \ell_2 m_2 | L M} .
\end{equation}
The term $\Y_{(i_1 \hdots i_{\ell_1}}^{\ell_1 m_1}\Y_{j_1\hdots j_{\ell_2})}^{\ell_2 m_2 }$ is a symmetric tensor of order $\ell_1+\ell_2$. Thus, we can expand it as a linear combination of basis tensors $\Y^{\ell' m'}_{i_1\hdots i_{\ell_1} j_1 \hdots j_{\ell_2}}$, Eq.~(\ref{Eq:Ylm_deltas}), with degrees $\ell'=0,\,2,\,\hdots ,\ell_1+\ell_2$. 
When computing the tensor products on the left side of Eq.~(\ref{Eq:CG_w_STF_general_ell12_wI}), only the basis term $\ell' = L$ of the expansion of $\Y_{(i_1 \hdots i_{\ell_1}}^{\ell_1 m_1}\Y_{j_1\hdots j_{\ell_2})}^{\ell_2 m_2 }$ returns a nonzero value. If we call $\xi$  the coefficient accompanying the $\ell' = L$ term of the expansion, then Eq.~(\ref{Eq:CG_w_STF_general_ell12_wI}) can be written as
%we can rewrite Eq.~(\ref{Eq:CG_w_STF_general_ell12_wI}) as
%multiplying this tensor with $\Y^{L M\, *}_{k_1 \hdots k_{\ell_1+\ell_2}}$ and tracing it with $I_{i_1\hdots i_{\ell_1}j_1 \hdots j_{\ell_2}k_1 \hdots k_{\ell_1+\ell_2}} $, only the factor coupled with the term  $\ell' = L$ survives. If we call $\beta$  the term accompanying $\ell' = L$, we can rewrite Eq.~(\ref{Eq:CG_w_STF_general_ell12_wI}) as
\begin{equation}\label{Eq:betaTerm}
	\Y_{(i_1 \hdots i_{\ell_1}}^{\ell_1 m_1}\Y_{j_1\hdots j_{\ell_2})}^{\ell_2 m_2 }\Y^{L M\, *}_{k_1 \hdots k_{\ell_1+\ell_2}}\, I_{i_1\hdots i_{\ell_1}j_1 \hdots j_{\ell_2}k_1 \hdots k_{\ell_1+\ell_2}} \!=\!  	\xi \,\Y^{LM}_{i_1\hdots i_{\ell_1}j_1 \hdots j_{\ell_2}} \Y^{L M\, *}_{k_1 \hdots k_{\ell_1+\ell_2}}\, I_{i_1\hdots i_{\ell_1}j_1 \hdots j_{\ell_2}k_1 \hdots k_{\ell_1+\ell_2}} \, ,
%	\!=\!	\beta \!=\!   \frac{2\ell_1+2\ell_2  +1}{2L+1}  \left\langle \ell_1 \,0 \,\ell_2\, 0 \mid L\, 0\right\rangle \la \ell_1 \,m_1 \,\ell_2\, m_2 |L\, M \ra,
\end{equation}
%\dn{$\beta$ was used as Euler angle}\sc{now $\beta \rightarrow \xi$ and $\alpha \rightarrow \zeta$}
which combined with Eq.~(\ref{eq:STForthogonalityEXT}) provides the  value 
$\xi = \braket{\ell_1 0 \ell_2 0 | L 0} \braket{\ell_1 m_1 \ell_2 m_2 | L M}$.
Using Eq.~(\ref{eq:alpha_STF}),  we can write the product
\begin{equation}\label{Eq:STF_self_contraction}
	\Y_{(i_1 \hdots i_{\ell_1}}^{\ell_1 m_1}\Y_{j_1\hdots j_{\ell_2})}^{\ell_2 m_2 }\Y^{L M\, *}_{i_1\hdots i_{\ell_1}j_1 \hdots j_{\ell_2}} 
	= \xi \cdot \zeta(\ell_1 +\ell_2,L)
	=  \zeta(\ell_1+\ell_2,L) 
	 \braket{\ell_1 0 \ell_2 0 | L 0} \braket{\ell_1 m_1 \ell_2 m_2 | L M} , 
	\quad \ell_1+\ell_2 \geq L \,, 
\end{equation}
since only the basis term $\ell' = L$ of the expansion of $\Y_{(i_1 \hdots i_{\ell_1}}^{\ell_1 m_1}\Y_{j_1\hdots j_{\ell_2})}^{\ell_2 m_2 }$ returns a nonzero value, and $\zeta(\ell_1+\ell_2,L)$ is the projection of the basis tensor onto itself,  Eq.~(\ref{eq:alpha_STF}). 
Substituting $\ell_1=\ell_2=2$ in Eq.~(\ref{Eq:STF_self_contraction}) yields Eq.~(\ref{eq:YYY_CGcoeffs}). 

\subsection{Relation between $\Q$ and $\T$ tensors and compartmental spherical tensor covariances}
\noindent
Clebsch–Gordan coefficients are sparse since only those satisfying $m_1 + m_2 = M$ are nonzero (19 out of 125 for $\ell_1 = \ell_2 = L =2$ and 25 out of 225 for $\ell_1 = \ell_2 = 2,\,L =4$). 
We can substitute these in Eq.~(\ref{TLM}), or equivalently Eq.~(\ref{eq:ellSYSTEM_QT}), and combine it with Eq.~(\ref{QLM}) to get the system:
\begin{equation}\label{eq:TQ_DD_system}
\begin{aligned}
	\textstyle\Q^{00} & =    \lla (\Dc^{00})^2\rra \,, \\
	\textstyle \Q^{2m} & =  2\,  \lla \Dc^{00} \Dc^{2m}\rra  \quad \text{for} \quad m=-2,-1,\hdots,2\,;\\
	\textstyle \T^{00} &=  \tfrac{1}{\sqrt5}\,  \Big ( 2 \tfrac1{\sqrt5}\lla \Dc^{2-2}\Dc^{22}\rra - 2 \tfrac1{\sqrt5} \lla \Dc^{2-1} \Dc^{21}\rra +  \tfrac1{\sqrt5}\lla({\Dc^{20}})^2\rra  \Big ) \,, \\%  \tfrac25 \lla \Dc^{2-2}\Dc^{22}\rra - \tfrac25  \lla \Dc^{2-1} \Dc^{21}\rra + \tfrac15 \lla ({\Dc^{20}})^2\rra  \,, \\
	\textstyle  \T^{2-2}  &=  -\sqrt{\tfrac27}\,  \Big(2 \sqrt{\tfrac27} \lla \Dc^{2-2} \Dc^{20}\rra - \sqrt{\tfrac37} \lla ({\Dc^{2-1}})^2\rra \Big) \,,\\
	\textstyle \T^{2-1}  &= -\sqrt{\tfrac27}\,   \Big( 2 \sqrt{\tfrac37} \lla \Dc^{2-2} \Dc^{21}\rra - 2 \sqrt{\tfrac1{14}} \lla \Dc^{2-1} \Dc^{20}\rra \Big) \,,\\
	\textstyle  \T^{20}  &= -\sqrt{\tfrac27}\,   \Big(2 \sqrt{\tfrac27} \lla \Dc^{2-2} \Dc^{22}\rra + 2 \sqrt{\tfrac1{14}} \lla \Dc^{2-1} \Dc^{21}\rra - \sqrt{\tfrac27} \lla ({\Dc^{20}})^2\rra  \Big) \,,\\
	\textstyle  \T^{21}  &= -\sqrt{\tfrac27}\,   \Big( 2 \sqrt{\tfrac37} \lla \Dc^{2-1} \Dc^{22}\rra - 2 \sqrt{\tfrac1{14}} \lla \Dc^{20} \Dc^{21}\rra \Big) \,,\\
	\textstyle  \T^{22}  &=  -\sqrt{\tfrac27}\,  \Big ( 2 \sqrt{\tfrac27} \lla \Dc^{20} \Dc^{22}\rra - \sqrt{\tfrac37}\lla ({\Dc^{21}})^2\rra \Big) \,,\\
	\textstyle \T^{4-4} & = \sqrt{\tfrac{18}{35}}\,  \lla ({\Dc^{2-2}})^2\rra  \,,\\
	\textstyle \T^{4-3} & =\sqrt{\tfrac{18}{35}}\, 2\tfrac1{\sqrt2} \, \lla \Dc^{2-2} \Dc^{2-1}\rra  \,,\\
	\textstyle \T^{4-2} & = \sqrt{\tfrac{18}{35}}\,  \Big(2 \sqrt{\tfrac3{14}} \lla \Dc^{2-2} \Dc^{20}\rra + \sqrt{\tfrac47} \lla ({\Dc^{2-1}})^2\rra \Big) \,,\\
	\textstyle \T^{4-1} & = \sqrt{\tfrac{18}{35}}\,   \Big( 2\sqrt{\tfrac1{14}} \lla \Dc^{2-2} \Dc^{21}\rra + 2 \sqrt{\tfrac37}\lla \Dc^{2-1} \Dc^{20}\rra \Big) \,,\\
	\textstyle \T^{40} &  = \sqrt{\tfrac{18}{35}}\, \Big(2 \sqrt{\tfrac1{70}} \lla \Dc^{2-2} \Dc^{22}\rra + 2 \sqrt{\tfrac8{35}} \lla \Dc^{2-1} \Dc^{21}\rra + \sqrt{\tfrac{18}{35}} \lla ({\Dc^{20}})^2\rra  \Big) \,,\\
	\textstyle \T^{41} &  =  \sqrt{\tfrac{18}{35}}\, \Big( 2 \sqrt{\tfrac1{14}} \lla \Dc^{2-1} \Dc^{22}\rra + 2 \sqrt{\tfrac37} \lla \Dc^{20} \Dc^{21}\rra \Big) \,,\\
	\textstyle \T^{42} & =  \sqrt{\tfrac{18}{35}}\,  \Big ( 2 \sqrt{\tfrac3{14}} \lla \Dc^{20} \Dc^{22}\rra + \sqrt{\tfrac47}\lla ({\Dc^{21}})^2\rra \Big) \,,\\
	\textstyle \T^{43} & = \sqrt{\tfrac{18}{35}}\,  2\tfrac1{\sqrt2} \,  \lla \Dc^{21} \Dc^{22}\rra  \,,\\
	\textstyle \T^{44} & = \sqrt{\tfrac{18}{35}}\, \lla ({\Dc^{22}})^2\rra  \,,\\
\end{aligned}
\end{equation}
where Clebsch-Gordan coefficients are written explicitly for the reader (see page 737 in Ref.~\cite{PDG2022} for a complete table). 
%which together with Eq.~(\ref{QLM}) 
This shows how all covariances contribute to $\T$ and $\Q$ irreducible components. Solving for all the covariances gives
\begin{equation}\label{eq:DDsolved}
\begin{aligned}
	\lla ({\Dc^{00} })^2 \rra & =  \Q^{00} \, ,
&&	\lla \Dc^{00}\Dc^{2m} \rra  = \tfrac12 \, \Q^{2m}  \quad \text{for} \quad m=-2,\hdots,2\,; \\
	\lla ({\Dc^{2-2}})^2    \rra & = \tfrac{\sqrt{70}}{6} \,\T^{4-4} \, ,
&&	\lla \Dc^{2-2}\Dc^{2-1} \rra =\tfrac{\sqrt{35}}{6} \,\T^{4-3} \, ,\\ 
	\lla \Dc^{2-2}\Dc^{20}  \rra & = \tfrac{\sqrt{15}}{6}  \, \T^{4-2}  - \T^{2-2}\,,
&&	\lla \Dc^{2-2}\Dc^{21}  \rra  =  \tfrac{\sqrt{5}}{6} \,\T^{4-1} - \tfrac{\sqrt{6}}{2}\, \T^{2-1},\\ 
	\lla \Dc^{2-2}\Dc^{22}  \rra & =  \tfrac16 \,\T^{40}- \T^{20} + \T^{00}   \, ,
&&	\lla ({\Dc^{2-1}})^2    \rra  =  \tfrac{\sqrt{10}}{3}  \,\T^{4-2}  +  \tfrac{\sqrt{150}}{10} \, \T^{2-2}\, ,\\ 
	\lla \Dc^{2-1}\Dc^{20}  \rra & =  \tfrac{\sqrt{30}}{6}\,  \T^{4-1} + \tfrac12\, \T^{2-1}\, ,
&&	\lla \Dc^{2-1}\Dc^{21}  \rra  =  \tfrac23 \,\T^{40}- \tfrac12 \,\T^{20} - \T^{00}   \, ,\\
	\lla \Dc^{2-1}\Dc^{22}  \rra &= \tfrac{\sqrt{5}}{6} \, \T^{41} - \tfrac{\sqrt{6}}{2} \,   \T^{21}  \, ,
&&	\lla ({\Dc^{20}})^2     \rra  =  \T^{40} +\T^{20} + \T^{00}   \, ,\\
	\lla \Dc^{20}\Dc^{21}   \rra & = \tfrac{\sqrt{30}}{6} \, \T^{41} + \tfrac12\, \T^{21}\, ,
&&	\lla \Dc^{20}\Dc^{22}   \rra  = \tfrac{\sqrt{15}}{6}\, \,\T^{42} - \T^{22} \, ,\\
	\lla ({\Dc^{21}})^2     \rra & = \tfrac{\sqrt{10}}{3}\,  \,\T^{42}   +  \tfrac{\sqrt{150}}{10}\,  \T^{22} \, ,
&&	\lla \Dc^{21}\Dc^{22}   \rra  =\tfrac{\sqrt{35}}{6} \,\T^{43} \, ,\\ 
	\lla ({\Dc^{22}})^2     \rra &=\tfrac{\sqrt{70}}{6} \,\T^{44}\, .
\end{aligned}
\end{equation}
The sparsity of the QT decomposition is seen in all covariances depending on few $\T$ and $\Q$ elements. The simplicity of Eq.~(\ref{eq:DDsolved}) relies on the usage of the canonical, complex-valued spherical tensor basis. As the $\C$ tensor is real-valued, we can go back and forth from complex to real spherical tensor representations. Specifically,  to obtain real spherical harmonics covariances $\lla \Dc^{\ell m}\Dc^{\ell' m'} \rra$, one can take $\T^{\ell m}$ and $\Q^{\ell m}$ in the complex spherical tensor basis,  compute $\lla \Dc^{\ell m}\Dc^{\ell' m'} \rra$ using Eqs.~(\ref{eq:DDsolved}), and then convert the latter to the real spherical tensor basis, as spelled out in the following Supplementary Section~\ref{SM:SH_real_complex}.

\section{Relations between real and complex spherical tensors/harmonics}
\label{SM:SH_real_complex}
\noindent
Since $\C$ is real-valued one may want to work with the real spherical tensor basis and, thus, write the solution (\ref{eq:DDsolved}) for it. This is cumbersome because in the real basis Eq.~(\ref{eq:YYY_CGcoeffs}) includes multiple combinations of Clebsch-Gordan coefficients $\la 2 \pm\! m_1 \,2\pm \! m_2 |2\pm \!M \ra$. An alternative is to compute complex-valued $\lla \Dc^{\ell m}\Dc^{\ell' m'} \rra$ using Eq.~(\ref{eq:DDsolved}) and then transform these to real-valued ones. This can be done using the relation between complex and real spherical basis tensors:
\begin{equation}
        \Y_\mathbb{C}^{\ell m}= \begin{cases}
        \frac{1}{\sqrt{2}}\left(\Y^{\ell -m}_\mathbb{R} - i \,\Y^{\ell m}_\mathbb{R} \right)
         & \text { if } m<0 \\
        \Y^{\ell 0}_\mathbb{R} & \text { if } m=0 \\
        \frac{(-1)^m}{\sqrt{2}}\left(\Y^{\ell m}_\mathbb{R} +  i \,\Y^{\ell -m}_\mathbb{R} \right) & \text { if } m>0 
        \end{cases},
        \quad
        \Y^{\ell m}_\mathbb{R}= \begin{cases}
        	(-1)^m\sqrt{2}\, \mathrm{Im} (\Y_\mathbb{C}^{\ell -m}) & \text { if } m<0 \\
        	\Y_\mathbb{C}^{\ell 0} & \text { if } m=0 \\
        	(-1)^m\sqrt{2} \, \mathrm{Re} (\Y_\mathbb{C}^{\ell m}) & \text { if } m>0 
        \end{cases},
\end{equation}
%and substituting $\Y_\mathbb{C}^{\ell m}$ into Eq.~(\ref{eq:YYY_CGcoeffs}) which will involve multiple combinations of $\la 2 \pm\! m_1 \,2\pm \! m_2 |2\pm \!M \ra$.  Here
where $\Y_\mathbb{C}^{\ell m}$ and $\Y_\mathbb{R}^{\ell m}$ are the complex and real STF basis tensors (identical relations apply to the spherical harmonics basis). Note that, unlike the complex  spherical tensors, the real  spherical tensor basis (following conventions) does {\it not} include the Condon-Shortley phase. 
%Although compartmental diffusion tensors $\Dc_{ij}$ are real-valued, it is simpler to first decompose $\C$ into complex-valued $\lla \Dc^{\ell m} \Dc^{\ell' m'} \rra$ and then map these to the corresponding real-basis coefficients for computing rotational invariants. To solve this we can use the following relations:
Since compartmental diffusion tensors $\Dc_{ij}$ are real-valued, we can map complex-valued $\Dc^{\ell m}$ into real-valued ones using the following relations:
\begin{equation}
\Dc^{\ell m}_\mathbb{R}= \begin{cases}
(-1)^{m+1}\sqrt{2}\, \mathrm{Im} (\Dc_\mathbb{C}^{\ell -m}) & \text { if } m<0 \\
\Dc_\mathbb{C}^{\ell 0} & \text { if } m=0 \\
(-1)^m\sqrt{2} \, \mathrm{Re} (\Dc_\mathbb{C}^{\ell m}) & \text { if } m>0 
\end{cases},
\quad \text{and}\quad 
{\Dc^{\ell m}_\mathbb{C}}^*= (-1)^m {\Dc^{\ell -m}_\mathbb{C}},
\end{equation}
where $\Dc^{\ell m}_\mathbb{R}$ and $\Dc^{\ell m}_\mathbb{C}$ are  spherical tensor coefficients using either real or complex basis. Thus, the mapping $ \lla \Dc_\mathbb{C}^{\ell m}\Dc_\mathbb{C}^{\ell' m'} \rra \rightarrow  \lla \Dc_\mathbb{R}^{\ell m}\Dc_\mathbb{R}^{\ell' m'} \rra$ becomes:
\begin{equation}
\lla \Dc_\mathbb{R}^{\ell m}\Dc_\mathbb{R}^{\ell' m'} \rra =
\begin{cases}
 \lla \Dc_\mathbb{C}^{00}\Dc_\mathbb{C}^{00} \rra 
& \text { if } \ell = 0,\, m= 0,\, \ell'=0,\, m'=0 \,,\\
 \sqrt{2}\, \mathrm{Im} \left( \lla \Dc_\mathbb{C}^{00} \Dc_\mathbb{C}^{2m'} \rra \right) 
& \text { if } \ell = 0,\, m= 0,\, \ell'=2,\, m'<0 \,,\\
 \lla \Dc_\mathbb{C}^{00}\Dc_\mathbb{C}^{20} \rra 
& \text { if } \ell = 0,\, m= 0,\, \ell'=2,\, m'=0\,,\\
 (-1)^{m'} \sqrt{2}\, \mathrm{Re} \left( \lla \Dc_\mathbb{C}^{00} \Dc_\mathbb{C}^{2m'} \rra \right) 
& \text { if } \ell = 0,\, m= 0,\, \ell'=2,\, m'>0\,,\\
- \mathrm{Re} \left( \lla   \Dc_\mathbb{C}^{2m}  \Dc_\mathbb{C}^{2m'}  \rra \right ) + (-1)^{m'}  \mathrm{Re} \left( \lla  \Dc_\mathbb{C}^{2m}  \Dc_\mathbb{C}^{2-m'}  \rra \right )   
& \text { if } \ell = 2,\, m< 0,\, \ell'=2,\, m'<0 \,,\\
 \sqrt{2} \, \mathrm{Im} \left( \lla D^\mathbb{C}_{2m} D^\mathbb{C}_{20} \rra \right) 
& \text { if } \ell = 2,\, m< 0,\, \ell'=2,\, m'=0 \,,\\
(-1)^{m'} \mathrm{Im} \left( \lla  \Dc_\mathbb{C}^{2m} \Dc_\mathbb{C}^{2m'}  \rra \right ) +   \mathrm{Im} \left( \lla  \Dc_\mathbb{C}^{2m} \Dc_\mathbb{C}^{2-m'}  \rra \right ) 
& \text { if } \ell = 2,\, m< 0,\, \ell'=2,\, m'>0\,,\\
 \lla \Dc_\mathbb{C}^{20}  \Dc_\mathbb{C}^{20} \rra 
& \text { if } \ell = 2,\, m= 0,\, \ell'=2,\, m'=0 \,,\\
 (-1)^{m'} \sqrt{2} \, \mathrm{Re} \left( \lla  \Dc_\mathbb{C}^{20} \Dc_\mathbb{C}^{2m'} \rra \right ) 
& \text { if } \ell = 2,\, m= 0,\, \ell'=2,\, m'>0 \,,\\
 (-1)^{m+m'} \mathrm{Re} \left( \lla  \Dc_\mathbb{C}^{2m} \Dc_\mathbb{C}^{2m'}  \rra \right ) +  (-1)^{m}\, \mathrm{Re} \left( \lla  \Dc_\mathbb{C}^{2m} \Dc_\mathbb{C}^{2-m'}  \rra \right )  
& \text { if } \ell = 2,\, m> 0,\, \ell'=2,\, m'>0 \,.\\
\end{cases}
\end{equation}

\section{Second-order 6D representation for fully symmetric fourth-order 3D tensors}\label{SM:C6x6}
\noindent
\begin{equation} \label{S4_6x6}
\Sf_{6\times6}  = \left(
\scalemath{0.78}{
\begin{array}{cccccc}
	\frac{\sqrt{35} }{8}\S^{44}-\frac{\sqrt{5} }{4}\S^{42}+\frac{3 }{8} \S^{40}& -\frac{\sqrt{35} }{8}\S^{44}+\frac{1}{8} \S^{40}& \frac{\sqrt{5} }{4}\S^{42}-\frac{1}{2}\S^{40} &	\frac{\sqrt{70}}{8} \S^{4-4}-\frac{\sqrt{10} }{8} \S^{4-2}& \frac{\sqrt{35} }{8}\S^{43}-\frac{3 \sqrt{5} }{8}\S^{41} & \frac{\sqrt{35}}{8} \S^{4-3}-\frac{\sqrt{5}}{8}  \S^{4-1} \vspace{1mm} \\ \vspace{1mm} 
	-\frac{\sqrt{35} }{8}\S^{44}+\frac{1}{8}\S^{40} & \frac{\sqrt{35} }{8}\S^{44}+\frac{\sqrt{5} }{4}\S^{42}+\frac{3 }{8}\S^{40} & -\frac{\sqrt{5} }{4}\S^{42}-\frac{1}{2}\S^{40} &	-\frac{\sqrt{70} }{8}\S^{4-4}-\frac{\sqrt{10} }{8}\S^{4-2} & -\frac{\sqrt{35} }{8}\S^{43}-\frac{\sqrt{5} }{8}\S^{41} & -\frac{\sqrt{35} }{8}\S^{4-3}-\frac{3 \sqrt{5} }{8} \S^{4-1}\\ \vspace{1mm}
	\frac{\sqrt{5} }{4}\S^{42}-\frac{1}{2}\S^{40} & -\frac{\sqrt{5} }{4}\S^{42}-\frac{1}{2}\S^{40} & \S^{40} 	& \frac{\sqrt{10} }{4}\S^{4-2} & \frac{\sqrt{5} }{2}\S^{41} & \frac{\sqrt{5} }{2}\S^{4-1} \\ \vspace{1mm}
	\frac{\sqrt{70} }{8}\S^{4-4}-\frac{\sqrt{10} }{8}\S^{4-2} & -\frac{\sqrt{70} }{8}\S^{4-4}-\frac{\sqrt{10} }{8}\S^{4-2} & \frac{\sqrt{10} }{4}\S^{4-2} &	-\frac{\sqrt{35} }{4}\S^{44}+\frac{1}{4}\S^{40} & \frac{\sqrt{70} }{8}\S^{4-3}-\frac{\sqrt{10} }{8}\S^{4-1} & -\frac{\sqrt{70} }{8}\S^{43}-\frac{\sqrt{10} }{8}\S^{41}\\ \vspace{1mm}
	\frac{\sqrt{35} }{8}\S^{43}-\frac{3 \sqrt{5} }{8}\S^{41} & -\frac{\sqrt{35} }{8}\S^{43}-\frac{\sqrt{5} }{8}\S^{41} & \frac{\sqrt{5} }{2}\S^{41} &	\frac{\sqrt{70} }{8}\S^{4-3}-\frac{\sqrt{10} }{8}\S^{4-1} & \frac{\sqrt{5} }{2}\S^{62}-\S^{40} & \frac{\sqrt{5} }{2}\S^{4-2} \\ \vspace{1mm}
	\frac{\sqrt{35} }{8}\S^{4-3}-\frac{\sqrt{5} }{8}\S^{4-1} & -\frac{\sqrt{35} }{8}\S^{4-3}-\frac{3 \sqrt{5} }{8}\S^{4-1} & \frac{\sqrt{5} }{2}\S^{4-1} &
-\frac{\sqrt{70} }{8}\S^{43}-\frac{\sqrt{10} }{8}\S^{41}& \frac{\sqrt{5} }{2}\S^{4-2} & -\frac{\sqrt{5} }{2}\S^{42}-\S^{40}
\end{array}
}
\right).
\end{equation}
Direct inspection of Eq.~(\ref{S4_6x6}) makes it evident that:
(i) $\tr \Sf_{6\times6} = 0$; 
(ii) one of the eigenvalues is zero ($\lambda_{a_0} = 0$); and
(iii) its associated eigenvector is $\hat{\mathbf{v}}^{(a_0)}=\tfrac{1}{\sqrt{3}}\,(1,\,1,\,1,\,0,\,0,\,0)^t$.

\newpage
\section{Minimal spherical designs for $L=2$ and $L=4$ with antipodal symmetry}\label{SM:SphDesigns}
\noindent
Spherical designs are sets of unit vectors that fulfill Eq.~(\ref{eq:SphDesign}). One can check that for any degree-2 function $f^{(2)}(\n)$ with antipodal symmetry (i.e., involving $\ell=0$ and $\ell=2$ spherical harmonics), the 2-design 
\[
\{\n^\alpha\}  = \{\e_1,\,\e_2,\,\e_3\}= \{(1,0,0),(0,1,0),(0,0,1)\}
\]
yields 
%\dn{formal problems: see new transition in the second $=$, which means the next step is incorrect, since $Y^{00}_{ij}$ does not exist. $ij$ only for $\ell=2$.}
\begin{equation}\begin{aligned} \nonumber
    \tfrac13 \sum_{\alpha=1}^3 f^{(2)}(\n^{\alpha}) &= \tfrac13  \sum_{\ell=0,2}\, \sum_{m=-\ell}^\ell \, f^{\ell m} \, \Y^{\ell m}_{ij} \,(n^{1}_i n^{1}_j +n^{2}_i n^{2}_j +n^{3}_i n^{3}_j  )
%    = \tfrac13 \,  \sum_{\ell,m} f^{\ell m} \, \sum_{ij}\Y^{\ell m}_{ij} \, (\delta^1_i\delta^1_j + \delta^2_i\delta^2_j + \delta^3_i\delta^3_j )
%    = \tfrac13 \,  \sum_{\ell,m} f^{\ell m} \, \sum_{ij}\Y^{\ell m}_{ij} \, (\delta_{11} + \delta_{22} + \delta_{33} )
    = \tfrac13 \,  \sum_{\ell=0,2}\, \sum_{m=-\ell}^\ell f^{\ell m} \,\Y^{\ell m}_{ij} \, (\e_1^{\otimes 2} + \e_2^{\otimes 2} + \e_3^{\otimes 2} )_{ij} \\
    &= \tfrac13 \,  \sum_{\ell=0,2}\, \sum_{m=-\ell}^\ell f^{\ell m} \, (\Y^{\ell m}_{11} +\Y^{\ell m}_{22} + \Y^{\ell m}_{33} ) 
    = \tfrac13 \sum_{\ell=0,2}\, \sum_{m=-\ell}^\ell f^{\ell m} \Y^{\ell m}_{ij} \delta_{ij} 
    = f^{00} \,, 
\end{aligned}\end{equation}
fulfilling Eq.~(\ref{eq:SphDesign}). 
Here we used $\Y^{00}_{ij} = \delta_{ij}^{}$  that follows from Eq.~(\ref{Eq:Ylm_deltas}).

Likewise, for any degree-4 function $f^{(4)}(\n)$ with antipodal symmetry (involving $\ell=0$, $2$ and $4$ spherical harmonics),  the 4-design 
\begin{align} \nonumber
\{\n^\alpha\} & =\tfrac{1}{\sqrt{1+\varphi^2}}  \{\e_1 + \varphi \e_2,\, \e_2 + \varphi \e_3,\, \e_3 + \varphi \e_1,\, \e_1 - \varphi \e_2,\, \e_2 - \varphi \e_3,\, \e_3 - \varphi \e_1\} 
\\  & = 
\tfrac{1}{\sqrt{1+\varphi^2}} \{(1,\varphi,0),(0,1,\varphi),(\varphi,0,1),(1,-\varphi,0),(0,1,-\varphi),(-\varphi,0,1)\}
\nonumber
\end{align}
where  $\varphi = (1+\sqrt{5})/2$, yields
\begin{equation}
\begin{aligned} \nonumber
    \tfrac16 \sum_{\alpha=1}^6 f^{(4)}(\n^{\alpha}) &=  \tfrac16 \sum_{\ell,m}  \, f^{\ell m} \, \Y^{\ell m}_{ijkl} \,
     (n^{1}_i n^{1}_j n^{1}_k n^{1}_l
    + n^{2}_i n^{2}_j n^{2}_k n^{2}_l
    + n^{3}_i n^{3}_j n^{3}_k n^{3}_l
    + n^{4}_i n^{4}_j n^{4}_k n^{4}_l
    + n^{5}_i n^{5}_j n^{5}_k n^{5}_l
    + n^{6}_i n^{6}_j n^{6}_k n^{6}_l ) \\
%    &= \tfrac16 \, \sum_{\ell,m} f^{\ell m} \, \sum_{ijkl} \Y^{\ell m}_{ijkl} \, \tfrac65
%     \Big(\delta_{11}\delta_{11} + \delta_{22} \delta_{22} + \delta_{33}\delta_{33} + 2 \, (\delta_{(11}\delta_{22)} + \delta_{(11} \delta_{33)} + \delta_{(22}\delta_{33)}) \Big) \\
& =  \tfrac16 \, \sum_{\ell,m} f^{\ell m} \, \Y^{\ell m}_{ijkl} \,\tfrac1{(1+\varphi^2)^2} \Big( (\e_1 + \varphi \, \e_2)^{\otimes4} +(\e_2 + \varphi \, \e_3)^{\otimes4} + \hdots + (\e_3 - \varphi \, \e_1)^{\otimes4} \Big)_{ijkl} \\
& =  \tfrac16 \, \sum_{\ell,m} f^{\ell m} \, \Y^{\ell m}_{ijkl} \,\tfrac{6\varphi^2}{(1+\varphi^2)^2} \Big (\e_1^{\otimes 4} + \e_2^{\otimes 4} + \e_3^{\otimes 4}  + 2\, \e_1^{\otimes 2}\,\e_2^{\otimes 2}+ 2\, \e_2^{\otimes 2}\,\e_3^{\otimes 2}+ 2\, \e_3^{\otimes 2}\,\e_1^{\otimes 2} \Big )_{ijkl} \\
& =  \tfrac16 \, \sum_{\ell,m} f^{\ell m} \,\tfrac65 \Big(\Y^{\ell m}_{1111} + \Y^{\ell m}_{2222} + \Y^{\ell m}_{3333} + 2 \, (\Y^{\ell m}_{1122} + \Y^{\ell m}_{1133} + \Y^{\ell m}_{2233} \Big)  \\
 &= \tfrac15 \, \sum_{\ell,m} f^{\ell m}\, \Y^{\ell m}_{ijkl} \delta_{ij}\delta_{kl}
   = \tfrac15  \sum_{\ell,m} f^{\ell m} \Y^{\ell m}_{ijkl} \delta_{(ij}\delta_{kl)} 
   = f^{00} \,, 
\end{aligned}
\end{equation}
fulfilling Eq.~(\ref{eq:SphDesign}). 
Here we used 
$\varphi^2 = 1+\varphi$, $1+\varphi^2 = \sqrt{5}\, \varphi$, $1+\varphi^4 = 3\varphi^2$, as well as 
$\Y^{00}_{ijkl} = \delta_{(ij}\delta_{kl)}$ that follows from Eq.~(\ref{Eq:Ylm_deltas}), with $\zeta(4,0)=5$ in Eq.~(\ref{eq:alpha_STF}), cf. Eqs.~(\ref{eq:rank24_CART2STF}).

\newpage
\section{SA irreducible decomposition: all rotational invariants}\label{SM:SA_invariants}
\noindent
%RICE maps for SA irreducible decomposition.
\begin{figure*}[htbp]
	\centering
	\includegraphics[width=\textwidth]{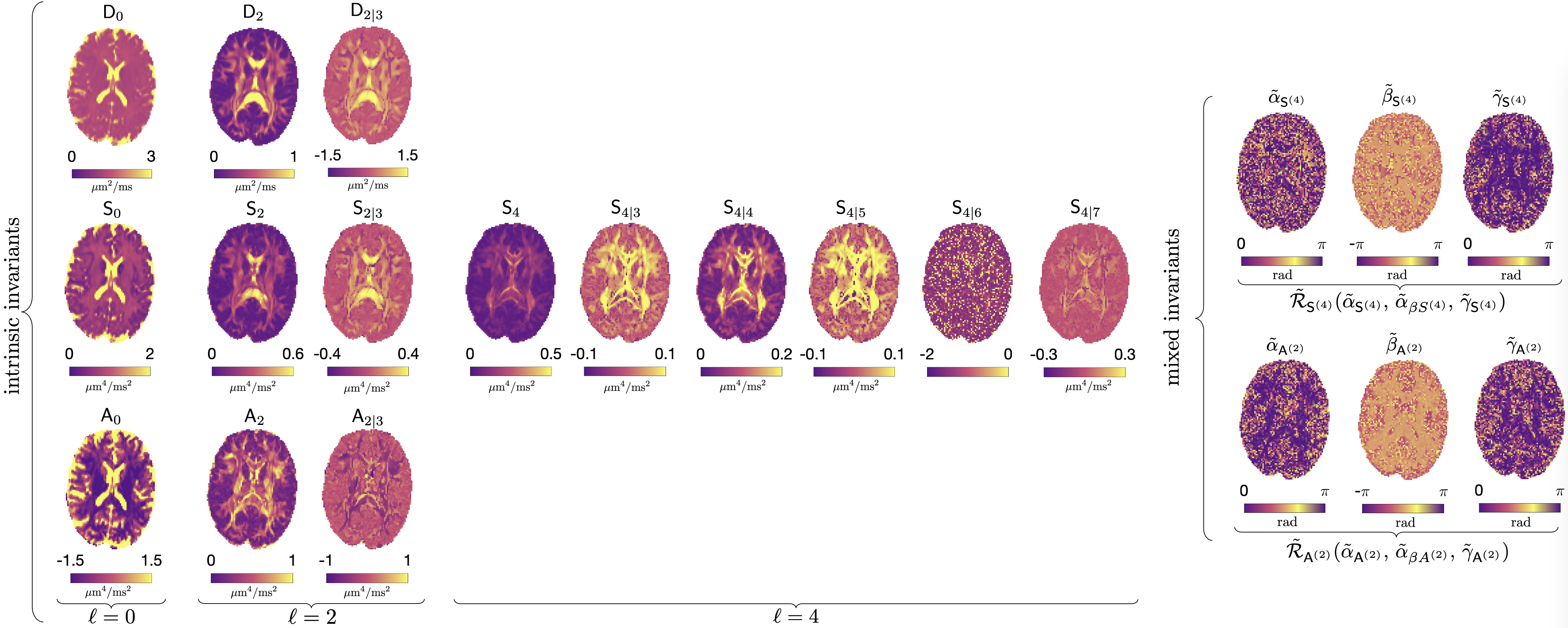}
	\caption[caption FIG]{RICE maps for a normal brain (33 y.o. male). Intrinsic invariants for each irreducible decomposition of $\D$, $\S$ and $\A$ are shown as powers of corresponding traces, to match units of $\D$ and $\C$ tensors.  
		Combinations of just 6 intrinsic invariants (out of $3+12$) generate all previously used model-independent dMRI contrasts, Eqs.~(\ref{FA})-(\ref{axsym}). The 6 mixed invariants correspond to Euler angles of eigenframes of $\Sf$ and $\At$ relative to that of $\St$ (see text). The underlying tissue microstructure introduces correlations between invariants. For example, small relative angles $\tilde\beta$ in white matter tracts exemplify the alignment of eigenframes of different representations of SO(3) with the tract.}\label{fig:maps_SA}
\end{figure*}

%\newpage
%\clearpage 
\section{Minimal protocols validation with magnitude denoising preprocessed DWI}\label{SM:MinimalDesigns}
\noindent
\begin{figure*}[th!!]
	\centering
	\includegraphics[width=\textwidth]{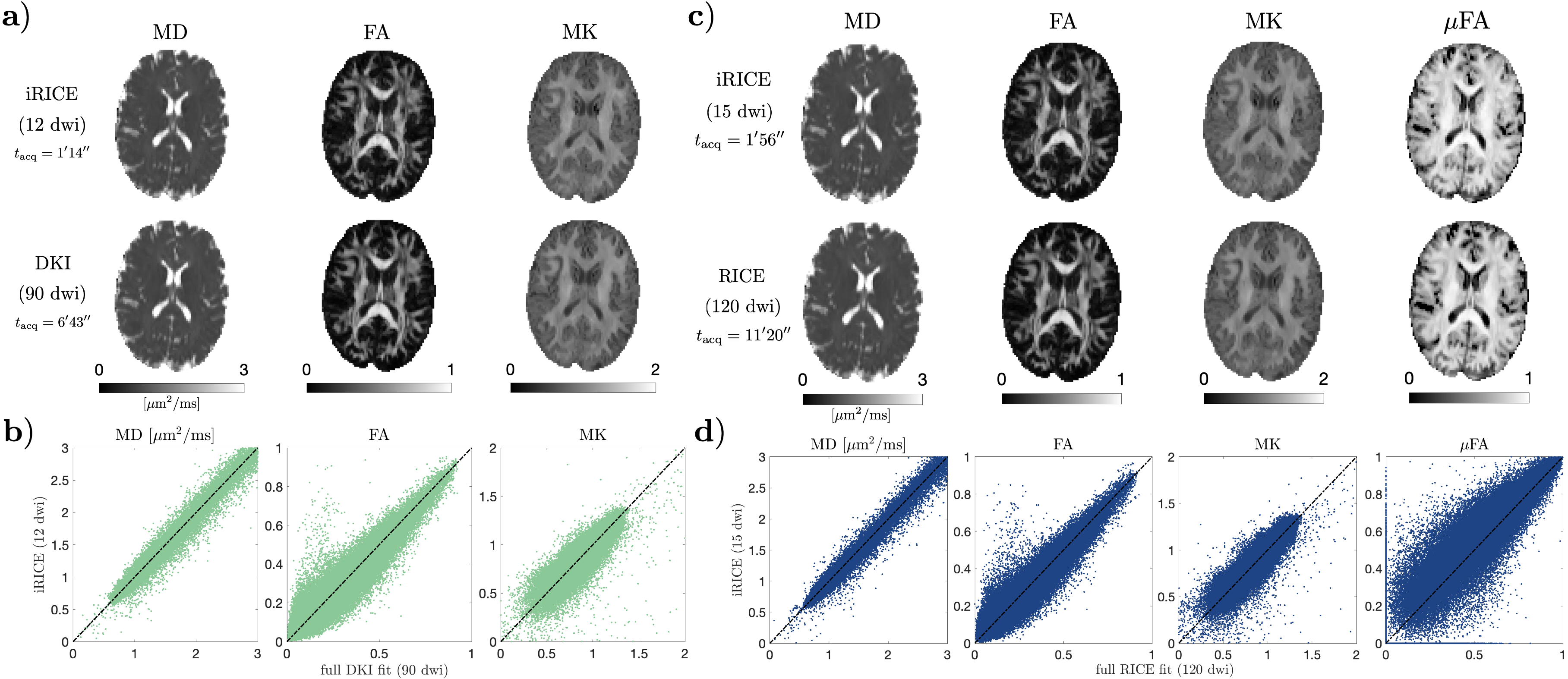}
	\caption[caption FIG]{Comparison of iRICE (1-2 minutes) with fully sampled acquisitions (6-11 minutes) for magnitude-denoised data. 
		(a,c): iRICE maps (top) vs fully sampled DKI maps (bottom) for a healthy volunteer. Panels (b) and (d) show scatter plots for all brain voxels in a normal volunteer.
	}\label{fig:fastRICEcomparison_magnMP}
\end{figure*}

\update{
\section{Derivation of time-dependent cumulants (\ref{eq:SH_DDE}) via DDE}\label{SM:DDE_WignerD}
}
\noindent
%The DDE log-signal $\mathcal{L}_{\mathrm{DDE}}(\g_1,\g_2)$ depends on the two gradient directions $\g_1$ and $\g_2$ separately. However, when considering a shell of measurements obtained by applying all global rotations to a gradient pair at a fixed angle $\psi=\arccos(\g_1\!\cdot\!\g_2)$, the resulting signals may be viewed as a scalar function on $\mathrm{SO}(3)$. Here $\R$ represents a global rotation of the DDE gradient pair $\{\g_1,\g_2\}$, so that $\mathcal{F}(\R)\equiv\mathcal{L}_\mathrm{DDE}(\R^{-1}\g_1,\R^{-1}\g_2)$ is a scalar function on $\mathrm{SO}(3)$. Projecting this $\mathrm{SO}(3)$-valued signal onto irreducible representations yields components whose \emph{scalar amplitudes} depend solely on $\psi$, while their angular structure reflects the transformation properties under global rotations.
%
%
Consider the logarithm  $\mathcal{L}_{\mathrm{DDE}}(\g_1,\g_2)$ of the DDE signal, that depends on the two unit gradient directions $\g_1$ and $\g_2$. A shell of DDE measurements (for a given set of scalar parameters, such as $b$-values and timings) is obtained by applying a sufficient number of  SO(3) rotations $\R$ to a gradient pair $\g_1$ and $\g_2$, keeping angle $\psi=\arccos(\g_1\!\cdot\!\g_2)$ between them fixed. On a shell, $\mathcal{L}_{\mathrm{DDE}}$  can be viewed as a scalar function $\mathcal{F}(\R)\equiv\mathcal{L}_\mathrm{DDE}(\R\g_1,\R\g_2)$ on the $\mathrm{SO}(3)$ group manifold $\mathbb{S}^3/\mathbb{Z}_2$, 
with the pair $\g_1$ and $\g_2$ setting the shell's ``origin" (initial orientation).  

The rationale behind Eq.~(\ref{eq:DDE_g1proj}) comes from the fact that for any function on a 2-dimensional sphere, 
\begin{equation} \label{Fg}
F(\g) =  \sum_{\ell=0}^\infty\sum_{m=-\ell}^\ell F^{\ell m}\, Y^{\ell m}(\g) \,, \quad \g \in \mathbb{S}^2 \,, 
\end{equation}
its spherical harmonics coefficients $F^{\ell m}$ can be found by lifting the integration from $\mathbb{S}^2$ to the SO(3) manifold:  
\begin{equation} \label{SH_from_SO3}
F^{\ell m} = (2\ell+1) \int_{\mathbb{S}^2} \! \d\g \, Y^{\ell m *}(\g)\, F(\g) \equiv (2\ell+1) \int_{\mathrm{SO}(3)}\! \d\R\, Y^{\ell m *}(\R\g) \, F(\R\g)  \,.
\end{equation}
To prove, we expand $F(\R\g)$, use Eqs.~(\ref{eq:wigner_ortho}) and (\ref{Y=DY}), and Uns\"old theorem 
$\sum_{m_1=-\ell}^\ell Y^{\ell m_1 *}(\g) Y^{\ell m_1}(\g) \equiv 1$ [cf. Eq.~(\ref{addition_th}) below]: 
\[
 (2\ell+1) \int\! \d\R\, Y^{\ell m *}(\R\g) Y^{\ell m'}(\R\g) 
 = 
 \sum_{m_1, m_2}
 Y^{\ell m_1 *}(\g) Y^{\ell m_2}(\g) \,  (2\ell+1) \int\! \d\R\, 
 \mathcal{D}^{\ell*}_{m_1 m}(\R^{-1})  \mathcal{D}^{\ell}_{m_2 m'}(\R^{-1}) 
% = \sum_{m_1} Y^{\ell m_1 *}(\g) Y^{\ell m_1}(\g) \,  \delta_{mm'} 
 =  \delta_{mm'} \,.
\]

The spherical harmonics (\ref{eq:SH_DDE}) can be derived directly by a projection (\ref{eq:DDE_g1proj}) in the spirit of Eq.~(\ref{SH_from_SO3}), while averaging over all SO(2) rotations of $\g_2$ around every $\g_1 \in \mathbb{S}^2$, exemplifying the Hopf mapping $\mathbb{S}^2 = \mathrm{SO(3)/SO(2)}.$\cite{Hopf1931}  
Here, we find it instructive to come to the solution by remaining on the SO(3) group manifold, developing the harmonic analysis there by employing the complete Fourier basis (\ref{eq:wigner_ortho}) of Wigner matrices, and then ``descending" onto $\mathbb{S}^2$ [cf. Eq.~(\ref{eq:DtoSH}) below]. 

For any $\mathcal{F}(\R)$ on SO(3), define the forward and inverse Fourier transforms as 
%Thus, taking the $\mathrm{SO}(3)$ Fourier transform yields components whose amplitudes that depend solely on the relative angle $\psi$. Because the $\mathrm{SO}(3)$ Fourier transform is invariant under a choice of reference orientation, we are free to fix a convenient frame for the gradient pair without loss of generality. This freedom is analogous to the arbitrary choice of origin in the Euclidean Fourier transform $e^{-ik(x-x_0)}$, which only affects the phase of Fourier coefficients. We exploit this invariance below when evaluating the Wigner coefficients.
%Any square-integrable function $\mathcal{F}(\R)$ with $\R\in\mathrm{SO}(3)$ admits the Wigner expansion
\begin{equation}
	\mathcal{F}(\R)=
	\sum_{\ell=0}^{\infty} \, \sum_{m,m'=-\ell}^{\ell}
	\mathcal{F}^{\ell}_{m m'}\,\mathcal{D}^{\ell *}_{m m'}(\R) 
	\quad \Leftrightarrow \quad
	\mathcal{F}^{\ell}_{m m'} =(2\ell+1) \int_{\mathrm{SO}(3)}\! \d \R\; \mathcal{D}^{\ell}_{m m'}(\R)\, \mathcal{F}(\R)\,.
	\label{eq:wigner_expansion}
\end{equation}
We adopt the convention that uses 
%Without the loss of generality, we choose the 
 ${\cal D}^{\ell *}_{mm'}$ for the SO(3) basis, and ${\cal D}^{\ell}_{mm'}$ for the forward  transform, given the  correspondence with the spherical harmonics expansion (\ref{Fg}) that involves the complex-conjugates (\ref{SH=Wigner}) of Wigner matrix elements. This convention is analogous to  $e^{-i\omega t}$ for the temporal Fourier expansion on $\mathbb{R}$, and is opposite to the phase $e^{+ikx}$ of the plane wave expansion.

%Therefore, we need to calculate To obtain Eq.~(\ref{eq:SH_DDE}), and generalize Eq.~(\ref{eq:DDE_g1proj}), 
Consider the forward SO(3) Fourier transform, with the ``shell origin" DDE pair $\g_{1}$ and $\g_2$ setting the ``initial phase": 
\begin{equation}\label{eq:DDE_Dproj}
	{\mathcal{L}^{\ell}_\mathrm{DDE}}_{mm'} =(2\ell+1)\!\int_\mathrm{SO(3)}\!\!\d\R\,\mathcal{D}^{\ell}_{mm'}(\R)\,\mathcal{L}_\mathrm{DDE}(\R\g_1,\R\g_2)\,.
\end{equation}
The terms of Eq.~(\ref{eq:DDE_cumExp}) of the form (\ref{Fg}) involving a single $\g = \g_1$ or $\g_2$, are straightforward:  on a shell, 
\begin{equation} \label{Fg_wigner}
F(\R\g) =  \sum_{\ell,m} F^{\ell m} \, Y^{\ell m} (\R\g) 
=  \sum_{\ell,m,m'} F^{\ell m}\, Y^{\ell m'}(\g)\, \mathcal{D}^{\ell }_{m'm} (\R^{-1})
=  \sum_{\ell,m,m'} F^{\ell m}\, Y^{\ell m'}(\g)\, \mathcal{D}^{\ell *}_{mm'} (\R) 
\quad \Rightarrow \quad F^{\ell}_{mm'} = F^{\ell m}\, Y^{\ell m'}(\g) \,,
\end{equation}
where we used Eq.~(\ref{Y=DY}) and  Eq.~(\ref{eq:wigner_expansion}). Here and below the sums over $m$ run between the corresponding $-\ell$ and $\ell$. 

The terms $\lla D(\R\g_1) D(\R\g_2)\rra = \sum \lla D^{\ell_1 m_1} D^{\ell_2 m_2}\rra Y^{\ell_1 m_1}(\R\g_1)\, Y^{\ell_2 m_2}(\R\g_2)$ depending on both DDE vectors involve 
\begin{equation}\label{eq:DDE_Dproj_aux3}
\begin{aligned}
	\lb Y^{\ell_1 m_1}(\R\g_1)\, Y^{\ell_2 m_2}(\R\g_2)\rb^\ell_{mm'} \! &=
	(2\ell+1)\!\int_\mathrm{SO(3)}\!\!\d\R\,\mathcal{D}^{\ell}_{mm'}(\R)\,Y^{\ell_1 m_1}(\R\g_1)Y^{\ell_2 m_2}(\R\g_2) 
	\\
&=	
	(2\ell+1)\!\int_\mathrm{SO(3)}\!\!\d\R\,\mathcal{D}^{\ell}_{mm'}(\R)\, 
	\sum_{m_1',m_2'}\!\!\mathcal{D}^{\ell_1}_{m_1'm_1}(\R^{-1})\mathcal{D}^{\ell_2}_{m_2'm_2}(\R^{-1}) \, 		Y^{\ell_1 m_1'}(\g_1)Y^{\ell_2 m_2'}(\g_2)
	\\
&=	
	(2\ell+1)\!\int_\mathrm{SO(3)}\!\!\d\R\,\mathcal{D}^{\ell}_{mm'}(\R)\, 
	\sum_{m_1',m_2'}\!\!\mathcal{D}^{\ell_1 *}_{m_1m_1'}(\R)\mathcal{D}^{\ell_2 *}_{m_2m_2'}(\R) \, 
	Y^{\ell_1 m_1'}(\g_1)Y^{\ell_2 m_2'}(\g_2)
	\\
&=		
	\braket{\ell_1 m_1 \ell_2 m_2| \ell m} \mathbb{Y}^{\ell m'}_{\ell_1 \ell_2}(\g_1,\g_2) \,, 
\end{aligned}
\end{equation}
where we used Eq.~(\ref{eq:DDD_int}) for a triple product of Wigner matrices and the fact that Clebsch-Gordan coefficients are real, and introduced the {\it bipolar spherical harmonics}\cite{VARSHALOVICH1988}, the states with definite $\ell$ and $m'$ that depend on two unit vectors: 
\begin{equation} \label{Ybipolar}
\mathbb{Y}^{\ell m'}_{\ell_1 \ell_2}(\g_1,\g_2)  \equiv	
\sum_{m_1'+m_2'=m'} \!\!\braket{\ell_1 m_1' \ell_2 m_2'|\ell m'}  Y^{\ell_1 m_1'}(\g_1)Y^{\ell_2 m_2'}(\g_2) \,.
\end{equation}
Note that for the coinciding arguments $\g_2=\g_1$, using Eq.~(\ref{YY=Y}), 
\[
\mathbb{Y}^{\ell m'}_{\ell_1 \ell_2}(\g,\g)  
%= \sum_{m_1',m_2'} \!\!\braket{\ell_1 m_1' \ell_2 m_2'|\ell m'}  Y^{\ell_1 m_1'}(\g_1)Y^{\ell_2 m_2'}(\g_1)
= \sum_{L,M}  \braket{\ell_1 0 \ell_2 0 | L 0}  \sum_{m_1',m_2'} \!\!\braket{\ell_1 m_1' \ell_2 m_2'|\ell m'} 
\braket{\ell_1 m_1' \ell_2 m_2' | L M} {Y^{LM}}(\g)
%=  \braket{\ell_1 0 \ell_2 0 | L 0} \delta_{L\ell} \delta_{Mm'} Y^{LM} (\g) 
=  \braket{\ell_1 0 \ell_2 0 | \ell 0} Y^{\ell m'} (\g) 
\]
since the sum over $m_1',m_2'$ spans all orthogonal states, %and acts as a resolution of unity, 
yielding $\sum_{m_1' m_2'} \braket{\ell_1 m_1' \ell_2 m_2'|\ell m'} 
\braket{\ell_1 m_1' \ell_2 m_2' | L M} = \delta_{L\ell} \delta_{Mm'}$.  
Therefore,  Eq.~(\ref{eq:DDE_Dproj_aux3}) for the coinciding arguments equals % yields - results in  
\begin{equation} \label{eq:DDE_Dproj_aux4}
\lb Y^{\ell_1 m_1}(\R\g)\, Y^{\ell_2 m_2}(\R\g)\rb^\ell_{mm'}
= \braket{\ell_1 0 \ell_2 0 | \ell 0} \braket{\ell_1 m_1 \ell_2 m_2| \ell m} Y^{\ell m'} (\g) \,.
\end{equation}

We can now use Eqs.~(\ref{Fg_wigner})--(\ref{eq:DDE_Dproj_aux4}) to compute the Wigner coefficients (\ref{eq:DDE_Dproj}): 
\begin{equation} \label{LDDE_Wigner}
\begin{aligned}
	{\mathcal{L}^{\ell}_\mathrm{DDE}}_{mm'} &= -b_1 \D^{\ell m}(t) Y^{\ell m'}(\g_1)-b_2 \D^{\ell m}(t) Y^{\ell m'}(\g_2) +\tfrac12{b_1^2}\, \Smu^{\ell m}(t) Y^{\ell m'}(\g_1) +\tfrac12{b_2^2} \, \Smu^{\ell m}(t) Y^{\ell m'}(\g_2) \\
	& + \tfrac12 \sum_{\ell_1,\ell_2,m_1 ,m_2}\!\!\! \lla \Dc^{\ell_1 m_1}(t) \Dc^{\ell_2 m_2}(t)\rra  \braket{\ell_1 m_1 \ell_2 m_2 |\ell m}\left[  \braket{\ell_1 0 \ell_2 0 |\ell 0} \left( {b_1^2}\, Y^{\ell m'}(\g_1) + {b_2^2}\, Y^{\ell m'}(\g_2) \right)+  2b_1 b_2\, \mathbb{Y}^{\ell m'}_{\ell_1 \ell_2}(\g_1,\g_2) \right] .
\end{aligned}
\end{equation}
Generalizing the QT-decomposition (\ref{QTLM}) onto $t$-dependent $\lla \Dc^{\ell_1 m_1}(t) \Dc^{\ell_2 m_2}(t)\rra$, Eq.~(\ref{LDDE_Wigner}) yields the SO(3) Fourier components:  
\begin{equation}\label{eq:W_DDE}
\begin{aligned}
	{\mathcal{L}^{0}_\mathrm{DDE}}_{00}&=-\left(b_1+b_2\right)\D^{00}(t)  + \tfrac{1}{2}\left(b_1+b_2\right)^2\Q^{00}(t) +\tfrac12 \left(b_1^2+b_2^2+2p_2\, b_1b_2   \right)\T^{00}(t)  +\tfrac12 \left(b_1^2+b_2^2\right)\Smu^{00}(t)\,,
	\\
	{\mathcal{L}^{2}_\mathrm{DDE}}_{mm'}&=-\D^{2m}(t) \left(b_1 Y^{2m'}(\g_1) +b_2 Y^{2m'}(\g_2) \right) 
	+ \tfrac12\,  \Q^{2m}(t)\left[ b_1^2 Y^{2m'}(\g_1) + b_2^2 Y^{2m'}(\g_2)+ b_1 b_2  \left( Y^{2m'}(\g_1) + Y^{2m'}(\g_2) \right) \right] \\
	&+ \tfrac12\, \T^{2m}(t) \left(b_1^2 Y^{2m'}(\g_1) + b_2^2 Y^{2m'}(\g_2)+ 2b_1 b_2  \frac{\mathbb{Y}^{2m'}_{22}(\g_1,\g_2)}{\braket{2020|20}} \right)   + \tfrac12\, \Smu^{2m}(t) \left( b_1^2\, Y^{2m'}(\g_1) + b_2^2\, Y^{2m'}(\g_2) \right) \,,\\
	{\mathcal{L}^{4}_\mathrm{DDE}}_{mm'}&=
	\,\, \tfrac12\,\T^{4m}(t) \left(b_1^2 Y^{4m'}(\g_1) + b_2^2 Y^{4m'}(\g_2)+ 2b_1 b_2 \frac{\mathbb{Y}^{4m'}_{22}(\g_1,\g_2)}{\braket{2020|40}} \right) 
	+ \tfrac12 \,\Smu^{4m}(t) \left( b_1^2\, Y^{4m'}(\g_1) + b_2^2\, Y^{4m'}(\g_2) \right) . 
\end{aligned}
\end{equation}
Here, for $\ell=0$, we used $\mathbb{Y}^{00}_{00}(\g_1,\g_2) = \braket{0000|00} = 1$,  
as well as (setting $\ell' = 2$ in the $\sim b_1 b_2$ term) 
\begin{equation} \label{Ybipolar0}
\mathbb{Y}^{0 0}_{\ell' \ell'}(\g_1,\g_2)  
=\sum_{m} \!\!\braket{\ell' m \ell'  (-m) | 0 0}  Y^{\ell' m}(\g_1)Y^{\ell',-m}(\g_2) 
=\braket{\ell' 0 \ell' 0|0 0} \sum_{m} (-1)^m  Y^{\ell' m}(\g_1)Y^{\ell', -m}(\g_2) 
= \braket{\ell' 0 \ell' 0|0 0} p_{\ell'}\,,
\end{equation}
coming from $\braket{\ell m \ell (-m) | 0 0}\! =\! (-1)^{m} \braket{\ell 0 \ell 0| 0 0}$, cf. Ref. \cite{VARSHALOVICH1988}, Eqs.~(\ref{condon-shortley}) and (\ref{eq:T00}), and  the  {\it addition theorem} in Racah normalization: 
%that comes from $\braket{\ell m \ell (-m) | 0 0} = (-1)^{m} \braket{\ell 0 \ell 0| 0 0}$, cf. Ref. \cite{VARSHALOVICH1988}, Eqs.~(\ref{condon-shortley}) and (\ref{eq:T00}), and  the  {\it addition theorem} in the Racah normalization: 
\begin{equation} \label{addition_th}
\sum_m Y^{\ell m *}(\g_1) Y^{\ell m}(\g_2) = P_\ell(\g_1\!\cdot\!\g_2) \equiv p_\ell (\psi) \,. 
\end{equation}
For $\ell=2$, in the $\Q^{2m}(t)$ part
we used $\mathbb{Y}^{\ell m'}_{\ell 0}(\g_1,\g_2) = Y^{\ell m'} (\g_1)$ and $\mathbb{Y}^{\ell m'}_{0 \ell}(\g_1,\g_2) = Y^{\ell m'} (\g_2)$. 

The SO(3) Fourier coefficients (\ref{eq:W_DDE}) are general yet  {\it redundant}. 
All tensor components are contained in harmonics (\ref{eq:SH_DDE}) on $\mathbb{S}^2$: 
\begin{equation}\label{eq:DtoSH}
\mathcal{L}_\mathrm{DDE}^{\ell m}(\g_1,\g_2)
=(2\ell+1)\!\int_\mathrm{SO(3)}\!\!\!\d\R\, Y^{\ell m*}(\R\g_1)\, \sum_{m_1,m_2}{\mathcal{L}^{\ell}_\mathrm{DDE}}_{m_1 m_2} \mathcal{D}^{\ell *}_{m_1 m_2}(\R) 
%=\sum_{m',m_1,m_2}\! Y^{\ell m'*}(\g_1)\, (2\ell+1)\!\int_\mathrm{SO(3)}\!\!\!\d\R\,  \mathcal{D}^{\ell }_{mm'}(\R) {\mathcal{L}^{\ell}_\mathrm{DDE}}_{m_1 m_2} \mathcal{D}^{\ell *}_{m_1 m_2}(\R) 
=  \sum_{m'} {\mathcal{L}^{\ell}_\mathrm{DDE}}_{mm'}(\g_1,\g_2) \, Y^{\ell m'\,*}(\g_1) \,, 
\end{equation}
where we rotated $Y^{\ell m*}(\R\g_1) = \sum_{m'} Y^{\ell m'*}(\g_1)  \mathcal{D}^{\ell }_{mm'}(\R)$ via Eq.~(\ref{Y=DY}) and used the unitarity of $\mathcal{D}$. 
The sums in (\ref{eq:DtoSH}) with single $Y^{\ell m'}$ in Eq.~(\ref{eq:W_DDE}) are evaluated via the addition theorem (\ref{addition_th}). 
The sums involving bipolar harmonics can be found by noting that both $Y^{\ell m'}(\g_1)$ and $\mathbb{Y}^{\ell m'}_{\ell_1 \ell_2}(\g_1,\g_2)$ 
transform according to the same $\ell$-irreducible representation of SO(3)\cite{VARSHALOVICH1988}. 
Hence, their contraction (\ref{eq:DtoSH}) is a Hermitian inner product  in the unitary representation space, invariant under global rotations $\R$.  
Thus, $\sum_{m'} \mathbb{Y}^{\ell m'}_{\ell_1\ell_2}(\R\g_1,\R\g_2)  Y^{\ell m' *}(\R\g_1) = \sum_{m'} \mathbb{Y}^{\ell m'}_{\ell_1\ell_2}(\g_1,\g_2)  Y^{\ell m' *}(\g_1)$ 
%\begin{equation}\label{eq:bipolarYSO2avg}
%	\begin{aligned}
%%	\sum_m Y^{\ell m}(\g_1)Y^{\ell m *}(\g_2) & = 	\sum_m Y^{\ell m}(\R\g_1)Y^{\ell m *}(\R\g_2) = p_\ell\,,\\
%%	\sum_{m} \mathbb{Y}^{\ell m}_{\ell_1\ell2}(\g_1,\g_2)  Y^{\ell m}(\g_1) & = \sum_{m} \mathbb{Y}^{\ell m}_{\ell_1\ell2}(\R\g_1,\R\g_2)  Y^{\ell m}(\R\g_1)\,,
%	\sum_m Y^{\ell m}(\g_1)Y^{\ell m *}(\g_2)  = 	\sum_m Y^{\ell m}(\R\g_1)Y^{\ell m *}(\R\g_2) \,,\quad\sum_{m} \mathbb{Y}^{\ell m}_{\ell_1\ell2}(\g_1,\g_2)  Y^{\ell m *}(\g_1)  = \sum_{m} \mathbb{Y}^{\ell m}_{\ell_1\ell2}(\R\g_1,\R\g_2)  Y^{\ell m *}(\R\g_1)\,,
%	\end{aligned}
%\end{equation}
is independent of the global rotation $\R$. 
%\dn{probably possible not to choose basis and use triple products}
Therefore, we select the frame where $\g_1=\z$ and $\g_2= \R_{21}(\psi)\z$, such that $Y^{\ell m'}(\g_1) = \delta_{m' 0}$, $Y^{\ell 0}(\g_2)= p_\ell$, and the only relevant bipolar harmonics 
\begin{equation}\label{eq:bipolarProj}
%	\begin{aligned}
		\sum_{m'} \mathbb{Y}^{\ell m'}_{\ell_1\ell_2}(\g_1,\g_2)  Y^{\ell m' *}(\g_1) 
		= \mathbb{Y}^{L0}_{\ell_1\ell_2}(\g_1,\g_2) 
		%\sum_{m_1,m_2} \braket{\ell_1 m_1 \ell_2 m_2|L0}  Y^{\ell_1 m_1}(\z)Y^{\ell_2 m_2}(\R_y(\psi)\z)  
		= \sum_{m} \braket{\ell_1 m \ell_2 (-m) |L0}  \delta_{m0} Y^{\ell_2, -m}(\g_2)  
		=   \braket{\ell_1 0 \ell_2 0 |L0} Y^{\ell_2 0}(\g_2)=   \braket{\ell_1 0 \ell_2 0 |L0} p_{\ell_2}(\psi)\,.
%	\end{aligned}
\end{equation}
In other words, the result of summation (\ref{eq:DtoSH}) is equivalent to ${\mathcal{L}^{\ell m}_\mathrm{DDE}} = {\mathcal{L}^{\ell}_\mathrm{DDE}}_{m0}$ when $\g_1=\z$, which yields  Eq.~(\ref{eq:SH_DDE}). Alternatively,  Eq.~(\ref{eq:bipolarProj}) can be derived introducing an auxiliary rotation $\R_1$, such that $\g_1=\R_1\z$, without fixing a particular reference frame. Using Eqs.~(\ref{Y=DY})--(\ref{SH=Wigner}), the product $YY$ can be reduced to $YY\rightarrow\mathcal{D}\mathcal{D}\rightarrow\mathcal{D}$. For brevity, this derivation is omitted.
%

%====================================================
%========END OF SUPPLEMENTARY INFORMATION============
%====================================================

\end{document}